\documentclass[11pt]{article}
\usepackage{amsmath,amssymb,amsthm}
\usepackage{array}
\usepackage{fullpage}
\usepackage[colorlinks]{hyperref}
\hypersetup{linkcolor=cyan,filecolor=cyan,citecolor=cyan,urlcolor=cyan,hypertexnames=false}
\usepackage{xspace}
\usepackage{thm-restate,color,xcolor}
\usepackage{boxedminipage}
\usepackage[boxed]{algorithm}
\usepackage{epigraph}
\usepackage{amsmath}
\usepackage{mathtools}
\usepackage{framed}
\usepackage[framemethod=tikz]{mdframed}
\usepackage{titlesec}
\usepackage{lipsum}%
\usepackage{cleveref,aliascnt}
\usepackage{tikz}
\usepackage{graphicx}
\usepackage{float}
\usepackage{enumitem}

\usepackage{wrapfig}
\usepackage{framed}
\usepackage[framemethod=tikz]{mdframed} 

\usetikzlibrary{shapes.geometric}
\usetikzlibrary{arrows}
\usetikzlibrary{arrows.meta}
\usetikzlibrary{patterns}
\usetikzlibrary{shapes.misc}
\usepackage{algorithm,algpseudocode}
\algrenewcommand{\algorithmicrequire}{\textbf{Input:}}

\newcommand{\Enc}{\operatorname{Enc}}
\newcommand{\Sketch}{\operatorname{Sketch}}

\theoremstyle{plain}
\makeatletter
\@ifundefined{newcounteralias}{}{%
  \renewcommand\thmt@autorefsetup{%
    \@xa\def\csname\thmt@envname autorefname\@xa\endcsname
      \@xa{\thmt@thmname}%
  }%
}
\makeatother

\newtheorem{theorem}{Theorem}[section]
\newtheorem{corollary}{Corollary}[section]
\newcommand{\BTHM}{\begin{theorem}} \newcommand{\ETHM}{\end{theorem}}
\newtheorem{lemma}{Lemma}[section]

\newtheorem{remark}{Remark}[section]
\newtheorem{observation}{Observation}[section]
\theoremstyle{definition}
\newtheorem{definition}{Definition}[section]
\newtheorem{question}{Question}[section]

\theoremstyle{plain}
\newtheorem{claim}[theorem]{Claim}

\Crefname{claim}{Claim}{Claims}

\Crefname{observation}{Observation}{Observations}

\theoremstyle{definition}

\newenvironment{rem}{\noindent{\bf Remark:~~}}{}
\newcommand{\BREM}{\begin{rem}} \newcommand{\EREM}{\end{rem}}
\newenvironment{discussion}{\noindent{\bf Discussion:~~\\}}{}
\newcommand{\BDIS}{\begin{discussion}} \newcommand{\EDIS}{\end{discussion}}

\numberwithin{equation}{section}

\usepackage{wrapfig}
\usepackage{xcolor}

\def\blackslug
     {\hbox{\hskip 1pt\vrule width 8pt height 8pt depth 1.5pt\hskip 1pt}}
\def\qed{\quad\blackslug\lower 8.5pt\null\par}

\newcommand{\poly}{{\rm poly}}

\newtheorem{exmp}[theorem]{Example}
\newtheorem{fact}{Fact}[section]
\newcommand{\BEX}{\begin{exmp}} \newcommand{\EEX}{\end{exmp}}
\newcommand{\BF}{\begin{fact}}   \newcommand{\EF}{\end{fact}}
\newcommand{\Bcr}{\begin{techcorr}}
\newcommand{\Ecr}{\end{techcorr}}
\newcommand{\BDS}{\begin{description}}
\newcommand{\EDS}{\end{description}}
\newcommand{\BE}{\begin{enumerate}}
\newcommand{\EE}{\end{enumerate}}
\newcommand{\BI}{\begin{itemize}}
\newcommand{\EI}{\end{itemize}}
\newcommand{\BPF}{\begin{proof}}
\newcommand{\EPF}{\end{proof}}

\newcommand{\BB}{\begin{enumerate}}
\newcommand{\EB}{\end{enumerate}}

\usepackage{geometry}
\title{Deterministic Fault-Tolerant Connectivity Oracles \\ with Subquadratic Space}

\author{
    Merav Parter \thanks{Weizmann Institute, Israel. Email: \texttt{merav.parter@weizmann.ac.il}. Supported by the European Research Council (ERC) under the European Union’s Horizon 2020 research and
    innovation programme, grant agreement No. 949083.}
    \and
    Elad Tzalik\thanks{Weizmann Institute, Israel. Email: \texttt{elad.tzalik@weizmann.ac.il}.
    }
    }
\date{}

\begin{document}
\maketitle

\begin{abstract}
A fault-tolerant (FT) connectivity oracle for an $n$-vertex graph $G=(V,E)$ determines whether $G- F$ is connected for any query set $F\subseteq E$. When the number of failures is unrestricted, prior constructions exhibited a sharp contrast between randomization and determinism: randomized oracles use only $\widetilde{O}(n)$ space [Duan and Pettie, SICOMP 2020], whereas no deterministic construction was known to improve upon storing $G$ explicitly, requiring $\Theta(n^2)$ bits in the worst case.

The promise that $F\subseteq E$ is essential for subquadratic space: allowing arbitrary queries $F\subseteq V\times V$ forces quadratic space. For queries containing at most $f$ failures, Khanna, Konrad, and Putterman [SODA 2027] recently constructed deterministic oracles using $\widetilde{O}(\sqrt{f}n)$ space that also report approximate distances with polylogarithmic stretch. Their result improves upon the classical $O(fn)$ space bound for deterministic connectivity oracles [P{\u{a}}tra{\c{s}}cu and Thorup, FOCS 2007], but offers no space savings over storing $G$ when $f=\Theta(n^2)$.

We present the first deterministic connectivity oracles with subquadratic space for failure sets of unrestricted size, using $\widetilde{O}(n^{7/4})$ space. We also establish the first polynomial separation between deterministic and randomized connectivity oracles, proving that every deterministic oracle requires $\widetilde{\Omega}(n^{3/2})$ bits in the worst case. For queries containing at most $f$ failures, we obtain deterministic connectivity oracles using $\widetilde{O}(f^{2/5}n)$ space, improving the dependence on $f$ in the bound of Khanna et al., and prove a lower bound of $\widetilde{\Omega}(f^{1/4}n)$ bits.
\end{abstract}

\hypersetup{pageanchor=false}
\pagenumbering{gobble}

\newpage
\setcounter{tocdepth}{2}
{\small\tableofcontents}
\newpage

\hypersetup{pageanchor=true}
\pagenumbering{arabic}

\section{Introduction}
\label{sec:introduction}

A fault-tolerant connectivity oracle compresses a graph while storing
enough information to determine its connectivity after edge failures.
Let $G=(V,E)$ be a simple undirected graph with $n$ vertices and $m$ edges. 
Given a failure set $F\subseteq E$, a \emph{connectivity oracle}
determines whether $G-F$ is connected (and possibly also reports the corresponding connected components). In this paper, we focus on the \emph{space}
complexity of such oracles: how many bits of information
must be stored to answer every valid failure query? Here, a \emph{valid} query supplies
the failed edges by their endpoints, with the promise that they belong
to $G$. In particular, it does not supply any additional
labels attached to the failed edges. We write $f$ for
an upper bound on $|F|$.

\smallskip \noindent \textbf{Beyond sparse subgraphs.}
A natural approach is to store a subgraph that preserves connectivity
under every set of at most $f$ edge failures. Classical sparse
certificates, due to Nagamochi and Ibaraki~\cite{NagamochiI92}, use
$O(fn)$ edges and directly give a deterministic oracle using
$\widetilde O(fn)$ bits. The $fn$ dependence is unavoidable for
subgraph certificates when $1\le f\le n$. Indeed, every edge incident
to a vertex of degree at most $f+1$ must be stored: otherwise, failing
the at most $f$ stored incident edges isolates that vertex in the
certificate while leaving a surviving incident edge in the original
graph. Applying this argument to graphs of degree $\Theta(f)$ gives
an $\Omega(fn)$-edge lower bound.

The promise $F\subseteq E$ is the key for breaking the $fn$ space barrier. If a query may name
arbitrary vertex pairs, including nonedges of $G$, then even randomized
oracles require $\Omega(\min\{f,n\}n)$ bits; see Khanna, Konrad, and
Putterman~\cite[Claim~B.1]{KKP206}.\footnote{While the proof of \cite{KKP206}
is stated for distance oracles, the same approach holds for connectivity oracles that only report whether $G-F$ is connected. This can be obtained by adding a universal vertex to their construction
and including its edge to the tested vertex in the failure query. All
other vertices remain connected, so the answer still tests the presence
of a single edge.} The reason is that,
in a graph family with a bounded set of candidate neighbors per vertex,
a query can delete all candidate incident edges except one, thereby
testing whether that remaining edge belongs to the graph.
Thus unrestricted queries can force an oracle to encode the graph
itself. Overall, a smaller oracle must exploit both the freedom to store
information other than a subgraph and the promise that every queried
edge is present.

Recently, \cite{KKP206} broke the $fn$ barrier with a deterministic
oracle using $\widetilde O(\sqrt f\,n)$ bits. Their oracle also reports
approximate distances with polylogarithmic stretch. Their analysis
rederives a related robustness guarantee for expanders with large
minimum degree, previously studied by Bodwin, Haeupler, and
Parter~\cite{BodwinHP24} under bounded faulty degree. In the bounded faulty degree setting,
the restriction on $F$ is on the degree of $F$, namely, the largest number of failed edges incident to a
vertex. Their subgraph certificates can handle $\Theta(fn)$ faults (with degree at most $f$) using $\tilde{O}(fn)$ edges. The construction
of~\cite{KKP206} instead handles an arbitrary set of $f$ failures.
Its space bound is subquadratic for $f\le n^{2-\varepsilon}$, for any
constant $\varepsilon>0$, but gives no savings over storing the adjacency
matrix when $f=\Theta(n^2)$.

Handling arbitrary failures also goes beyond preserving \emph{minimum cuts}.
For a connected graph of edge connectivity $\lambda$, queries with
$|F|=\lambda$ admit a deterministic $\widetilde O(n)$-bit 
oracle. Indeed, Lo, Schmidt, and Thorup~\cite{LoST19} give a graph contraction 
that has $O(n)$ edges and preserves every nontrivial minimum cut (namely, non-singleton cuts);
see also the expander-based contraction of
Saranurak~\cite{Saranurak20}. Storing the contraction map, the edges of
the contracted graph, and the vertex degrees suffices: the degrees detect singleton
cuts, and the contracted graph detects the remaining minimum cuts.
Our challenge is to handle failure sets of arbitrary size, which may
expose any of the exponentially many cuts of the graph. This leads to
our first key question.
\vspace{-5pt}
\begin{question}\label{q:subquadratic}
Is there a deterministic connectivity oracle with subquadratic space
for any number of edge failures?
\end{question}
\vspace{-5pt}
The challenge is to build an oracle that is smaller than the queries
it must answer: a single failure set may contain $\Theta(n^2)$ edges.
At first glance, such large queries seem to demand a comparably large
representation of the graph. The related problem of cut approximation
reinforces this intuition: given a subset $S\subseteq V$, approximating
the number of edges leaving $S$ even to within a $(1 \pm 1/\sqrt{n})$ multiplicative factor requires $\Theta(n^2)$ bits of space \cite{AndoniCKQWZ16}. 

\smallskip \noindent \textbf{The role of randomization.}
Randomization yields much smaller representations through the linear
graph sketches of Ahn, Guha, and McGregor~\cite{AhnGM12}.
Each vertex sketches its edge-incidence vector. Summing these
sketches over a set $S$ cancels internal edges, and subtracting the
contributions of failed edges (that must be in $G$) leaves a sketch of the edges leaving
$S$ in $G-F$. If this cut is nonempty, $\ell_0$-sampling recovers a
surviving outgoing edge with high probability. Repeatedly merging
components along recovered edges gives a Bor\r{u}vka-style procedure
for constructing a spanning forest.

This sketching approach underlies small randomized representations
in connectivity oracles, fault-tolerant labeling, and distributed
graph sketching. In particular, Duan and Pettie~\cite{DuanP20} give
a Monte Carlo connectivity oracle using $\widetilde O(n)$ bits,
independently of the number of failed edges. In the oracle and
labeling settings, the randomized guarantee is \emph{per query}:
each query fixed independently of the random choices is answered
correctly with probability $1-n^{-\Omega(1)}$. The guarantee does
not require a single random encoding to answer all queries correctly
simultaneously. Whether deterministic oracles can match this space
bound is less clear: two closely related models suggest different
answers.

Fault-tolerant labeling suggests that randomness might be avoidable.
In this model, each vertex and edge receives a label, and a decoder
determines connectivity in $G-F$ from the labels of the two queried
vertices and the labels of the failed edges.
Dory and Parter~\cite{DoryP21} used linear graph sketches to obtain
randomized labels of $O(\log^3 n)$ bits, independent of $f$.
A line of deterministic constructions, including those of
Izumi, Emek, Wadayama, and Masuzawa~\cite{IzumiEWM23} and
Long, Pettie, and Saranurak~\cite{LongPS25}, yields a polynomial
dependence on $f$. The latter achieved the previous best bound of
$\widetilde O(\sqrt f)$ bits per label, echoing the $\sqrt f$
dependence in the oracle bound of~\cite{KKP206}.
Very recently, Long, Pettie, and Saranurak~\cite{LongPS26} removed
this dependence entirely: deterministic labels of $O(\log^2 n)$
bits suffice for an arbitrary number of edge failures.
Thus polylogarithmic labels do not require randomness.
This does not immediately settle the oracle question, however:
the labeling decoder receives a label for each failed edge,
whereas an oracle receives only its endpoints.
Storing all vertex and edge labels would use
$\widetilde O(n+m)$ bits, giving no subquadratic bound for dense graphs.

Distributed graph sketching points in the other direction.
Here each vertex sends one message, computed from its neighborhood
and any shared randomness, to a referee who must determine
connectivity from these messages alone.
The sketches of~\cite{AhnGM12} use $O(\log^3 n)$ bits per vertex
and recover a spanning forest with high probability.
Nelson and Yu~\cite{NelsonYu19} and Yu~\cite{Yu21} proved matching
lower bounds for spanning-forest construction and connectivity,
respectively, establishing optimality for inverse-polynomial
failure probability.
In contrast, Robinson and Tan~\cite{RobinsonTan2026} proved an
$\Omega(n^{1/3})$ worst-case message-length lower bound for
deterministic connectivity protocols.
This separation does not directly imply an oracle lower bound:
an oracle's preprocessing algorithm sees the entire graph,
whereas each distributed message depends only on one vertex's
neighborhood. These contrasting results motivate the following question.
\vspace{-5pt}
\begin{question}\label{q:randomization}
Is there a polynomial separation between the space required by
randomized and deterministic connectivity oracles?
\end{question}
\vspace{-5pt}
We focus on the space required to store the encoding. Decoding
takes exponential time in $n$ but uses working space of the
same asymptotic order as the oracle size.

\smallskip
\noindent \textbf{Our Results.}
We answer both questions affirmatively: our oracles are deterministic and also report the connected components in $G-F$ for a given query $F \subseteq E(G)$.
In Section~\ref{sec:upper-bounds},
we show that deterministic connectivity oracles admit subquadratic space of $\tilde{O}(n^{7/4})$
even when the number of failures is unrestricted. For bounded failure
sets, we improve the dependence on $f$ from $f^{1/2}$ to $f^{2/5}$.  Our key observation is that a failure query $F$ provides
\emph{information} about the graph: the promise $F\subseteq E$
certifies the presence of every edge in $F$. This point of view treats the query as an \emph{opportunity} to learn something about the graph. In contrast,
a cut query specifies only a subset $S$ of the known vertex
set, whose labels can already be stored in $O(n\log n)$ bits.
It reveals nothing about the graph's edges. We show:

\begin{theorem}\label{thm:intro-upper}
For every $n\ge2$ and integer $1\le f\le\binom n2$, every simple
undirected $n$-vertex graph $G$ admits a deterministic connectivity oracle using
\[
  \widetilde O\!\left(\min\{f^{2/5},n^{3/4}\}n\right)
\]
bits that reports the connected-component partition of $G-F$ for every
$F\subseteq E(G)$ with $|F|\le f$. In particular, there is a
deterministic oracle using $\widetilde O(n^{7/4})$ bits that supports
failure sets of arbitrary size. Obtaining the oracle in polynomial time introduces an extra factor of $n^{o(1)}$ in the space bounds.
\end{theorem}
This addresses Question~\ref{q:subquadratic} and should be compared with the state-of-the-art $\tilde{O}(\sqrt{f}n)$-space bound by \cite{KKP206}. Turning to lower bounds, in Section~\ref{sec:lower-bound}, we show that the near-linear
randomized space bound \emph{cannot} be achieved deterministically.

\begin{theorem}\label{thm:intro-lower}
Every deterministic connectivity oracle for simple
undirected $n$-vertex graphs requires $\widetilde\Omega(n^{3/2})$ bits in the worst case. For at most $f$ failures, the lower bound becomes  $\widetilde\Omega(f^{1/4}n)$. 
\end{theorem}

Together with the $\widetilde O(n)$-bit randomized oracle
of~\cite{DuanP20}, this establishes a polynomial separation and answers
Question~\ref{q:randomization}. Thus deterministic connectivity
oracles admit subquadratic representations, but require polynomially
more space than randomized ones, even though both kinds of
fault-tolerant labeling schemes have polylogarithmic label length. 

In Section~\ref{sec:fast-query}, we show how to obtain fast queries
at the $\widetilde O(\sqrt f\,n)$-bit
space bound of~\cite{KKP206}. 

\begin{theorem}\label{thm:fast}
There is a deterministic connectivity oracle using
$\widetilde O(\sqrt f\,n)$ bits and polynomial preprocessing time
that processes any $F\subseteq E(G)$, $|F|\le f$, in
$\widetilde O(f)$ time, after which component identifiers can be queried in $\widetilde O(1)$ time.

For positive integer edge weights bounded by $n^{O(1)}$, there is
also a deterministic distance oracle using $\sqrt f\,n^{1+o(1)}$
bits and $(m+n)^{1+o(1)}$ preprocessing time. Given $(s,t,F)$ and
the lengths of the failed edges, it returns a
$\operatorname{polylog} n$ approximation to $d_{G-F}(s,t)$ in
$fn^{o(1)}$ time.
\end{theorem}
For connectivity, this improves the $O(fn)$-word space bound obtained
by applying the oracle of P\u{a}tra\c{s}cu and
Thorup~\cite{PatrascuT07}. It also improves the
polynomial-in-$n$ query time stated by Khanna, Konrad, and
Putterman~\cite{KKP206}, while matching their
$\widetilde O(\sqrt f\,n)$-bit space bound.
The key is to apply the $O(m)$-space oracle of~\cite{PatrascuT07}
to the sparse edge set kept by~\cite{KKP206}, augmented with
auxiliary stars representing the expander pieces. Fast deterministic
sketches recover the surviving edges needed to repair connectivity.
Each failure set induces only $O(f)$ deletions and insertions in
this auxiliary graph, allowing us to complete the identification
of connectivity in $\widetilde O(f)$ time.

\smallskip \noindent \textbf{Open problems.}
Can our $\widetilde O(f^{2/5}n)$ and $\widetilde O(n^{7/4})$ space
bounds be achieved with polynomial decoding time? More ambitiously,
what is the minimum space needed for optimal $O(f)$ query time for
pairwise connectivity, matching the time needed to read the failure
set? Our fast-query construction achieves this time bound up to
polylogarithmic factors using $\widetilde O(\sqrt f\,n)$ bits,
leaving open whether the space can be reduced. Another direction is
to understand the tradeoff between space and stretch for deterministic
fault-tolerant distance oracles. Can we obtain subquadratic space while providing
$\operatorname{polylog} n$ distance approximation in $G-F$ for any $F$?

\smallskip
\noindent \textbf{AI disclosure.} The main ideas behind both the upper- and lower-bound space arguments were found by ChatGPT-6 Astra. The $f^{2/5}n$ upper bound emerged over many rounds of interaction, in which we posed the problem, steered the research direction, and suggested simplifications of intermediate versions. The lower-bound argument took fewer rounds and required little input from us, but the version it produced was considerably harder to follow. We verified both arguments in full and then substantially reworked them. We reinterpreted key steps and reorganized the proofs into modular, combinatorial components, so that the ideas can be used beyond this particular problem. We then reworked the approach behind the $n^{7/4}$ bound so that it rests on the same ideas as the $f^{2/5}n$ argument. The encoding-coloring equivalence and the techniques behind the fast query time were developed several months earlier, largely by us and independently of the space bounds, with Sol-5.6 and Claude assisting in formulating the sketches. Finally, we used Claude Opus 5.5 and ChatGPT-6 Astra in writing and auditing the paper and in literature research.

\noindent \textbf{Acknowledgments.} We thank Asaf Petruschka for useful conversations in the early stages of this work and Eliran Subag for clarifying to us connections between the result of \Cref{sec:lower-bound} and the Ising model.

\section{Preliminaries}\label{sec:prelim}

\paragraph{Model and notation.}
Throughout, $G=(V,E)$ is a simple undirected graph on the labeled vertex
set $V=[n]=\{1,\ldots,n\}$, with $n\ge 2$. For $S\subseteq V$, let
\[
  \delta_G(S)=\{(u,v)\in E:u\in S,\ v\notin S\}
\]
denote the set of undirected edges crossing $(S,V\setminus S)$.
Write $A_G$ for the adjacency matrix of $G$ and
$\operatorname{cc}(G-F)$ for its connected-component partition.
A valid query is an edge set $F\subseteq E(G)$.
Throughout, an oracle consists of two algorithms, $\operatorname{Enc}$ and  $\operatorname{Dec}$. For a graph $G$, a vertex $v$
and $U\subseteq[n]$, write $N_G(v)$ for $v$'s neighbors in $G$, define $N_G(U)=\bigcup_{u\in U}N_G(u)$ and
$d_G(v,U)=|N_G(v)\cap U|$. The encoding algorithm takes a graph $G$ and returns an encoding of it as a bit string, whereas the decoding algorithm takes as input $\operatorname{Enc}(G)$ and $F$, and returns whether $G-F$ is connected in the case of Boolean connectivity oracles. We also consider the setting of component oracles, in which the decoder returns $\operatorname{cc}(G-F)$ for valid queries, and not just Boolean connectivity, i.e., deciding whether $G-F$ is connected for $F \subseteq E(G)$. 

The next theorem shows that a Boolean encoding scheme can be converted to a
component-reporting scheme at an additive cost of $O(n\log n)$ bits.

\begin{theorem}\label{thm:fromconnectedtocomponents}
Fix an integer $f\ge 1$. Suppose a deterministic encoding scheme uses
$s(n,f)$ bits and determines whether $G-F$ is connected for every
$n$-vertex graph $G$ and every $F\subseteq E(G)$ with $|F|\le f$.
Then there is a deterministic component-reporting scheme using
$s(n,f)+O(n\log n)$ bits that reports the components of $G-F$ for
every such $G$ and $F$.
\end{theorem}

We defer a proof of this equivalence to Appendix \ref{sec:missing}. In light of this equivalence, we focus on component oracles throughout the paper.

\subsection{Coloring and space complexity}\label{subsec:coloring}

We spell out the equivalence between deterministic space complexity and graph coloring. Earlier uses of this connection in data-structure lower bounds include Bloomier filters~\cite[Theorem~5.1]{ChazelleKilianRubinfeldTal2004} and monotone minimal perfect hashing~\cite[Sections~1 and~3.1]{AssadiFarachColtonKuszmaul2023}.\footnote{Related coding--coloring formulations appear in zero-error source coding with side information~\cite{Witsenhausen1976} and in index coding~\cite[Section~2]{AlonEtAl2008}.} We describe it here explicitly for clarity, as this viewpoint is essential to both our upper and lower bounds.

Let $\mathcal I$ be a finite nonempty family of inputs, let $\mathcal Q$ be a finite set of queries, and let $\mathcal A$ be a nonempty set of answers. We assume each input $x\in\mathcal I$ has a set $\mathcal Q(x)\subseteq\mathcal Q$ of valid queries, each with a prescribed answer $a(x,Q)\in\mathcal A$ for every $Q\in\mathcal Q(x)$. A deterministic encoding with at most $M$ states consists of maps $\operatorname{Enc}:\mathcal I\to[M]$ and $\operatorname{Dec}:[M]\times\mathcal Q\to\mathcal A$ such that $\operatorname{Dec}(\operatorname{Enc}(x),Q)=a(x,Q)$ for every $x\in\mathcal I$ and $Q\in\mathcal Q(x)$. The important detail is that queries outside $\mathcal Q(x)$ are unsupported: if such a query is supplied to $\operatorname{Dec}$ together with $\operatorname{Enc}(x)$, no restriction is imposed on its output.

The \emph{confusion graph} $X$ has vertex set $\mathcal I$. Distinct inputs $x,y$ are adjacent if there is a query $Q\in\mathcal Q(x)\cap\mathcal Q(y)$ with $a(x,Q)\ne a(y,Q)$. 

\begin{claim}\label{cl:coloring-equivalence}
For every integer $M\ge1$, a correct deterministic encoding with at most $M$ states exists if and only if $X$ has a proper coloring with at most $M$ colors. Consequently, if all encodings have the same length in binary, then every valid encoding uses at least $\lceil\log_2\chi(X)\rceil$ bits.
\end{claim}

The proof is deferred to Appendix \ref{sec:missing}. We remark that the claim above is purely information-theoretic. In practice, we would like computational guarantees: encoding and decoding algorithms with low space usage and running time. For Las Vegas oracles, which answer all valid queries correctly but may use a random seed for the encoding, one can prove equivalence between encoding length and fractional chromatic number, see Assadi, Farach-Colton, and Kuszmaul~\cite{AssadiFarachColtonKuszmaul2023}.

\smallskip
\noindent \textbf{Specialization to connectivity.}
We now specialize the preceding framework to connectivity under edge failures. Let $\mathcal I_n$ be the family of all labeled simple graphs on $[n]$. For $G\in\mathcal I_n$, the valid queries are $\mathcal Q(G)=\{F\subseteq E(G):|F|\le f\}$, and the answer to $F$ is $\operatorname{cc}(G-F)$, the partition of $[n]$ into connected components of $G-F$. Thus the confusion graph $X_{n,f}$ has vertex set $\mathcal I_n$, and distinct graphs $G,H$ are adjacent if some $F\subseteq E(G)\cap E(H)$ with $|F|\le f$ satisfies $\operatorname{cc}(G-F)\ne\operatorname{cc}(H-F)$.

Both our upper and lower bounds are influenced by \Cref{cl:coloring-equivalence}, as it suffices to lower bound $\log_2\chi(X_{n,f})$ in order to prove a lower bound, and to construct an encoding which always distinguishes adjacent inputs in $X_{n,f}$, to establish an upper bound. We first consider the setting of an unbounded number of faults, i.e., $f=\binom{n}{2}$, and set $X_n=X_{n,\binom{n}{2}}$.

We now introduce two combinatorial notions: the first characterizes when $\operatorname{Enc}$ defines a proper coloring of the confusion graph, and the second gives a sufficient condition for two graphs $G,H\in\mathcal I_n$ to receive distinct encodings under every correct encoding scheme.

\begin{definition}[Cut-compatible encodings]\label{def:compatibleencd}
An encoding $\operatorname{Enc}$ is
\emph{$f$-cut compatible} if, for every ordered pair of graphs $G,H$ on
$[n]$ with $\operatorname{Enc}(G)=\operatorname{Enc}(H)$ and every
$S\subseteq[n]$,
\begin{equation}\label{eq:cut-compatibility}
  \delta_G(S)\subseteq E(H)
  \quad\text{and}\quad |\delta_G(S)|\le f
  \quad\Longrightarrow\quad
  \delta_H(S)=\delta_G(S).
\end{equation}

We say simply \emph{cut compatible} when the condition holds without
the restriction on $|\delta_G(S)|$.
\end{definition}

\begin{observation}\label{obs:cut-compatible}
An encoding admits an exact component decoder for failures of at most $f$ edges if and only if it is $f$-cut compatible.
\end{observation}
\begin{proof}
By \Cref{cl:coloring-equivalence}, an encoding admits a decoder
if and only if graphs with the same encoding have identical component
partitions after every common valid query. Fix $G,H$ with
$\operatorname{Enc}(G)=\operatorname{Enc}(H)$.
If cut compatibility fails for $S$, then setting $F=\delta_G(S)$ we have that $F$ is a common
valid query (since $F=\delta_G(S) \subseteq \delta_H(S)$), and $H-F$ has an edge crossing $(S,[n]\setminus S)$, whereas $G-F$ has none. Thus their component partitions differ, which contradicts the assumption that $\operatorname{Enc}$ admits a decoder.

Conversely, suppose the encoding is $f$-cut compatible, and let $F$ be
a common valid query. For every component $S$ of $G-F$, we have
$\delta_G(S)\subseteq F\subseteq E(H)$ and $|\delta_G(S)|\le f$.
Hence $\delta_H(S)=\delta_G(S)\subseteq F$, so every component of
$H-F$ lies in a component of $G-F$. Interchanging $G$ and $H$ gives
the reverse containment, so the two partitions into connected components coincide.
\end{proof}

\begin{definition}[Cut domination]
Let $G,H$ be graphs on $[n]$ and let $S\subseteq[n]$. We say that $G$
\emph{$S$-dominates} $H$ if $\delta_G(S)\supsetneq\delta_H(S)$.
We say that $G$ \emph{dominates} $H$ if it $S$-dominates $H$ for some
$S\subseteq[n]$.
\end{definition}

\begin{observation}\label{obs:domination-encoding}
Let $G,H$ be graphs on $[n]$. If $G$ dominates $H$, then
$\operatorname{Enc}(G)\ne\operatorname{Enc}(H)$ for every correct
deterministic component encoding $\operatorname{Enc}$ in the
\emph{unbounded-fault setting}.
\end{observation}
\begin{proof}
Choose $S\subseteq[n]$ with $\delta_H(S)\subsetneq\delta_G(S)$, and set
$F=\delta_H(S)$. Since $F\subseteq E(G)\cap E(H)$, this query is valid
for both graphs. No component of $H-F$ meets both $S$ and its complement,
whereas $G-F$ contains an edge crossing this cut. Hence
$\operatorname{cc}(G-F)\ne\operatorname{cc}(H-F)$, so $G,H$ are adjacent
in $X_n$ and must receive distinct encodings.
\end{proof}

\vspace{-10pt}\section{Technical overview}

\vspace{-8pt}\subsection{Space upper bounds}

We explain the $\widetilde O(f^{2/5}n)$- and
$\widetilde O(n^{7/4})$-bit constructions through the encoding
viewpoint of \Cref{cl:coloring-equivalence} and \Cref{obs:cut-compatible}.
Our goal is to construct an encoding $\Enc$ such that graphs with
the same encoding are $f$-cut compatible. We focus here on the
information stored and the argument for correctness, leaving
decoding to the full constructions.

Suppose that two graphs $G,G'$ on $[n]$ have the same encoding
but violate $f$-cut compatibility. Thus, some $S\subseteq[n]$
satisfies
\[
  \delta_G(S)\subsetneq\delta_{G'}(S),
  \qquad |\delta_G(S)|\le f.
\]
We call the edges in $\delta_{G'}(S)\setminus\delta_G(S)$
\emph{hidden edges} and denote their number by $b>0$.
Deleting $F=\delta_G(S)$ leaves no edge crossing $S$ in $G-F$,
whereas the hidden edges still cross $S$ in $G'-F$.

Both constructions combine two ingredients: expander sketches
and auxiliary cut information. Together, these ingredients
exclude such a pair by imposing conflicting bounds on $b$:
the sketches force any nonempty set of hidden edges to be large,
while the auxiliary cut information forces it to be small.
Choosing the parameters so that these bounds cannot hold
simultaneously establishes correctness and determines the space.

\smallskip
\noindent\textbf{Lossless-expander sketches.}
For an integer parameter $1\le t<n/4$, fix a bipartite graph
with left side $[n]$, right side $[r]$, and left degree
$\ell=O(\log n)$, where $r=\widetilde O(t)$, such that
\[
  |\Gamma(U)|\ge \frac34\ell|U|
  \qquad\text{for every }U\subseteq[n]\text{ with }|U|\le 2t.
\]
Here $\Gamma(U)$ is the set of right neighbors of $U$. Throughout $\Gamma$ is reserved for neighborhoods in the bipartite expander $B$, whereas $N$ denotes the neighborhoods of the graph under consideration.
The expansion is \emph{lossless} because the neighborhood
retains a $(1-\eta)$ fraction of the maximum possible
$\ell|U|$ distinct neighbors; we use $\eta=1/4$. 
Such expanders are standard in sparse recovery; see Berinde,
Gilbert, Indyk, Karloff, and Strauss~\cite{BerindeGIKS08}.

Let $B\in\{0,1\}^{r\times n}$ be the incidence matrix,
with column $b_v$ for each $v\in[n]$. We store
\[
  s_{v,G}=\sum_{u\in N_G(v)}b_u,
  \qquad
  \Sketch(G)=\{(b_v,s_{v,G})\}_{v\in[n]}.
\]
Since each coordinate of $s_{v,G}$ is an integer in $[0,n]$,
the sketches occupy $\widetilde O(tn)$ bits.
Whereas \cite{KKP206} uses (different) sketches to \emph{recover} small neighborhoods
after failures, we use this particular definition of sketches to exploit their \emph{cut-amplification} property rather than their sparse recovery property.

The following consequence of lossless expansion drives the argument.

\begin{fact}\label{fact:overview-equal-nbhd}
If $X,Y\subseteq[n]$ are disjoint and nonempty with
$\Gamma(X)=\Gamma(Y)$, then $|X|,|Y|>t$.
\end{fact}

Indeed, suppose $|X|\le |Y|$ and $|X|\le t$, and choose
$Y'\subseteq Y$ with $|Y'|=|X|$. Expansion gives
$|\Gamma(X\cup Y')|\ge \frac32\ell|X|$, whereas
$\Gamma(X\cup Y')=\Gamma(X)$ has size at most $\ell|X|$.

For the encoding argument, we use the following form of
\emph{gap cut amplification}.

\begin{lemma}\label{lem:overview-gap-amplification}
Let $G,G'$ be simple undirected graphs on $[n]$ with
$\Sketch(G)=\Sketch(G')$. If
$\delta_G(S)\subsetneq\delta_{G'}(S)$ for some $S\subseteq[n]$,
then
\[
  |\delta_{G'}(S)\setminus\delta_G(S)|>\frac34t^2.
\]
\end{lemma}
\begin{proof}[Proof sketch]
Let $K$ be the graph of \emph{hidden edges},
$E(K)=\delta_{G'}(S)\setminus\delta_G(S)\ne\varnothing$. Each hidden
edge has exactly one endpoint in $S$.
Let $X\subseteq S$ be the set of $S$-endpoints of hidden edges, and
let $P=\Gamma(X)$. Thus $P$ consists of the right vertices in $B$ that are adjacent to at least one vertex in $X$. We call these right vertices \emph{participating}. For each $i \in [r]$ let $T_i$ be the set of its left neighbors in $B$ and put
\[
  X_i=N_K(T_i\cap S^c)\subseteq S,
  \qquad
  Y_i=N_K(T_i\cap S)\subseteq S^c.
\]
The core of the argument is the following claim.

\smallskip\noindent
\textbf{Claim.} $|X_i|,|Y_i|>t$ for every $i\in P$.

\smallskip
We first derive the lemma from the claim in two amplification
steps. Then we prove the claim.

\emph{First amplification: from one participating right vertex
to $\Omega(\ell t)$ of them.}
Since $X\ne\varnothing$, also $P\ne\varnothing$. Pick $i\in P$.
By the claim and since $X_i\subseteq X$, we have $|X|\ge|X_i|>t$.
In particular, there are already more than $t$ hidden edges.
Expansion of a $t$-element subset of $X$ then gives
$|P|\ge\frac34\ell t$.

\emph{Second amplification: from participating right vertices
to $\Omega(t^2)$ hidden edges.}
Fix $i\in P$. Every $v\in Y_i$ has a $K$-neighbor in $T_i\cap S$.
So $T_i$ contains the $S$-endpoints of more than $t$ distinct hidden
edges. Conversely, the $S$-endpoint $u\in X$ of a hidden edge lies
in exactly $\ell$ sets $T_i$, namely those with
$i\in\Gamma(u)\subseteq P$. Double counting gives
\[
  \ell|E(K)|
  \ge \sum_{i\in P}|Y_i|
  > t|P|
  \ge \frac34\ell t^2,
\]
which proves the lemma.

\emph{Proof of the claim.}
Fix $i\in[r]$. Equality of the sketches of $G$ and $G'$ gives
$\sum_{u\in T_i\cap S}(s_{u,G'}-s_{u,G})=0$. Expanding this sum over
neighbors, each $b_v$ is counted once for each of its neighbors in
$T_i\cap S$, that is, each neighbor in $S$ that is connected to the
right vertex $i$ in $B$. So the coefficient of $b_v$ is
$d_{G'}(v,T_i\cap S)-d_G(v,T_i\cap S)$.

For $v\in S^c$, the only edges between $v$ and $S$ that change are
the added hidden edges, since $\delta_G(S)\subseteq\delta_{G'}(S)$.
So the coefficient is $d_K(v,T_i\cap S)$.

For $v\in S$, edges inside $S$ may change as well. However, equality
of the $i$th coordinates of the sketches of $v$ gives
$d_{G'}(v,T_i)=d_G(v,T_i)$. Hence the gain $d_K(v,T_i\cap S^c)$
across the cut is offset by an equal net loss inside $S$, and the
coefficient is $-d_K(v,T_i\cap S^c)$.

This yields the \emph{local balance} identity
\begin{equation}\label{eq:overview-balance}
  \sum_{v\in S^c}d_K(v,T_i\cap S)\,b_v
  =
  \sum_{v\in S}d_K(v,T_i\cap S^c)\,b_v.
\end{equation}
Both sides are nonnegative combinations of the vectors $b_v$. The
coefficients in the identity are positive exactly on $Y_i$ and on
$X_i$, respectively. Hence the two sides have supports $\Gamma(Y_i)$
and $\Gamma(X_i)$, and so $\Gamma(Y_i)=\Gamma(X_i)$.

Now let $i\in P$. Choose $x\in X$ with $i\in\Gamma(x)$, that is,
$x\in T_i\cap S$. Any hidden neighbor of $x$ lies in $Y_i$, so
$Y_i\ne\varnothing$. Since $\ell\ge1$, we get
$\Gamma(X_i)=\Gamma(Y_i)\ne\varnothing$, so $X_i\ne\varnothing$.
The sets $X_i\subseteq S$ and $Y_i\subseteq S^c$ are disjoint, so
Fact~\ref{fact:overview-equal-nbhd} gives $|X_i|,|Y_i|>t$.
\end{proof}

\smallskip
\noindent\textbf{Cut sparsifiers.}
Our second ingredient is a cut sparsifier, introduced by
Bencz\'ur and Karger~\cite{BenczurK96}.
For $0<\varepsilon\le 1/2$, a $(1+\varepsilon)$-cut sparsifier
$H_G$ is a weighted subgraph of $G$ satisfying
\[
  (1-\varepsilon)|\delta_G(S)|
  \le c_{H_G}(S)
  \le (1+\varepsilon)|\delta_G(S)|
  \qquad\text{for every }S\subseteq[n],
\]
where $c_{H_G}(S)$ is the total weight crossing $S$.
Batson, Spielman, and Srivastava~\cite{BatsonSS09} give a
deterministic construction with $O(n/\varepsilon^2)$ edges.
With suitably rounded weights, this sparsifier occupies
$\widetilde O(n/\varepsilon^2)$ bits.

\paragraph{The $\widetilde O(f^{2/5}n)$-bit oracle.}
Set $\varepsilon=\frac{1}{4f^{1/5}}$ and $t=\left\lceil\sqrt{8\varepsilon f}\right\rceil$, and store
\[
  \Enc(G)=\langle\Sketch(G),H_G\rangle,
\]
where $\Sketch(G)$ is the collection of the lossless expander sketches and $H_G$ is a $(1+\varepsilon)$-cut sparsifier. 
Since $t,\varepsilon^{-2}=\Theta(f^{2/5})$, the total space is
\[
  \widetilde O\bigl(n(t+\varepsilon^{-2})\bigr)
  =\widetilde O(f^{2/5}n).
\]

To prove correctness, suppose that $G,G'$ share the encoding
but violate $f$-cut compatibility. Choose $S\subseteq[n]$ with
$\delta_G(S)\subsetneq\delta_{G'}(S)$, and write
\[
  a=|\delta_G(S)|\le f,
  \qquad
  b=|\delta_{G'}(S)\setminus\delta_G(S)|>0.
\]
The common sketches force the hidden-edge set to be large:
by \Cref{lem:overview-gap-amplification}, $b>\frac34t^2\ge 6\varepsilon f$.
The common sparsifier $H=H_G=H_{G'}$ forces it to be small:
\[
  (1-\varepsilon)(a+b)\le c_H(S)\le(1+\varepsilon)a,
\]
and hence $ b\le\frac{2\varepsilon}{1-\varepsilon}a
  \le 4\varepsilon f$, by using $\varepsilon\le 1/2$ and $a\le f$.
These bounds on $b$ contradict each other, establishing
$f$-cut compatibility.

\paragraph{The $\widetilde O(n^{7/4})$-bit oracle.}
We again combine expander sketches with auxiliary cut information,
now obtained from a hierarchy of additive cut sparsifiers.
Specifically, we construct
\[
  G=G_0\supseteq G_1\supseteq\cdots\supseteq G_L=\varnothing
\]
such that, for every $i<L$,
\[
  |E(G_{i+1})|\le \frac12|E(G_i)|,
  \qquad
  \bigl||\delta_{G_i}(S)|-2|\delta_{G_{i+1}}(S)|\bigr|\le D
  \quad\text{for every }S\subseteq[n],
\]
where $D=Cn^{3/2}\sqrt{\log n}$ for a sufficiently large
constant $C$. Thus, assigning weight $2$ to each edge of
$G_{i+1}$ gives an additive-$D$ cut sparsifier of $G_i$.

The key is to represent this hierarchy without storing its graphs
explicitly. We construct a fixed family $\mathcal M_n$ of graphs
on $[n]$, called \emph{masks}, with $|\mathcal M_n|=2^{O(n)}$,
such that every graph $G_i$ admits a mask $M_{i+1}\in\mathcal M_n$
for which $G_{i+1}=G_i\cap M_{i+1}$ satisfies both guarantees.
Each transition is therefore specified by an $O(n)$-bit mask
identifier.

The family has an implicit description: fix a decomposition of
$K_n$ into $O(n)$ matchings and let $\mathcal M_n$ consist of all
unions of these matchings. Applying Zouzias's deterministic
matrix-balancing algorithm~\cite{Zouzias2012} to the adjacency
matrices of the matchings restricted to $G_i$ finds, in polynomial
time, a mask satisfying the cut-approximation guarantee; see \Cref{lem:mask-polynomial}.

We store the expander sketches at each level together with the
identifier of the next mask. Equivalently, for nonempty $G$,
\[
  \Enc(G)=
  \bigl\langle
    \Sketch(G),\operatorname{ID}(M_1),\Enc(G_1)
  \bigr\rangle,
  \qquad G_1=G\cap M_1.
\]
Set $t=\left\lceil 2\sqrt D\right\rceil
    =\widetilde O(n^{3/4})$.
Since each transition halves the number of edges, the hierarchy
has $L=O(\log n)$ levels. Each level uses
$\widetilde O(tn)$ bits for its sketches and $O(n)$ bits for
its mask identifier, giving total space
$\widetilde O(tn)=\widetilde O(n^{7/4})$.
As before, if $t\ge n/4$, storing the adjacency matrix fits
within the same bound.

For correctness, we prove cut compatibility by induction on
the recursion depth. The base case is immediate from the
distinguished encoding of the empty graph.
Suppose $\Enc(G)=\Enc(G')$ but some $S\subseteq[n]$ satisfies $\delta_G(S)\subsetneq\delta_{G'}(S)$. 
As before, write $a=|\delta_G(S)|$ and $b=|\delta_{G'}(S)\setminus\delta_G(S)|\geq 1$. 
Equality of the encodings gives identical sketches, a common first mask $M$, and
\[
  \Enc(G_1)=\Enc(G'_1),
  \qquad
  G_1=G\cap M,\quad G'_1=G'\cap M.
\]
Intersecting with the common mask preserves cut containment:
\[
  \delta_{G_1}(S)
  =\delta_G(S)\cap E(M)
  \subseteq
  \delta_{G'}(S)\cap E(M)
  =\delta_{G'_1}(S).
\]
By the inductive hypothesis, $G_1$ and $G'_1$ are cut compatible,
so these two cut sets must be equal. Consequently, their common
doubled cut value
\[
  q:=2|\delta_{G_1}(S)|=2|\delta_{G'_1}(S)|
\]
approximates the cut sizes in both parent graphs: $|a-q|\le D$ and $|a+b-q|\le D$. The triangle inequality now gives $b\le 2D$. On the other hand, the common sketches and
\Cref{lem:overview-gap-amplification} give $b>\frac34t^2\ge 3D$, a contradiction. Thus the encoding is cut compatible. 

\smallskip 
\noindent \textbf{Space-Efficient Decoding.}
The encoding viewpoint establishes correctness, but does not by
itself provide a space-efficient decoder. Given a failure query
$F$, a generic decoder could enumerate the $2^{\binom n2}$ labeled
graphs and recompute their encodings until finding a graph that
contains $F$ and matches the stored encoding. Even assuming
polynomial-time encoding, this exhaustive-search approach uses
$\Omega(n^2)$ bits of space merely to identify the current candidate graph.

To obtain decoding with $\widetilde O(n)$ additional workspace,
the full constructions develop a direct, single-graph formulation
of cut amplification. For a cut $S$ and a query $F\subseteq E(G)$,
we test the symmetry of an $r\times r$ matrix $Q_F(S)$, where
$r=\widetilde O(t)$, whose entries are computable directly from
the stored sketches and $F$. Surviving edges internal to $S$
contribute a symmetric matrix, so any asymmetry comes from the
surviving cut. We show that an empty cut always yields a symmetric
matrix, while
\[
  Q_F(S)=Q_F(S)^{\mathsf T}
  \quad\Longrightarrow\quad
  |\delta_{G-F}(S)|=0
  \ \text{or}\
  |\delta_{G-F}(S)|>\frac34t^2.
\]
Thus, the gap now concerns the surviving edges of a single graph,
and its premise can be checked from the encoding itself.
The auxiliary cut information resolves the remaining ambiguity
between an empty cut and a large one. Testing symmetry two entries
at a time avoids storing the matrix; enumerating cuts and refining
the component partition then requires only $\widetilde O(n)$
additional bits, with read-only access to the encoding and query.
This directly usable form of cut amplification is the basis of
the decoding procedures in the technical sections.

\subsection{Space lower bounds}

We prove that every deterministic oracle that reports the connected
components of $G-F$ for every failure set $F\subseteq E(G)$ with
$|F|\le f$ requires $\widetilde\Omega(f^{1/4}n)$ bits.
By \Cref{thm:fromconnectedtocomponents}, this also gives the same lower bound for oracles returning a Boolean answer of whether $G-F$ is connected, up to a space loss in the lower bound of $n\log(n)$\footnote{When $f=\tilde{O}(1)$, the trivial $\tilde{\Omega}(n)$ lower bound applies. For the other values of $f$ the lower bound for component-reporting oracles gives the same lower bound for Boolean oracles.}. The main
step in the proof is establishing a lower bound of $\Omega(n^{3/2}/\log(2n))$ bits
for unrestricted failures. Obtaining the bounded-failure result from the unbounded setting is routine: we take disjoint unions of $ \approx n/\sqrt f$ graphs on
$ \approx \sqrt f $ vertices each. Each block has at most $f$ possible edges,
so every failure query confined to one block respects the budget.
Applying the unrestricted lower bound within each block implies the desired $\tilde{\Omega}\left((\sqrt{f})^{3/2}\right) \cdot \Omega\left(\tfrac{n}{\sqrt{f}}\right)=\tilde{\Omega}(f^{1/4}n)$. We therefore focus on the $\tilde{\Omega} (n^{3/2})$ lower bound in the technical overview. 

\smallskip 
\noindent \textbf{The natural approach, and where it fails.}
A natural argument, appearing in many works in the area, is to design a family
$\mathcal F$ of instances (in our case, graphs),  for which every element of $\mathcal F$ can be reconstructed by querying the oracle and viewing its answers; see, e.g., \cite{thorup2005approximate,BDPW18}. This forces giving each member of $\mathcal{F}$ a different encoding, and therefore yields a
$\log_2|\mathcal F|$-bit lower bound.

As an example, suppose the oracle supports arbitrary deletion sets
$F\subseteq\binom{[n]}2$, so deleting a nonexistent edge is allowed. In this case, we may take $\mathcal F=\mathcal I_n$, the family of
all labeled simple graphs on $[n]$. Given an unknown graph
$G\in\mathcal I_n$, we query $F_e=\binom{[n]}2\setminus\{e\}$
for every pair $e\in \binom{[n]}{2}$. The endpoints of $e$ lie in the same component
of $G-F_e$ exactly when $e\in E(G)$. Thus the answers reconstruct
$G$, yielding a lower bound of $\log_2|\mathcal I_n|=\binom n2$ bits.

The key distinction in our model is that a query containing a nonedge
is invalid, and the oracle's answer to it is unconstrained. The
queries above may therefore produce identical answer sequences on
distinct graphs in $\mathcal F$, so the answers to queries need not identify the input.
Consequently, the reconstruction argument no longer applies.

\smallskip 
\noindent \textbf{A first attempt: guaranteeing legal queries.}
A natural first attempt is to fix queries $F_1,\ldots,F_k$ in
advance and choose a family $\mathcal F$ such that every graph in
the family contains every queried edge: $F_i\subseteq E(G)$ for
all $i$ and all $G\in\mathcal F$. All queries are then legal, so
the oracle must answer them correctly. The goal is to choose the
family so that (1) the answers to the queries
uniquely determine any member of $\mathcal F$, and (2) $\mathcal{F}$ is as large as possible.

The problem is that in such a case, one cannot guarantee property (2) - $\mathcal F$ must be small. Setting $U=\bigcup_iF_i$, observe that all answers are determined by the partition
$\operatorname{cc}(G-U)$, since
$G-F_i=(G-U)\cup(U\setminus F_i)$ and the edges
$U\setminus F_i$ are known and independent of $G$. Therefore, if requirement (1) is satisfied, two members of $\mathcal F$ can't share the same component partition after failing $U$. As there are at most $n^n$ partitions of
$[n]$, one gets $|\mathcal F| \le n^n$. Hence this approach cannot yield a lower bound better than $\Omega(\log(n^n))=\Omega(n\log n)$ bits. One may wonder if the reason the approach failed is that adaptivity was not allowed (the queries are oblivious to the partition provided in each answer). One can show by a counting argument that adaptivity can't help improve upon the $\Omega(n \log n)$ lower bound; we therefore need another approach.  

\smallskip 
\noindent \textbf{Passing to the confusion graph.}
This limitation leads us to examine the requirements of correctness
directly. As explained in \Cref{sec:prelim}, it suffices to
lower-bound $\log_2\chi(X_n)$, where $X_n$ is the \emph{confusion
graph}. Families $\mathcal F$ reconstructed by a single strategy using only legal
queries correspond to a special kind of cliques in $X_n$: any two inputs are separated by the
common valid query at which their reconstruction paths first diverge.
Our barrier bounds only these reconstructible cliques, and determining
the asymptotic growth of $\omega(X_n)$, and whether larger cliques
yield a lower bound beyond $O(n\log n)$ bits, remains open.

To lower-bound $\chi(X_n)$, we use Schrijver's nonnegative variant
$\vartheta'$~\cite{Schrijver1979} of the Lov\'asz theta
function~\cite{Lovasz1979}, which satisfies
$\vartheta'(\overline{X_n})\le\chi(X_n)$.
In our setting, this bound admits a simple probabilistic formulation,
which we now describe. We say a symmetric distribution $\mu$ on pairs $(\widehat G,\widehat H)$
is \emph{confusing} if
\begin{equation}\label{eq:overview-confusing}
    \Pr_\mu\bigl[\widehat G=\widehat H
        \text{ or }(\widehat G,\widehat H)\in E(X_n)\bigr]=1.
\end{equation}
In other words, a confusing distribution either outputs the same graph in both marginals, or outputs two graphs adjacent in $X_n$, i.e. graphs that can't share an encoding. Let $K_{G,H}=\mu(G,H)$ define its \emph{coupling matrix}, and let
$\pi$ be the distribution of one of the marginals of $\mu$. The key requirement we need from $K$ to obtain a lower bound is that $K$ is positive semidefinite. The reason is that it is not hard to show that when $K$ is PSD, the distribution $\mu$ necessarily has positively correlated marginals, namely:
\begin{equation}\label{eq:overview-psd-correlation}
    K\succeq0\quad\Longrightarrow\quad
    \Pr_\mu[\widehat G\in C,\widehat H\in C]\ge\pi(C)^2
    \quad\text{for every }C.
\end{equation}
If $K\succeq0$ and $C_1,\ldots,C_M$ are the classes of a proper
coloring of $X_n$, then
\[
    \Pr_\mu[\widehat G=\widehat H]
    \overset{\eqref{eq:overview-confusing}}{=}
    \sum_{i=1}^M\Pr_\mu[\widehat G\in C_i,\widehat H\in C_i]
    \overset{\eqref{eq:overview-psd-correlation}}{\ge}
    \sum_{i=1}^M\pi(C_i)^2
    \overset{\text{Cauchy--Schwarz}}{\ge}\frac1M.
\]
Thus, $\log_2\chi(X_n)\ge\log_2(1/\Pr_\mu[\widehat G=\widehat H])$.
We therefore seek a confusing distribution $\mu$ with a positive semidefinite
coupling matrix and equality probability as small as possible.
The important thing is that unlike the clique approach,
$\operatorname{supp}(\pi)$ is not required to be a clique: only distinct graphs that occur together with positive probability under $\mu$ must be adjacent
in $X_n$.

\smallskip 
\noindent \textbf{The confusing distribution.}
The main idea is that, given graphs $R,P$ on the same vertex set $V$
and a set $S\subseteq V$, we can construct two instances
$\widehat G,\widehat H$ that answer \emph{some} common valid query differently
whenever $P$ is nonempty. We use $P$ to \emph{plant} their
disagreements according to $S$. Specifically, both graphs inherit
the edges and nonedges of $R$ outside $E(P)$. The edges of $P$ are
split between the two instances: $\widehat G$ gets
$P^{\mathrm{cross}}$, the edges crossing $S$, while $\widehat H$
gets $P^{\mathrm{in}}$, the edges with both endpoints in $S$ or both
in $S^c$. If $P^{\mathrm{cross}}$ is nonempty, $\widehat G$
$S$-dominates $\widehat H$. If $P^{\mathrm{cross}}$ is empty, and $P^{\mathrm{in}}$ is nonempty, then $\hat H$ $v$ -dominates $\hat G$ where $v$ is a vertex of positive degree in $P^{\mathrm{in}}$. Therefore,
\[P\neq \emptyset \implies \hat G \sim_{X_n} \hat{H}.\] We obtain $\mu$ by applying the construction above to random $R,P,S$ - the last equation explains that any $\mu$ obtained by the planting process is confusing. 

We now describe the distribution $\mu$ used in the lower bound. Assume $n$ is even and fix a balanced bipartition $V=A\sqcup B$.
Independently sample a uniformly random cut $S\subseteq V$, a
uniformly random graph $R \sim G(A,B,1/2)$, and an edge set
$P\sim  G(A,B,p)$, where $G(A,B,r)$ denotes a random graph with parts $A,B$ where each edge between $A$ and $B$ is included independently
with probability $r$. See \Cref{fig:planted-cut} for an illustration. For the rest of the overview, $(\widehat G, \widehat H)\sim \mu$, and $K_p$ is the coupling matrix where $P\sim  G(A,B,p)$ with $p$ a parameter.

Restricting $R$ and $P$ to bipartite graphs serves two purposes:
(1) it ensures the coupling matrix $K$ is symmetric, a preliminary to being PSD, and (2) it provides a bilinear structure which is essential in our proof of positive semidefiniteness below.

The sampled graphs are identical exactly when $P$ is empty, an event of probability $(1-p)^{n^2/4}$. Hence the collision argument
gives a space lower bound of $\log_2(1/\Pr_\mu[\widehat G=\widehat H])=\log_2(1/(1-p)^{n^2/4})=\Omega(pn^2)$ bits, provided that
$K$ is positive semidefinite. The main technical task is to prove this property for $p=\Theta(1/(\sqrt n\log(2n)))$, which yields
the desired $\Omega(n^{3/2}/\log(2n))$ bound.

\smallskip 
\noindent \textbf{Positivity of $K_p$.}
We next find an invertible matrix $U$ such that $D_p:=U^*K_pU$
is real and diagonal, after omitting the zero rows and columns
outside the bipartite support of $\pi$. Here $*$ denotes conjugate transpose.
Since $U$ is invertible, $K_p\succeq0$ if and only if all diagonal
entries of $D_p$ are nonnegative. To construct $U$, we exploit the
fact that, conditioned on any cut $S$, the pairs of edge indicators
in $\widehat G$ and $\widehat H$ are independent across $A\times B$. This allows us to conjugate all matrices $K_{p,\gamma}$, where $K_{p,\gamma}$ is the coupling matrix conditioned on the uniform cut $\gamma$, simultaneously using the same $U$. The diagonal values of $D_p$ will therefore be certain averages over cuts which we now describe. Each diagonal entry of $D_p$  is indexed by
$\varepsilon\in\{\pm1\}^{B\times A}$. For a parameter $\theta\approx p$,
the entry indexed by $\varepsilon$ is a positive multiple of
\[
    Z(\theta\varepsilon)
    :=\mathbb E_{\alpha,\beta}
      e^{\mathrm i\theta\beta^{\mathsf T}\varepsilon\alpha}.
\]
Here $\alpha\in\{\pm1\}^{A}$ and $\beta\in\{\pm1\}^{B}$ are
independent and uniform (where $1$/$-1$ represents whether a vertex is in $S$ or outside it). The main challenge is to prove that these diagonal entries are positive for our choice of $p$. 

Fix $\varepsilon$ and write $(\varepsilon_b)_{b\in B}$ for its rows. Our goal is therefore to prove positivity of $Z(\theta\varepsilon)$ for all sign matrices $\varepsilon$, and $\theta$ as large as possible. 

\smallskip 
\noindent \textbf{Positivity on the diagonal of $D_p$.}
Define $K(x):=\prod_{b\in B}\cos(x_b)$ for $x\in\mathbb R^B$.
Averaging over $\beta$ gives
\[
    Z(\theta\varepsilon)
    =\mathbb E_\alpha K(\theta\varepsilon\alpha)
    =\mathbb E_\alpha\prod_{b\in B}
      \cos\bigl(\theta\langle\varepsilon_b,\alpha\rangle\bigr).
\]
Set $\theta=1/(t\sqrt n)$, where $t\ge1$. A natural first attempt
is to show that all cosine factors are positive with high probability.
Each $\langle\varepsilon_b,\alpha\rangle$ is distributed as the
endpoint of a simple symmetric random walk with $n/2$ steps, so after
division by $\sqrt n$ it satisfies a Gaussian tail bound.
Formally, Hoeffding's inequality and a union bound over the $n/2$
coordinates give
\[
\begin{aligned}
    \Pr_\alpha\Bigl[\max_{b\in B}
      |\theta\langle\varepsilon_b,\alpha\rangle|>1/\sqrt t\Bigr] \le\sum_{b\in B}\Pr_\alpha\bigl[
      |\theta\langle\varepsilon_b,\alpha\rangle|>1/\sqrt t\bigr]
    &\le\frac n2\cdot 2e^{-t}=ne^{-t}.
\end{aligned}
\]
Thus, for $t$ a sufficiently large multiple of $\log n$, all cosine
factors are positive and close to $1$ with high probability,
suggesting that $Z(\theta\varepsilon)>0$.

\smallskip 
\noindent \textbf{The naive argument, and why it fails.}
Let $\mathcal E$ be the event that
$|\theta\langle\varepsilon_b,\alpha\rangle|\le1/\sqrt t$ for all
$b\in B$, so that $\Pr[\mathcal E^c]\le ne^{-t}$ by the bound above.
On $\mathcal E$, the point $\theta\varepsilon\alpha$ lies in the box
$Q_0=(-\pi/2,\pi/2)^{n/2}$, where every cosine factor is positive.
Since we land in $Q_0$ with very high probability, the natural approach
is to bound $K$ from below on $\mathcal E$, and by $-1$ elsewhere.
Using $\cos x\ge1-x^2$ on $\mathcal E$, this gives
\[
    Z(\theta\varepsilon)\;\ge\;
    \underbrace{(1-ne^{-t})}_{\le\Pr[\mathcal E]}
    \cdot
    \underbrace{(1-1/t)^{n/2}}_{\le K\text{ on }\mathcal E}
    \;+\;
    \underbrace{ne^{-t}}_{\ge\Pr[\mathcal E^c]}
    \cdot
    \underbrace{(-1)}_{\le K}.
\]
Since $(1-1/t)^{n/2}\approx e^{-n/(2t)}$, keeping the right-hand side
positive forces $t-\log n>n/(2t)$, that is, $t=\Omega(\sqrt n)$. As
$\theta=1/(t\sqrt n)$ and $p\approx\theta$, this gives $p=O(1/n)$,
which is no better than the trivial $\Omega(n)$ lower bound!

Looking back at the lower bound on $Z(\theta \varepsilon )$ leaves hope: near the origin we expect $K$ to get its largest value, which is extremely small, yet on the other hand, on $\mathcal E^c$, the bound we use for $K$, $K \geq -1$, is extremely weak - $K$ is typically
tiny on $\mathcal E^c$ as well. To obtain better bounds, we must therefore compare the contributions of
points near and far from the origin, \emph{both weighted by $K$}. 

\smallskip 
\noindent \textbf{The Gaussian case.}
Positivity is easy to prove if $\theta\varepsilon\alpha$ is replaced by a
centered Gaussian. Indeed, applying
$\cos a\cos b=\frac12(\cos(a+b)+\cos(a-b))$ repeatedly gives
$K(x)=\mathbb E_s\cos\langle s,x\rangle$, where $s$ is uniform in
$\{\pm1\}^B$. Let $g\sim N(0,C)$ on $\mathbb R^B$. For each fixed $s$, the variable $\langle s,g\rangle$ is a centered Gaussian with variance $s^{\mathsf T}Cs$. Moreover,
$\mathbb E\cos X=e^{-\operatorname{Var}X/2}$ for every centered Gaussian
$X$. Hence
\[
    \mathbb E K(g)
    =\mathbb E_s\,\mathbb E_g\cos\langle s,g\rangle
    =\mathbb E_s e^{-s^{\mathsf T}Cs/2}>0,
\]
for any covariance matrix $C$. Every term in the average over $s$ is
positive, and we could conclude the proof. Unfortunately, $\theta \varepsilon \alpha$ is not Gaussian. 

\smallskip 
\noindent \textbf{Proving positivity.}
Inspired by this, we show that a small Gaussian perturbation of
$\theta\varepsilon\alpha$ is close enough to Gaussian to force
positivity. Let $\rho$ be the density of $\theta\varepsilon\alpha+\phi$,
where $\phi\sim N(0,\tau I)$ is independent of $\alpha$ and
$\tau=\Theta(1/\log n)$. One may check that $\int\rho K=e^{-n\tau/4}Z(\theta\varepsilon)$, hence showing $\int\rho K>0$ will conclude the lower bound. The proof has two steps.

\emph{Step 1: $\rho$ inherits a Gaussian property.} The density $f$ of a Gaussian
$N(0,C)$ satisfies $f(u+v)f(u-v)=f(u)^2e^{-v^{\mathsf T}C^{-1}v}$. We show that
$\rho$ satisfies this Gaussian-like midpoint inequality up to a factor of $2$, namely:
\begin{equation}\label{eq:overview-midpoint}
    \rho(u+v)\,\rho(u-v)\le2\rho(u)^2e^{-v^{\mathsf T}C^{-1}v}
    \qquad(u,v\in\mathbb R^B),
\end{equation}
for a covariance matrix $C$ whose diagonal entries are $O(1/\log n)$, depending on $\theta \varepsilon$.
The factor $2$ appears in the analysis as $\rho$ is a mixture of Gaussians and need not be log-concave. Interestingly, the proof
reduces \eqref{eq:overview-midpoint} to a positivity statement for
genuine Gaussians, which Step~2 supplies.

\emph{Step 2: the $K$-weighted mass concentrates on $Q_0$.} Tile
$\mathbb R^B$ by the cubes $Q_k=\pi k+Q_0$ where $k \in \mathbb{Z}^B$, and set
$I_k=\int_{Q_k}\rho|K|$. By \eqref{eq:overview-midpoint}, the density
at a pair of points $x\in Q_k$ and $y\in-Q_k$ is controlled by its value
at their midpoint, which lies in $Q_0$. Using the Pr\'ekopa--Leindler
inequality, we turn this pointwise comparison into a comparison of
integrals. As a result, the $K$-weighted mass decays away from $Q_0$ at
least as fast as a Gaussian. Explicitly, in this step we prove:
\[
    \frac{I_k}{I_0}\le2\sqrt2\,\Pr[g\in Q_k],
    \qquad g\sim N(0,C).
\]
Summing over $k\ne0$, a union bound over the coordinates of $g$ and the
Gaussian tail bound give
\[
    \sum_{k\ne0}\frac{I_k}{I_0}
    \le2\sqrt2\,\Pr[g\notin Q_0]
    \le2\sqrt2\sum_{b\in B}\Pr\bigl[|g_b|\ge\tfrac\pi2\bigr]
    \le4\sqrt2\,|B|\,e^{-\pi^2/(8\max_bC_{bb})}
    <1,
\]
where the last inequality uses $\max_bC_{bb}\le1/(4\log n)$ which will always hold whenever $t$ is large enough as we explain below. Beforehand, notice that this gives
$\int\rho K\ge I_0 - \sum_{k\ne0} I_k>0$, and therefore $Z(\theta\varepsilon)>0$, concluding the proof.
We note that the bound on the diagonal values of $C$ holds once every row of
$\theta\varepsilon$ has Euclidean norm $1/(t\sqrt2)\leq \tau$, that is,
once $t$ is a sufficiently large multiple of $\log n$. Thus
$t=\Theta(\log n)$ suffices, compared with the $t=\Omega(\sqrt n)$ forced
by the naive argument. This gives $p=\Theta(1/(\sqrt n\log n))$ and the
$\Omega(n^{3/2}/\log n)$ bound.

\smallskip
\noindent \textbf{Roadmap.} In \Cref{sec:bounded-failures} and \Cref{sec:unrestricted-oracle}, we prove \Cref{thm:intro-upper} by presenting the upper bound constructions obtaining $\tilde{O}(f^{2/5}n)$ and $\tilde{O}(n^{7/4})$, respectively. \Cref{sec:lower-bound} proves \Cref{thm:intro-lower} by presenting the lower bound and \Cref{sec:fast-query} proves \Cref{thm:fast} by providing a near-optimal query time of $\tilde{O}(f)$ using $\tilde{O}(\sqrt{f}n)$ space.

\section{Upper bound}\label{sec:upper-bounds}

The goal of this section is to give two constructions of
deterministic connectivity oracles that, for a graph $G\in\mathcal I_n$ and a
query $F\subseteq E(G)$, return $\operatorname{cc}(G-F)$. The first
supports queries with $|F|\le f$ using $\widetilde O(nf^{2/5})$ bits;
the second supports an unbounded number of failures using
$\widetilde O(n^{7/4})$ bits.

While the constructions differ, they share two algorithmic components:
\begin{enumerate}
\item \emph{Lossless-expander sketches.} We use \emph{lossless-expander sketches} (or expander sketches, for short), namely linear sketches
defined by the adjacency matrices of lossless bipartite expanders.
These sketches are standard in sparse recovery; see Berinde,
Gilbert, Indyk, Karloff, and Strauss~\cite{BerindeGIKS08}.
 For a parameter $t$, we fix a lossless expander whose left side has $n$ vertices, and whose right side has $r=\widetilde O(t)$ vertices. We use the $r \times n$ incidence matrix $B$ to assign to the vertex $i$ the sketch $b_i$, the $i$th column of $B$. During preprocessing, we store for every vertex $v$ its sketch $b_v$ together with the sum of its neighbors' sketches, i.e. $s_v=\sum_{(u,v)\in E(G)} b_u$. 

\item \emph{Auxiliary cut information.} We store information for estimating cut sizes. The first construction uses a cut sparsifier $H$
of $G$. The second uses a hierarchy
$G=G_0\supseteq G_1\supseteq\cdots\supseteq G_L=\varnothing$.
Each $G_{i+1}$ contains at most half the edges of $G_i$, and its cut
sizes approximate half the corresponding cut sizes of $G_i$ up to
an \emph{additive error}. The crux is to avoid storing the graphs
$G_i$ explicitly. We construct a fixed family $\mathcal M_n$ of
graphs on $[n]$, called \emph{masks}, with
$|\mathcal M_n|=2^{\widetilde O(n)}$, such that for every $G_i$ there
is a mask $M_{i+1}\in\mathcal M_n$ for which
$G_{i+1}=G_i\cap M_{i+1}$ satisfies both properties above.
We therefore specify each transition by the index of $M_{i+1}$ in
$\mathcal M_n$, using $\widetilde O(n)$ bits, and store this index
together with the sketches at that level.
\end{enumerate}

The key to our approach is the \emph{cut amplification property}. We design a procedure $\operatorname{Test\text{-}Cut}$ that, for every valid query
$F\subseteq E(G)$ and set $S\subseteq[n]$, returns one whenever
$\delta_{G-F}(S)=\varnothing$ and satisfies
\[
 \begin{aligned}
 &\operatorname{Test\text{-}Cut}
 \bigl(S,F,\{(b_v,s_v)\}_{v\in S},\{(b_u,b_v)\}_{uv\in F}\bigr)=1
 \\
 &\qquad\Longrightarrow\quad
 |\delta_{G-F}(S)|=0
 \ \text{or}\ 
 |\delta_{G-F}(S)|>\tfrac34t^2.
 \end{aligned}
\]
Thus, an output of zero certifies that the cut is nonempty, whereas
an output of one leaves an ambiguity between two extremes: (1) an empty cut and (2) a cut
with more than $3t^2/4$ surviving edges. This is where we use the
auxiliary cut information. In each construction, we choose $t$ so that the gap supplied by cut amplification is large enough for the
cut-size estimates to distinguish these two cases. Together, the sketches and auxiliary information determine which cuts of $G-F$ are empty, and hence recover the component partition $\operatorname{cc}(G-F)$.

\subsection{Cut amplification}\label{sec:sparse-recovery}

We use the sketching framework of Berinde, Gilbert, Indyk, Karloff, and Strauss~\cite{BerindeGIKS08} based on lossless expanders. We first recall the construction.

\subsubsection{The construction}\label{subsec:sketch-construction}

We use the following constructions of lossless expanders~\cite[Proposition~6]{BerindeGIKS08}.

\begin{lemma}\label{thm:sketch-graph}
For every $n\ge2$ and integer $1\le t<n/4$, there is a bipartite
graph $B$ with left side $[n]$ and right side $[r]$, where
$r=O(t\log(2n))$. All left vertices have degree $\ell$, where
$1\le\ell=O(\log(2n))$, and
\begin{equation}\label{eq:expansion}
 |\Gamma(U)|\ge\tfrac34\ell|U|
 \qquad\text{for every }U\subseteq[n]\text{ with }|U|\le2t.
\end{equation}
Here $\Gamma(U)$\footnote{Throughout the notation $\Gamma$ is reserved only for the neighbor set inside the graph $B$.} denotes the set of right neighbors of $U$. If one insists on computation of $B$ in time polynomial in $n$, then $\ell,r$ grow by a factor of $n^{o(1)}$.
\end{lemma}

We think of $t$ as a parameter to be chosen in each construction. We fix such a graph $B$ for each $n,t$ (and omit $n,t$ from $B$'s notation - they will be clear from context). We also write $B$ for its
$r\times n$ incidence matrix, with columns $b_v\in\{0,1\}^r$.
For $j\in[r]$, let $T_j=\{v\in[n]:B_{jv}=1\}$ be the set of left neighbors of the $j$th right vertex of $B$. 

\paragraph{Sketch.}
The sketch of $x\in\mathbb R^n$ is $Bx=\sum_vx_vb_v$.
For a graph $G$, we store for every vertex $v$ the sketch of its indicator vector, $b_v$, together with the neighbor sketches
$s_v=\sum_{(u,v)\in E(G)}b_u$. 

\paragraph{Storage.}
Each $s_v$ has $r$ integer coordinates in $[0,n]$, so all neighbor sketches use $O(nt\log^2(2n))$ bits. Storing the $\ell$ nonzero positions of each $b_v$ uses another $O(n\log^2(2n))$ bits.

We also recall the following simple fact.
\begin{fact}\label{fact:same-neighbors}
Let $X,Y\subseteq[n]$ be nonempty disjoint sets.
If $\Gamma(X)=\Gamma(Y)$, then $|X|>t$ and $|Y|>t$.
\end{fact}

\begin{proof}
Assume without loss of generality that $|X|\le|Y|$.
Suppose that $|X|\le t$, and choose $Y'\subseteq Y$ with
$|Y'|=|X|$. Since $X$ and $Y'$ are disjoint and
$\Gamma(Y')\subseteq\Gamma(Y)=\Gamma(X)$,
\eqref{eq:expansion} gives
\[
 \tfrac32\ell \cdot |X| = \tfrac34\ell \cdot (|X|+|Y'|)
 \le |\Gamma(X\cup Y')|
 = |\Gamma(X)|
 \le \ell|X|,
\]
a contradiction.
\end{proof}
\subsubsection{Cut amplification lemmas}

\begin{definition}
For a graph $G$ on $[n]$ and $S\subseteq[n]$, the \emph{sketched cut matrix} is
\begin{equation}\label{eq:sketched-cut-matrix}
 Q(S)=\sum_{\substack{(u,v)\in\delta_G(S)\\u\in S}}b_ub_v^{\mathsf T}.
\end{equation}
\end{definition}

The key observation is that, due to the lossless expansion of $B$, the matrix $Q(S)$ is not symmetric whenever the cut is small and nonempty. Concretely, we show that if $Q(S)$ is symmetric and $\delta_G(S)\ne\varnothing$, then $|\delta_G(S)|=\Omega(t^2)$; that is, many crossing edges are needed before $Q(S)$ can become symmetric. We call this property \emph{cut amplification}: symmetry of $Q(S)$ amplifies the guarantee that some edge crosses $S$ into the guarantee that $\Omega(t^2)$ edges do.

Since $(b_xb_y^{\mathsf T})_{i,j}=1$ exactly when $x\in T_i$ and $y\in T_j$, the entries of $Q(S)$ count paths of length $3$ that alternate between $B$ and $G$, which we now record:

\begin{observation}\label{obs:cut-paths}
For $i,j\in[r]$, let $P_{i,j}$ be the set of paths $i\xrightarrow{B}x\xrightarrow{G}y\xrightarrow{B}j$ with $x\in S$ and $y\in S^c$. Then $Q(S)_{i,j}=|P_{i,j}|$.
\end{observation}
We exploit this interpretation with the lossless expansion of $B$ to prove the cut amplification property.

\paragraph{Example: a single edge.}
Suppose $(u,v)$ is the only edge crossing $S$, with $u\in S$ and
$v\in S^c$. Expansion of the set $\{u,v\}$ in $B$, together with the left regularity of $B$,
ensures that there are vertices $i\in\Gamma(u)\setminus\Gamma(v)$
and $j\in\Gamma(v)\setminus\Gamma(u)$.
The path $i\xrightarrow{B}u\xrightarrow{G}v\xrightarrow{B}j$ is
the unique path in $P_{i,j}$, whereas $P_{j,i}=\varnothing$ because
$j\notin\Gamma(u)$. By \Cref{obs:cut-paths}, $Q(S)_{i,j}=1$ while
$Q(S)_{j,i}=0$, so $Q(S)$ is not symmetric.

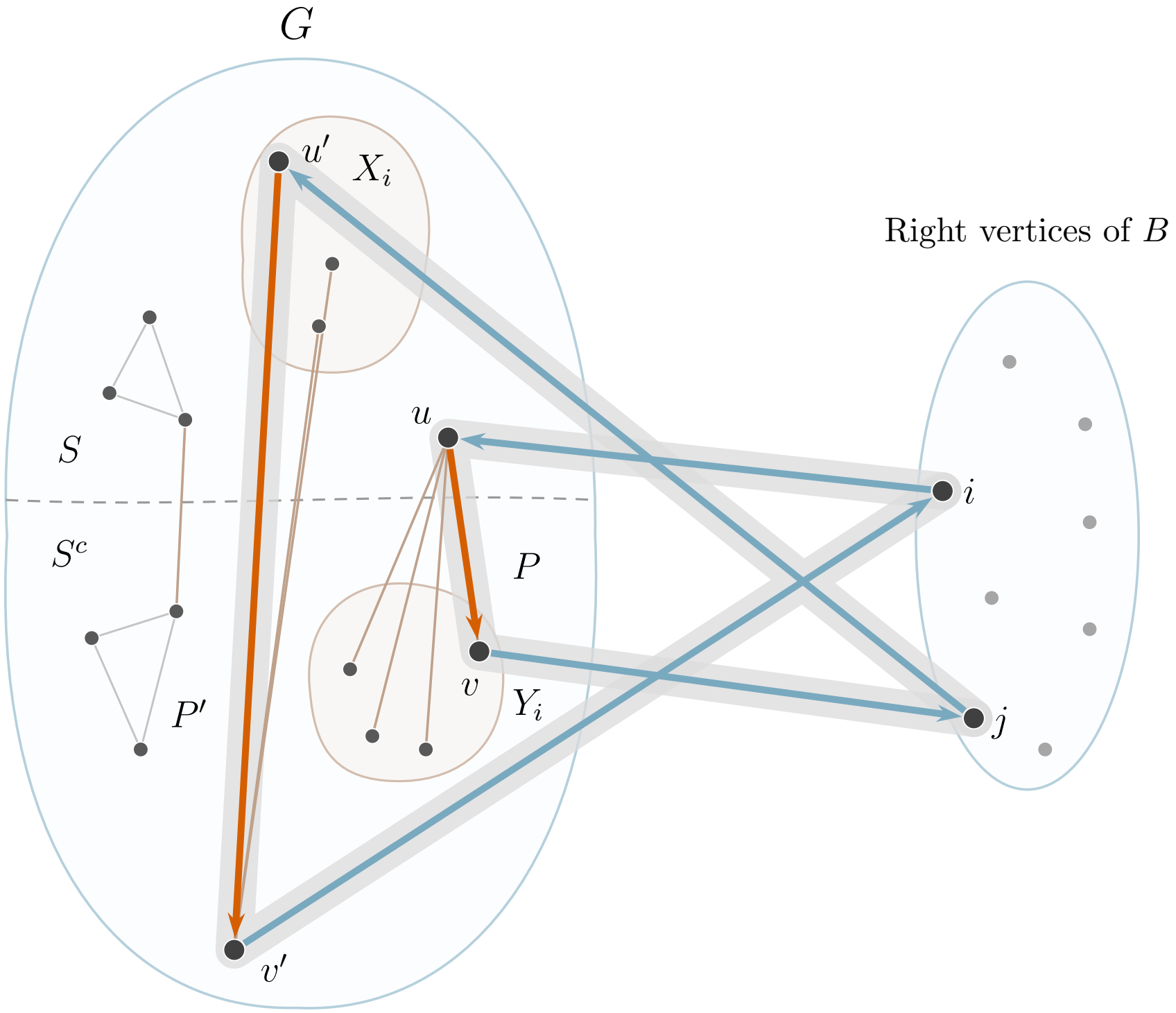
\begin{figure}[H]
\centering
\scalebox{0.85}{\begingroup
\definecolor{localblue}{HTML}{79A9BF}
\definecolor{localbrown}{HTML}{BFA18B}
\definecolor{pathorange}{HTML}{D55E00}
\definecolor{pathshade}{HTML}{DCDCDC}
\begin{tikzpicture}[x=1cm,y=1cm,font=\normalsize,
  vertex/.style={circle,fill=black!65,draw=white,line width=.4pt,
    minimum size=4.8pt,inner sep=0pt},
  marked vertex/.style={vertex,fill=black!75,minimum size=6.8pt},
  background edge/.style={draw=black!23,line width=.65pt},
  crossing/.style={draw=localbrown,line width=.8pt},
  path edge/.style={line width=2.2pt},
  directed path/.style={path edge,-{Stealth[length=2.8mm,width=1.9mm]},shorten <=3.8pt,shorten >=3.8pt},
  path halo/.style={line width=12pt,line cap=round,line join=round,opacity=.75},
  every node/.style={inner sep=2pt}]
\path[use as bounding box] (0,-.9) rectangle (13.6,10.85);
\path[draw=localblue!55,fill=localblue!2,line width=.8pt] (0.35,4.75)
  .. controls (0.18,7.75) and (1.55,10.1) .. (3.6,10.1)
  .. controls (5.6,10.15) and (7,7.9) .. (6.95,4.75)
  .. controls (7.1,1.35) and (5.6,-0.65) .. (3.6,-0.55)
  .. controls (1.55,-0.6) and (0.15,1.35) .. (0.35,4.75) -- cycle;
\path[draw=black!40,dash pattern=on 4pt off 3pt,line width=.7pt] (0.335,5.15)
  .. controls (2.5,5.05) and (4.9,5.25) .. (6.95,5.15);
\node[font=\Large] at (3.6,10.5) {$G$};
\node[font=\large] at (1.05,5.72) {$S$};
\node[font=\large] at (1.05,4.55) {$S^c$};
\path[draw=localblue!55,fill=localblue!2,line width=.8pt] (11.8,4.75) ellipse [x radius=1.25,y radius=2.85];
\node[font=\normalsize] at (11.8,8.15) {Right vertices of $B$};
\path[draw=localbrown!70,fill=localbrown!9,line width=.7pt] (3,7.85)
  .. controls (2.9,8.9) and (3.35,9.5) .. (4.05,9.45)
  .. controls (4.9,9.4) and (5.2,8.6) .. (5.05,7.8)
  .. controls (4.95,6.75) and (4.4,6.5) .. (3.75,6.6)
  .. controls (3.15,6.7) and (2.95,7.2) .. (3,7.85) -- cycle;
\path[draw=localbrown!70,fill=localbrown!9,line width=.7pt] (3.75,3)
  .. controls (3.65,3.85) and (4.25,4.3) .. (4.95,4.2)
  .. controls (5.7,4.1) and (6,3.6) .. (5.9,2.9)
  .. controls (5.8,2.2) and (5.2,1.95) .. (4.6,2)
  .. controls (4,2.05) and (3.8,2.4) .. (3.75,3) -- cycle;
\coordinate (x1) at (3.85,7.1);
\coordinate (x2) at (4,7.8);
\coordinate (x3) at (3.4,8.95);
\coordinate (u) at (5.3,5.85);
\coordinate (y1) at (4.2,3.25);
\coordinate (y2) at (4.45,2.5);
\coordinate (y3) at (5.05,2.35);
\coordinate (y4) at (2.9,0.1);
\coordinate (v) at (5.65,3.45);
\coordinate (i) at (10.85,5.25);
\coordinate (j) at (11.2,2.7);
\coordinate (a1) at (1.5,6.35);
\coordinate (a2) at (1.95,7.2);
\coordinate (a3) at (2.35,6.05);
\coordinate (a4) at (1.3,3.6);
\coordinate (a5) at (1.85,2.35);
\coordinate (a6) at (2.25,3.9);
\coordinate (r1) at (11.6,6.7);
\coordinate (r2) at (12.45,6);
\coordinate (r3) at (12.5,4.9);
\coordinate (r4) at (12.5,3.7);
\coordinate (r5) at (12,2.35);
\coordinate (r6) at (11.4,4.05);
\draw[path halo,draw=pathshade] (i) -- (u) -- (v) -- (j);
\draw[path halo,draw=pathshade] (j) -- (x3) -- (y4) -- (i);
\draw[background edge] (a1) -- (a2);
\draw[background edge] (a2) -- (a3);
\draw[background edge] (a3) -- (a1);
\draw[background edge] (a4) -- (a5);
\draw[background edge] (a5) -- (a6);
\draw[background edge] (a6) -- (a4);
\draw[crossing] (a3) -- (a6);
\draw[crossing] (u) -- (v);
\draw[crossing] (u) -- (y1);
\draw[crossing] (u) -- (y2);
\draw[crossing] (u) -- (y3);
\draw[crossing] (x1) -- (y4);
\draw[crossing] (x2) -- (y4);
\draw[crossing] (x3) -- (y4);
\draw[directed path,draw=localblue] (i) -- (u);
\draw[directed path,draw=pathorange] (u) -- (v);
\draw[directed path,draw=localblue] (v) -- (j);
\draw[directed path,draw=localblue] (j) -- (x3);
\draw[directed path,draw=pathorange] (x3) -- (y4);
\draw[directed path,draw=localblue] (y4) -- (i);
\node[vertex] at (x1) {};
\node[vertex] at (x2) {};
\node[marked vertex] at (x3) {};
\node[marked vertex] at (u) {};
\node[vertex] at (y1) {};
\node[vertex] at (y2) {};
\node[vertex] at (y3) {};
\node[marked vertex] at (y4) {};
\node[marked vertex] at (v) {};
\node[marked vertex] at (i) {};
\node[marked vertex] at (j) {};
\node[vertex] at (a1) {};
\node[vertex] at (a2) {};
\node[vertex] at (a3) {};
\node[vertex] at (a4) {};
\node[vertex] at (a5) {};
\node[vertex] at (a6) {};
\node[vertex,fill=black!35] at (r1) {};
\node[vertex,fill=black!35] at (r2) {};
\node[vertex,fill=black!35] at (r3) {};
\node[vertex,fill=black!35] at (r4) {};
\node[vertex,fill=black!35] at (r5) {};
\node[vertex,fill=black!35] at (r6) {};
\node[font=\large] at (4.45,8.85) {$X_i$};
\node[font=\large] at (6.2,2.85) {$Y_i$};
\node[font=\large] at (5,6.1) {$u$};
\node[font=\large] at (5.55,3.05) {$v$};
\node[font=\large] at (11.15,5.25) {$i$};
\node[font=\large] at (11.5,2.65) {$j$};
\node[font=\large] at (3.82,9.1) {$u'$};
\node[font=\large] at (3.35,-0.1) {$v'$};
\path (u) -- coordinate[pos=0.6] (path-label-p) (v);
\path (path-label-p) -- ++(0.36,0) node[right=2pt,font=\large,text=black] {$P$};
\path (x3) -- coordinate[pos=0.7] (path-label-pprime) (y4);
\path (path-label-pprime) -- ++(-0.3,0) node[left=2pt,font=\large,text=black] {$P'$};
\end{tikzpicture}
\endgroup}
\caption{The set $Y_i$ consists of vertices in $S^c$ reached from $i$
through $T_i\cap S$ by a $B$-edge followed by a crossing $G$-edge.
Every $B$-neighbor of $Y_i$ completes such a path to a $3$-path $P$ as depicted, showing $\Gamma(Y_i)=\{j\in[r]:Q(S)_{i,j}>0\}$. 
The symmetry of $Q(S)$ guarantees a path $P'\colon j\xrightarrow{B}u'\xrightarrow{G}v'\xrightarrow{B}i$.
Since $v'\in T_i\cap S^c$, we have $u'\in X_i$ and
$j\in\Gamma(X_i)$, implying $\Gamma(Y_i)=\Gamma(X_i)$. Only selected
vertices and edges are shown in the figure.}
\label{fig:local-amplification}
\end{figure}

We now proceed to analyze $Q(S)$, when there is at least one edge across the cut. We begin with some notation. For $i\in[r]$, let $X_i\subseteq S$ be the set of vertices of $S$ with a $G$-neighbor in $T_i\cap S^c$, and let $Y_i\subseteq S^c$ be the set of vertices of $S^c$ with a $G$-neighbor in $T_i\cap S$.
\begin{lemma}[Local amplification]\label{lem:vertex-crossing}
Let $G$ be a graph on $[n]$, $S\subseteq[n]$, and $i\in[r]$.\footnote{Throughout, $\Gamma$ is reserved for neighborhoods in $B$: for $U\subseteq[n]$, $\Gamma(U)$ is the set of right vertices adjacent to a vertex of $U$. The notation $\delta_G(S)$ denotes the set of edges of $G$ with exactly one endpoint in $S$.}
\begin{enumerate}
\item[(a)] $\Gamma(Y_i)=\{j\in[r]:Q(S)_{i,j}>0\}$ and $\Gamma(X_i)=\{j\in[r]:Q(S)_{j,i}>0\}$.
\item[(b)] If $Q(S)$ is symmetric, then $\Gamma(X_i)=\Gamma(Y_i)$. Consequently, either $X_i=Y_i=\varnothing$, or $|X_i|>t$ and $|Y_i|>t$.
\end{enumerate}
\end{lemma}
\begin{proof}
(a) A path $P\colon i\xrightarrow{B}x\xrightarrow{G}y\xrightarrow{B}j$ in $P_{i,j}$ is exactly a certificate that $y\in Y_i$ and $j\in\Gamma(y)$: its first two steps say that $y\in S^c$ has a neighbor $x\in T_i\cap S$, and its last step says $j\in\Gamma(y)$. Hence, by \Cref{obs:cut-paths}, $Q(S)_{i,j}>0$ iff $j\in\Gamma(Y_i)$. Similarly, a path $P'\colon j\xrightarrow{B}x\xrightarrow{G}y\xrightarrow{B}i$ in $P_{j,i}$, read backwards, is $i\xrightarrow{B}y\xrightarrow{G}x\xrightarrow{B}j$ with $y\in T_i\cap S^c$ and $x\in S$, which is exactly a certificate that $x\in X_i$ and $j\in\Gamma(x)$. Hence $Q(S)_{j,i}>0$ iff $j\in\Gamma(X_i)$.

(b) If $Q(S)$ is symmetric, then (a) gives
$\Gamma(X_i)=\Gamma(Y_i)$. Since $\ell\ge1$, the sets $X_i,Y_i$
are either both empty or both nonempty. In the latter case,
they are disjoint (as one is contained in $S$ and the other in $S^c$), so Fact \ref{fact:same-neighbors}
gives $|X_i|>t$ and $|Y_i|>t$. See \Cref{fig:local-amplification}.
\end{proof}

\begin{lemma}[Cut amplification]\label[lemma]{lem:many-negative-crossing}
Let $G$ be a graph on $[n]$ and $S\subseteq[n]$. If $Q(S)$ is symmetric, then $|\delta_G(S)|=0$ or $|\delta_G(S)|>3t^2/4$.
\end{lemma}
\begin{proof}
Assume that $Q(S)$ is symmetric and $\delta_G(S)\ne\varnothing$, and let $X\subseteq S$ be the set of $S$-endpoints of crossing edges. Choose a crossing edge $(u,v)$ with $u\in S$, and any $i\in\Gamma(v)$. Then $u\in X_i\subseteq X$, so \Cref{lem:vertex-crossing}(b) gives $|X|>t$, and \eqref{eq:expansion} applied to a $t$-element subset of $X$ gives $|\Gamma(X)|\ge\tfrac34\ell t$. 
For every $i\in\Gamma(X)$, any $u\in X\cap T_i$ certifies $|Y_i|>0$, so \Cref{lem:vertex-crossing}(b) gives $|X_i|>t$. Each vertex of $X_i$ has a neighbor in $T_i \cap S^c$, so more than $t$ crossing edges have their non-$S$ endpoint in $T_i$. Every crossing edge has its non-$S$ endpoint in exactly $\ell$ sets $T_i$ (by the left regularity of $B$), hence
\[
\ell|\delta_G(S)|\ \ge \sum_{i\in\Gamma(X)}|X_i|\ >\ t|\Gamma(X)|\ \ge\ \tfrac34\ell t^2.
\]
\end{proof}

\begin{remark}\label{rmk:sym-amp}
The proofs of \Cref{lem:vertex-crossing,lem:many-negative-crossing} give a stronger result. Whenever  $0<|\delta_G(S)|\le3t^2/4$, even the $0/+$ matrix
recording whether each entry of $Q(S)$ is zero or positive is not
symmetric, meaning there are $i,j\in[r]$ such that
$Q(S)_{i,j}>0$ and $Q(S)_{j,i}=0$.
\end{remark}

\paragraph{The cut test.}
In light of \Cref{lem:many-negative-crossing}, we now define
$\operatorname{Test\text{-}Cut}$. The algorithm takes a set $S\subseteq[n]$, a query $F\subseteq E(G)$,
the sketches $(b_v,s_v)$ for $v\in S$, and the sketches of the endpoints of $F$'s edges.

\begin{algorithm}[H]
\caption{$\operatorname{Test\text{-}Cut}$}\label{alg:test-cut}
\begin{algorithmic}[1]
\Require $S\subseteq[n]$, $F\subseteq E(G)$, $\{(b_v,s_v)\}_{v\in S}$, $\{(b_u,b_v)\}_{\{u,v\}\in F}$.
\State Compute $Q_F(S)\gets\displaystyle\sum_{u\in S}b_us_u^{\mathsf T}
-\sum_{\substack{u\in S,\,v\in[n]\\(u,v)\in F}}b_ub_v^{\mathsf T}$.
\State \Return $1$ if $Q_F(S)$ is symmetric, and $0$ otherwise.
\end{algorithmic}
\end{algorithm}

\begin{claim}\label[claim]{cl:empty-cut-test}
For every $F\subseteq E(G)$ and $S\subseteq[n]$, $\operatorname{Test\text{-}Cut}$
returns $1$ whenever $\delta_{G-F}(S)=\varnothing$.
\end{claim}
\begin{proof}
Let
\[
 I(S)=\sum_{u\in S}\sum_{\substack{v\in S\\(u,v)\in E(G)\setminus F}}
 b_ub_v^{\mathsf T}
\]
be the contribution of the surviving edges internal to $S$.
Observe that, since no edge of $G-F$ crosses $S$, we have
$Q_F(S)=I(S)$. Writing
\[
 I(S)=\sum_{\substack{(u,v)\in E(G)\setminus F\\u,v\in S,\,u<v}}
(b_ub_v^{\mathsf T}+b_vb_u^{\mathsf T}),
\]
we see that $I(S)$ is symmetric. Since $Q_F(S)=I(S)$, the test
returns $1$, proving the claim.
\end{proof}

The following lemma shows how cut amplification manifests itself
in the output of $\operatorname{Test\text{-}Cut}$.

\begin{lemma}\label{lem:cut-test}
Fix any $S\subseteq[n]$ and $F\subseteq E(G)$. Then
\[
 \operatorname{Test\text{-}Cut}\text{ returns }1
 \quad\Longrightarrow\quad
 |\delta_{G-F}(S)|=0
 \quad\text{or}\quad
 |\delta_{G-F}(S)|>\tfrac34t^2.
\]
\end{lemma}
\begin{proof}
If $\delta_{G-F}(S)=\varnothing$, the conclusion is immediate.
Henceforth, assume that the surviving cut is nonempty.
Let $I(S)$ be defined in the proof of
\Cref{cl:empty-cut-test}.

Let $Q(S)$ denote the sketched
cut matrix of $G-F$. Substituting the definition of $s_u$
into the expression for $Q_F(S)$ in \Cref{alg:test-cut} gives
\begin{equation}\label{eq:observable-cut-matrix}
 Q_F(S)=Q(S)+I(S).
\end{equation}
Since $I(S)$ is symmetric, $Q_F(S)$ is symmetric if and only if $Q(S)$ is symmetric. Thus, an output of one implies
$|\delta_{G-F}(S)|>3t^2/4$ by \Cref{lem:many-negative-crossing}.
\end{proof}

\begin{remark}\label{rmk:sym-test-cut}
In contrast to \Cref{rmk:sym-amp}, the symmetry test in $\operatorname{Test\text{-}Cut}$ cannot in general use only the $0/+$ pattern of $Q_F(S)$.
Indeed, adding the symmetric matrix $I(S)$ preserves asymmetry in the entry values, but not necessarily in the $0/+$ pattern.
\end{remark}

\subsection{Connectivity oracles with
\texorpdfstring{$\widetilde O(f^{2/5}n)$}{O-tilde(f to the two-fifths times n)} space}
\label{sec:bounded-failures}

In this section, we show:
\begin{theorem}\label{thm:boundedf-upperbound}
   For every $n\ge2$ and integer $1\le f\le\binom n2$, every simple
undirected $n$-vertex graph $G$ admits a deterministic connectivity oracle using $\tilde{O}(f^{2/5}n)$ bits that, given any query $F \subseteq E(G)$ with $|F|\leq f$, returns the connected components of $G-F$. The preprocessing time (i.e., encoding time) can be made polynomial at the cost of increasing the oracle space by an $n^{o(1)}$ factor. The decoding runs in $2^{\min\{f,n\}}\poly(n)$ time and uses $\tilde{O}(n)$ auxiliary bits. 
\end{theorem}

Before defining the construction, we recall basic definitions and
results on cut sparsification.

\smallskip 
\noindent \textbf{Cut Sparsifiers.} For a weighted
graph $H$ on $[n]$ with nonnegative edge weights
$w_H$, write $c_H(S)=\sum_{(u,v)\in\delta_H(S)}w_H(u,v)$. For
$0<\varepsilon<1$, a $(1+\varepsilon)$-cut sparsifier of an
unweighted graph $G$ is a weighted graph $H$ on the same vertex set
satisfying
\begin{equation}\label{eq:cut-sparsifier}
 (1-\varepsilon)|\delta_G(S)|\le c_H(S)
 \le(1+\varepsilon)|\delta_G(S)|
 \qquad\text{for every }S\subseteq[n].
\end{equation}
We use the following construction of Batson, Spielman, and
Srivastava~\cite{BatsonSS09} (which in fact provides the stronger notion of a spectral sparsifier).

\begin{theorem}\label{thm:sparsifier}
For every undirected graph $G$ with nonnegative edge weights,
$n$ vertices, and $m$ edges, and every $0<\varepsilon<1$, a
deterministic algorithm constructs  a $(1+\varepsilon)$ cut sparsifier for $G$, given as a reweighted subgraph $H$ with
$O(n/\varepsilon^2)$ edges. 
The algorithm uses $O(mn^3/\varepsilon^2)$ arithmetic operations.
\end{theorem}

For unweighted $G$, rounding the weights allows $H$ to be stored
using $O((n/\varepsilon^2)\log(n/\varepsilon))$ bits; see also
Carlson, Kolla, Srivastava, and Trevisan~\cite[Section~1]{CarlsonKST19}.\footnote{Apply
\Cref{thm:sparsifier} with accuracy $\varepsilon/2$ and round each
weight down to a multiple of a dyadic $\delta\in[\varepsilon/4,\varepsilon/2]$.
Since $H$ is a reweighted subgraph of $G$, rounding changes a cut
value by at most $\delta|\delta_G(S)|$. Singleton cuts bound each
weight by $2n$, so $O(\log(n/\varepsilon))$ bits per stored edge suffice.}

\smallskip
\noindent \textbf{The construction.}
Set
\begin{equation}\label{eq:teps-ftwofifthsol}
 \varepsilon=\frac1{4f^{1/5}},\qquad
 t=\left\lceil\sqrt{8\varepsilon f}\right\rceil.
\end{equation}
Thus $t=\Theta(f^{2/5})$ and $\varepsilon^{-2}=\Theta(f^{2/5})$.
If $t\ge n/4$, then $n^2=O(nf^{2/5})$, so we may store $G$ directly and there is nothing to prove. 
Henceforth assume $t<n/4$.

\paragraph{Encoding.}
We set $\operatorname{Enc}(G)=\langle H,\{(s_v,b_v)\}_{v\in[n]}\rangle$, where
\begin{itemize}
\item  $H$ is a $(1+\varepsilon)$-cut sparsifier as above.
\item For every $v\in[n]$, $b_v$ is its column in $B$ and $s_v=\sum_{(u,v)\in E(G)}b_u$.
\end{itemize}

We now show that, for any $S\subseteq[n]$, the encoding and a query
$F$ suffice to decide whether $G-F$ has an edge crossing $S$.

\subsubsection*{Decoding}
The decoding has the following structure. We first show how to determine if a specific cut $S$ has outgoing edges in $G-F$. This can be done in polynomial time. Then, we iterate over all possible $2^{\min\{n,f\}}$ subsets $S$.

\smallskip
\noindent \textbf{Deciding on a specific cut candidate $S$.} 
Set $d=|\delta_F(S)|$ to be the number of failing edges across the cut $S$ and return \emph{empty} if and only if
\begin{equation}\label{eq:bounded-cut-decoder}
 \operatorname{Test\text{-}Cut}\bigl(S,F,
 \{(b_v,s_v)\}_{v\in S},
 \{(b_u,b_v)\}_{\{u,v\}\in F}\bigr)=1
 \quad\text{and}\quad c_H(S)\le(1+\varepsilon)d.
\end{equation}

\begin{lemma}\label{cl:bounded-cut-decoding}
The cut decoder returns empty exactly when $\delta_{G-F}(S)=\varnothing$.
\end{lemma}
\begin{proof}
Let $b=|\delta_{G-F}(S)|$ be the number of surviving edges across
the cut. Since $F\subseteq E(G)$, we have
$|\delta_G(S)|=d+b$ and $d\le f$.
If $b=0$, then $\operatorname{Test\text{-}Cut}$ returns $1$ by
\Cref{cl:empty-cut-test}. By \eqref{eq:cut-sparsifier}, we also have
\[
 c_H(S)\le(1+\varepsilon)|\delta_G(S)|
 =(1+\varepsilon)(d+b)=(1+\varepsilon)d.
\]
Thus both conditions in \eqref{eq:bounded-cut-decoder} hold.

We now unleash the power of cut amplification. Suppose $b>0$ and
$\operatorname{Test\text{-}Cut}$ returns $1$, so $Q_F(S)$ is symmetric.
By the cut amplification guarantee of \Cref{lem:cut-test},
we have $b>3t^2/4\ge6\varepsilon f$.
We now show that the contribution of the $b$ surviving edges
outweighs the uncertainty due to the $d$ failing edges in the
sparsifier estimate. Concretely, we claim that
\[
 (1-\varepsilon)(b+d)>(1+\varepsilon)d,
\]
so \eqref{eq:cut-sparsifier} implies that the decoder cannot return
\emph{empty}. Indeed, it suffices to prove
$(1-\varepsilon)b>2\varepsilon d$.
Our choice $t^2\ge8\varepsilon f$ ensures $b>\tfrac34t^2\ge6\varepsilon f$,
and hence $(1-\varepsilon)b\ge\tfrac34b>\tfrac92\varepsilon f>2\varepsilon d$,
using $\varepsilon\le1/4$ and $d\le f$.
\end{proof}

\smallskip
\noindent \textbf{Iterating over candidate cuts.}
Applying the cut decoder to every $S\subseteq[n]$ recovers
$\operatorname{cc}(G-F)$ in $2^n\poly(n)$ time and requires only
$\widetilde O(n)$ additional bits of storage: maintain a partition,
initially $\{[n]\}$, and split its blocks along every cut declared
empty. The symmetry test can be evaluated two matrix entries at a time,
with $\poly(n)$ overhead.

We now explain how to reduce the running time to $2^f\poly(n)$
when $f<n$, using at most $2^{f+1}$ candidate cuts per connected component.

Assume first that $G$ is connected, and hence $H$ is connected by \eqref{eq:cut-sparsifier}. Consequently,
$H-F$ has components $C_1,\ldots,C_d$ with $d\le f+1$, since
deleting an edge increases the number of components by at most one.
As $H-F$ is a subgraph of $G-F$, each component of $G-F$ is obtained
by merging some of the $C_i$. It therefore suffices to examine the
at most $2^{d}\le2^{f+1}$ cuts obtained by partitioning
$C_1,\ldots,C_d$ into two sides, and for each partition, run the cut decoder.

For general $G$, apply the same procedure separately to each connected
component. These components are available from the encoding, since
\eqref{eq:cut-sparsifier} gives $\operatorname{cc}(H)=\operatorname{cc}(G)$.
There are at most $n$ components, each requiring at most $2^{f+1}$ cut
tests, so the total running time remains $2^f\poly(n)$.

We are now ready to complete the proof of \Cref{thm:boundedf-upperbound}.

\begin{proof}[Proof of \Cref{thm:boundedf-upperbound}]
\textbf{Space:}
We have $t=O(f^{2/5})$ and
$\varepsilon^{-2}=O(f^{2/5})$.
The neighbor sketches use $O(nt\log^2 n)$ bits, the sparsifier
uses $O(n\varepsilon^{-2}\log(n/\varepsilon))$ bits, and storing
the nonzero positions of \emph{all} columns $b_v$ uses
$O(n\log^2 n)$ bits.
Since $f\le n^2$, the total encoding size is
$O(nf^{2/5}\log^2 n)$ bits.

\medskip
\noindent\textbf{Encoding Guarantees:}
If encoding time is unrestricted, we can find $B$ by exhaustive
search, checking each candidate for the expansion property
in \eqref{eq:expansion}. Such a matrix exists by
\Cref{thm:sketch-graph}.
Alternatively, applying \cite[Proposition~7]{BerindeGIKS08}
with $k=2t$ and $\varepsilon=1/4$ gives an explicit construction
with the same expansion guarantee and parameters
$\ell=n^{o(1)}$ and $r=t\,n^{o(1)}$.
Using this construction increases the encoding size by at most
an $n^{o(1)}$ factor, giving $f^{2/5}n^{1+o(1)}$ bits.
The sparsifier can be constructed deterministically in
polynomial time by \Cref{thm:sparsifier}, and all stored sketches
can then be computed in polynomial time.
Hence this choice gives deterministic polynomial-time encoding.

\medskip
\noindent\textbf{Decoding time and working space:}
By \Cref{cl:bounded-cut-decoding}, each candidate cut can be
tested in polynomial time.
Enumerating all subsets when $f\ge n$, and the candidate cuts
described above when $f<n$, gives total query time
$2^{\min\{f,n\}}\poly(n)$.

The symmetry test can be evaluated two entries at a time,
using $O(\log n)$ bits for the running sums, so the decoder
need not store an entire $r\times r$ matrix.
The sparsifier cut value can also be computed by scanning
its edges. Storing the components of $H-F$, the current
candidate cut, and the partition being refined uses
$O(n\log n)$ bits.
Hence the decoder uses $\widetilde O(n)$ auxiliary bits,
with read-only access to the encoding and the query.
\end{proof}

The exponent $2/5$ comes from balancing the sketch space
$\widetilde O(nt)$ with the sparsifier space
$\widetilde O(n\varepsilon^{-2})$ under the requirement
$t^2=\Omega(\varepsilon f)$.

\subsection{Oracle for unbounded failures with $\tilde{O}(n^{7/4})$ space}
\label{sec:unrestricted-oracle}

We now prove:

\begin{theorem}\label{thm:any-upperbound}
   For every $n\ge2$, there is a deterministic encoding of $n$-vertex
graphs using $\tilde{O}(n^{7/4})$ bits that reports
$\operatorname{cc}(G-F)$ for every $F\subseteq E(G)$. The preprocessing time (i.e., encoding time) can be made polynomial at the cost of increasing the oracle space by a factor of $n^{o(1)}$. The decoding runs in $2^{n}\poly(n)$ time and uses $\tilde{O}(n)$ auxiliary bits. 
\end{theorem}

The following definition of mask plays a key role in our construction.

\begin{definition}
A \emph{mask} is a graph $M$ on $[n]$. Applying $M$ to a graph $H$
on $[n]$ means taking the graph $H\cap M$. For $D\ge0$, we say that
$M$ is \emph{$D$-good for $H$} if it satisfies the following properties:
\begin{enumerate}
\item \emph{Halving.} Applying $M$ retains at most half the edges
of $H$, that is, $|E(H\cap M)|\le\tfrac12|E(H)|$.
\item \emph{Cut approximation.} Counting each edge of $H\cap M$
twice approximates every cut of $H$ up to an additive error of $D$:
\begin{equation}\label{eq:mask-guarantees}
 \bigl||\delta_H(S)|-2|\delta_{H\cap M}(S)|\bigr|\le D
 \qquad\text{for every }S\subseteq[n].
\end{equation}
\end{enumerate}
\end{definition}

Applying a $D$-good mask and assigning weight $2$ to each retained edge gives an additive cut sparsifier of $H$ with at most half its edges. A related edge-selection step appears in the additive sparsification construction of Bansal, Svensson, and Trevisan~\cite[Lemma 4.3]{BST19}, with an error bound depending on the cut. Here we additionally require the mask to be part of a family with each mask having a short identifier, and for which a good mask is efficiently computable from $G$.  

We think of $M$ as a transition matrix, which, given $H$, points to an additive cut sparsifier $H \cap M$ of $H$.  \Cref{eq:mask-guarantees} exactly says that when edges of $H$ are given weight $1$ and edges of $H\cap M$ are given weight $2$, $H \cap M$ is a cut sparsifier of $H$ within additive error $D$. The halving condition implies $H\cap M$ is significantly sparser than $H$. 

\begin{definition}
A \emph{mask family} $\mathcal M_n$ is a collection of masks on
$[n]$, fixed independently of the graph being encoded. For $D\ge0$,
we call $\mathcal M_n$ \emph{$D$-good} if every graph $H$ on $[n]$
has a mask $M\in\mathcal M_n$ that is $D$-good for $H$.
\end{definition}
We highlight that the halving guarantee comes for free when $\mathcal{M}_n$ is closed under complements, since if $M$ satisfies the cut approximation property, then either $M$, or its complement, satisfies the approximation property \emph{and} halving. Therefore, the harder to obtain promise needed from $\mathcal M_n$ is a cut-approximation guarantee. We use the following lemma:

\begin{lemma}\label{lem:mask-polynomial}
For every even $n$, there exists a family of masks $\mathcal M_n$ satisfying:
\begin{enumerate}
\item $|\mathcal M_n|=2^{n-1}$, and each mask has a unique $(n-1)$-bit
identifier from which it can be computed in polynomial time.
\item $\mathcal M_n$ is $O(n^{3/2} \sqrt{\log n})$-good.
\item Given $G$, an $O(n^{3/2} \sqrt{\log n})$-good mask for $G$ can be found
deterministically in polynomial time.
\end{enumerate}
\end{lemma}

It is well known that the additive approximation of \Cref{lem:mask-polynomial} is optimal up to the $\sqrt{\log n}$ factor. Indeed, the theorem of Erd\H{o}s and Spencer~\cite{ES72} implies that every signing of $E(K_n)$ has an induced subgraph with signed edge sum of magnitude $\Omega(n^{3/2})$. A standard averaging argument converts this into a cut with signed edge sum of the same order, so every mask for $K_n$ incurs additive cut error $\Omega(n^{3/2})$.

The construction, and its proof, appear in Appendix \ref{sec:mask}. We now show how we can use the family to design the oracle.

Set $D=2 C n^{3/2} \sqrt{\log n}$ where $C$ is a large enough constant ensuring the mask family from \Cref{lem:mask-polynomial} is $(C n^{3/2} \sqrt{\log n})$-good, set $t= \Theta(n^{3/4} \log^{1/4} n)$ with a large enough constant so that
$3t^2/4>2D$, and fix the family $\mathcal M_n$ supplied by
the lemma. As before, fix a sketch matrix $B$
according to this choice of $n$ and
$t$, and write $b_v$ for its $v$th column.

\paragraph{Encoding.}
Starting with $G_0=G$, construct a sequence of graphs as follows.
For each $i\ge0$ with $G_i\ne\varnothing$, choose
a mask $M_{i+1}\in\mathcal M_n$ that is $D$-good for $G_i$ and set
$G_{i+1}=G_i\cap M_{i+1}$. Stop at the first index $L$ for which
$G_L=\varnothing$. For every $0\le i<L$ and $u\in[n]$, compute the sketch $ s_{i,u}=\sum_{(u,v)\in E(G_i)} b_v$. The encoding stores:
\begin{enumerate}
\item The vertex sketches $\{b_v\}_{v\in V(G)}$.
\item The number of levels $L$ and the identifiers of the masks
$M_1,\ldots,M_L$.
\item The sketches $s_{i,u}$ for every $0\le i<L$ and $u\in[n]$.
\end{enumerate}

\subsubsection*{Decoding}
As in the previous construction, we first describe how to determine whether $G-F$ has an edge crossing $S$ and then iterate over all candidate subsets.

\smallskip
\noindent\textbf{Deciding on a specific cut candidate.}
Given a query $F\subseteq E(G)$ and a set $S\subseteq[n]$, our goal
is to decide whether $G-F$ has an edge crossing $S$. For every
$0\le i\le L$, let $F_i=F\cap E(G_i)$ be the set of failed edges
at level $i$. Observe that since $F \subseteq E(G)$, these sets can be computed from the stored masks by
setting $F_0=F$ and $F_{i+1}=F_i\cap E(M_{i+1})$.

We prove the following stronger statement. Applying it at level
zero, where $G_0=G$ and $F_0=F$, gives the desired answer.

\begin{claim}\label[claim]{cl:mask-cut-decoding}
For every $0\le i\le L$, we can decide whether $G_i-F_i$ has an
edge crossing $S$ using only the encoding, the query $F$, and the
set $S$.
\end{claim}
\begin{proof}
We prove the claim by backward induction on $i$, starting from
$i=L$. In this case $G_L=\varnothing$, so $G_L-F_L$ has no edge
crossing $S$.

Suppose $i<L$ and we have already decided whether $G_{i+1}-F_{i+1}$
has an edge crossing $S$. We now explain how to decide whether
$G_i-F_i$ has such an edge. Observe that $G_{i+1}-F_{i+1}$ is a
subgraph of $G_i-F_i$. Hence, if $G_{i+1}-F_{i+1}$ has an edge
crossing $S$, then so does $G_i-F_i$, and we answer \textbf{nonempty}, meaning $\delta_{G_i-F_i}(S) \neq \emptyset$.

We may therefore assume that $G_{i+1}-F_{i+1}$ has no edge
crossing $S$. Let $b=|\delta_{G_i-F_i}(S)|$ be the number of surviving crossing
edges, and define
\[
 r_i=|\delta_{F_i\setminus F_{i+1}}(S)|,
 \qquad
 f_{i+1}=|\delta_{F_{i+1}}(S)|.
\]
Here $f_{i+1}$ counts the failed edges crossing $S$ in $G_{i+1}$,
and $r_i$ counts the residual failed crossing edges in $G_i$
outside $G_{i+1}$. Crucially, both $f_{i+1}$ and $r_i$ can be computed from $F$ and the
stored masks, and can be used to estimate $b$, as we now describe.

\begin{observation}\label[observation]{obs:mask-cut-estimate}
If $G_{i+1}-F_{i+1}$ has no edge crossing $S$, then
\[
 |b-(f_{i+1}-r_i)|\le D.
\]
\end{observation}
\begin{proof}
Since $M_{i+1}$ is $D$-good for $G_i$ and
$G_{i+1}=G_i\cap M_{i+1}$, the cut-approximation
property \eqref{eq:mask-guarantees} gives
\[
 \bigl||\delta_{G_i}(S)|-2|\delta_{G_{i+1}}(S)|\bigr|\le D.
\]

By assumption, $G_{i+1}-F_{i+1}$ has no edge crossing $S$, meaning every edge
crossing $S$ in $G_{i+1}$ has failed, so
$|\delta_{G_{i+1}}(S)|=f_{i+1}$. With regard to $G_i$, the cut formed by $S$ in $G_i$ consists of $r_i+f_{i+1}$ failed edges and $b$ surviving edges, so
$|\delta_{G_i}(S)|=r_i+f_{i+1}+b$. Substituting these two cut sizes
into the inequality above yields
\[
 |(r_i+f_{i+1}+b)-2f_{i+1}|=|b-(f_{i+1}-r_i)|\le D.
\]
\end{proof}

Therefore the quantity $f_{i+1}-r_i$ is an estimate of the unknown $b$
within additive error $D$. We use this estimate to make the
decoding decision. To decide whether the cut formed by $S$ in $G_i-F_i$ is empty or nonempty, we do as follows:

\medskip
\noindent\textbf{Case 1: $f_{i+1}-r_i>D$.}

\noindent We answer \textbf{nonempty}.

\noindent This is correct since \Cref{obs:mask-cut-estimate}
gives $b\ge f_{i+1}-r_i-D>0$.

\medskip
\noindent\textbf{Case 2: $f_{i+1}-r_i\le D$.}

To resolve this case, we run $\operatorname{Test\text{-}Cut}$ from \Cref{alg:test-cut} with
inputs $S$, $F_i$, the stored sketches
$\{(b_u,s_{i,u})\}_{u\in S}$, and 
$\{(b_u,b_v)\}_{(u,v)\in F_i}$.
\begin{itemize}
\item If the test returns $0$, we answer \textbf{nonempty}.

By \Cref{cl:empty-cut-test}, the test always returns $1$ on an
empty cut, so an output of $0$ certifies the cut is not empty.
\item If the test returns $1$, we answer \textbf{empty}.

By \Cref{obs:mask-cut-estimate}, we have
$b\le f_{i+1}-r_i+D\le2D$. The cut-amplification guarantee of \Cref{lem:cut-test}
therefore gives $b=0$ or $b>\tfrac34t^2\ge2D$.
Since $b\le2D$, only $b=0$ is possible, meaning $\delta_{G_i-F_i}(S)$ is empty.
\end{itemize}

\end{proof}
As before, applying the cut decoder to every $S\subseteq[n]$
recovers $\operatorname{cc}(G-F)$: start with the partition
$\{[n]\}$ and split its blocks, if necessary, along every cut declared empty.

\begin{proof}[Proof of \Cref{thm:any-upperbound}]
\textbf{Space:}
The masks halve the number of edges at each level, so
$L=O(\log(2n))$.
Each level stores $n$ sketches with $r=O(t\log(2n))$
coordinates, each using $O(\log(2n))$ bits.
The mask identifiers use $O(nL)$ bits in total, and storing
the nonzero positions of all columns $b_v$ uses
$O(n\log^2(2n))$ bits. Thus the total encoding size is
\[
 O(nt\log^3(2n))
 =O(n^{7/4}\log^{13/4}(2n)).
\]

\medskip
\noindent\textbf{Encoding Guarantees:}
If encoding time is unrestricted, we can find $B$ by exhaustive
search, checking each candidate for the expansion property
in \eqref{eq:expansion}. Such a matrix exists by
\Cref{thm:sketch-graph}.
Alternatively, applying \cite[Proposition~7]{BerindeGIKS08}
with $k=2t$ and $\varepsilon=1/4$ gives an explicit construction
with the same expansion guarantee and parameters
$\ell=n^{o(1)}$ and $r=t\,n^{o(1)}$.
Using this construction increases the encoding size by at most
an $n^{o(1)}$ factor, giving $n^{7/4+o(1)}$ bits.
The masks can be found deterministically in polynomial time
by \Cref{lem:mask-polynomial}, and all stored sketches can then
be computed in polynomial time. Hence this choice gives
deterministic polynomial-time encoding.

\medskip
\noindent\textbf{Decoding time and working space:}
By \Cref{cl:mask-cut-decoding}, each cut can be tested in
polynomial time. Enumerating all $2^n$ cuts and refining the
current partition along every cut declared empty therefore
recovers $\operatorname{cc}(G-F)$ in $2^n\poly(n)$ time.

The symmetry test can be evaluated two entries at a time,
using $O(\log(2n))$ bits for the running sums, so the decoder
need not store an entire $r\times r$ matrix.
Membership in $F_i$ can be determined by evaluating the first
$i$ masks on each edge of $F$, so the sets $F_i$ need not be
stored. Storing the current cut and maintaining the partition
uses $O(n\log(2n))$ bits. Hence the decoder uses
$\widetilde O(n)$ auxiliary bits, with read-only access to
the encoding and the query.

\end{proof}

\section{Lower bound}\label{sec:lower-bound}
The aim of this section is to prove the following theorem:

\begin{theorem}\label{thm:lower-bound}
For all sufficiently large $n$ and every integer $1\le f\le\binom n2$, every deterministic data structure that encodes an $n$-vertex graph $G$ and, given any $F\subseteq E(G)$ with $|F|\le f$, returns the connected-component partition of $G-F$ requires
\[
    \Omega\!\left(\frac{nf^{1/4}}{\log(2f)}\right)
\]
bits in the worst case.
\end{theorem}

By \Cref{thm:fromconnectedtocomponents}, we may prove the lower bound for oracles returning connected components, up to a negligible space loss in the lower bound of $n\log(n)$. We follow the coloring approach appearing in \Cref{sec:prelim} - it suffices to lower bound $\log \chi (X_{n,f})$. By \Cref{cl:bounded-fault-reduction} (see Appendix \ref{sec:LB-reduction}), it is enough to prove \Cref{thm:lower-bound} only in the case $f=n^2$. Specifically, it suffices to prove:
\begin{theorem}\label{thm:unbounded-bound}
For all sufficiently large $n$,
\[
    \log_2\chi(X_n)
    =\Omega\!\left(\frac{n^{3/2}}{\log(2n)}\right)
    =\widetilde\Omega(n^{3/2}).
\]
\end{theorem}

\subsection{Encoding lower bounds by collisions}\label{subsec:coupling-bound}

We now describe a way to obtain a lower bound on the length of the encoding by constructing a ``confusing'' distribution, defined below. For an alternative view of this approach in the language of Lov\'asz theta function, see \Cref{subsec:remarks}.

\begin{definition}
Let $\mu$ be a distribution on pairs $(\hat G,\hat H)\in\mathcal I_n^2$ whose left and right marginals coincide. Its \emph{coupling matrix} $K$, indexed by $\mathcal I_n\times\mathcal I_n$, is defined by $K_{G,H}=\Pr_{(\hat G,\hat H) \sim \mu}[(\hat G,\hat H)=(G,H)]$. For an encoding $\operatorname{Enc}$, an \emph{encoding collision} is the event $\operatorname{Enc}(\hat G)=\operatorname{Enc}(\hat H)$.
\end{definition}

\begin{definition}
We say that $\mu$ is \emph{confusing} if every $(G,H)\in\operatorname{supp}(\mu)$ satisfies either $G=H$ or $G\sim_{X_n}H$. Thus $\mu$ is supported on pairs consisting of the same instance or adjacent instances in the confusion graph.
\end{definition}

Fix a valid encoding $\operatorname{Enc}:\mathcal I_n\to[M]$. For each $c\in[M]$, write $p_c=\Pr[\operatorname{Enc}(\hat G)=c]=\Pr[\operatorname{Enc}(\hat H)=c]$.

\begin{claim}\label{cl:positive-color-collision}
If $K$ is positive semidefinite, then for every $c\in[M]$,
\[
    \Pr[\operatorname{Enc}(\hat G)=c=\operatorname{Enc}(\hat H)]\ge p_c^2.
\]
\end{claim}
\begin{proof}
Let $v_c$ be the indicator vector of the inputs with encoding $c$, and let $\mathbf1$ be the all-ones vector. We have $\mathbf1^{\mathsf T}K\mathbf1=1$, $v_c^{\mathsf T}K\mathbf1=p_c$, and $v_c^{\mathsf T}Kv_c=\Pr[\operatorname{Enc}(\hat G)=c=\operatorname{Enc}(\hat H)]$. Set $x=v_c-p_c\mathbf1$. Since $K$ is PSD, we have $x^{\mathsf T}Kx\ge0$. Since $K$ is symmetric $v_c^{\mathsf T}K\mathbf1 = \mathbf{1}^{\mathsf T}Kv_c$. Hence,
\begin{align*}
    0\le x^{\mathsf T}Kx
    &=v_c^{\mathsf T}Kv_c-2p_c v_c^{\mathsf T}K\mathbf1+p_c^2\mathbf1^{\mathsf T}K\mathbf1\\
    &=\Pr[\operatorname{Enc}(\hat G)=c=\operatorname{Enc}(\hat H)]-p_c^2.
\end{align*}
This proves the claim.
\end{proof}

\begin{claim}\label{cl:collision-diagonal}
If $\mu$ is confusing, then
\[
    \Pr[\operatorname{Enc}(\hat G)=\operatorname{Enc}(\hat H)]
    =\Pr[\hat G=\hat H].
\]
\end{claim}
\begin{proof}
The two graphs in any unequal pair in the support of $\mu$ are adjacent and therefore have different encodings, whereas equal inputs have equal encodings.
\end{proof}

\begin{claim}\label{cl:collision-lower-bound}
If $\mu$ is confusing and its coupling matrix $K$ is positive semidefinite, then $\Pr[\hat G=\hat H]\ge1/M$. Consequently,
\[
    \chi(X_n)\ge\frac1{\Pr[\hat G=\hat H]}.
\]
\end{claim}
\begin{proof}
By \Cref{cl:positive-color-collision,cl:collision-diagonal},
\[
    \Pr[\hat G=\hat H]
    =\sum_{c=1}^M\Pr[\operatorname{Enc}(\hat G)=c=\operatorname{Enc}(\hat H)]
    \ge\sum_{c=1}^M p_c^2
    \ge\frac1M,
\]
where the last inequality follows from Cauchy--Schwarz and $\sum_{c=1}^M p_c=1$. Taking an encoding with $M=\chi(X_n)$, which exists by \Cref{cl:coloring-equivalence}, proves the desired lower bound on $\chi(X_n)$.
\end{proof}

To prove \Cref{thm:unbounded-bound}, we will construct a confusing distribution $\mu$ with equal marginals whose coupling matrix satisfies $K\succeq0$ and, for some absolute constant $c>0$,
\[
    \Pr[\hat G=\hat H]\le 2^{-c n^{3/2}/\log(2n)}.
\]
\Cref{cl:collision-lower-bound} then gives $\chi(X_n)\ge2^{c n^{3/2}/\log(2n)}$, as required.

\subsection{The planted cut distribution}\label{subsec:hard-coupling}

Assume that $n$ is even, and fix a partition $[n]=A\sqcup B$ with $|A|=|B|=n/2$. The graphs in the support of $\mu$ will all be bipartite, with edges only between $A$ and $B$. Write $m=|A||B|=n^2/4$, and fix a parameter $p\in(0,1)$. Let $G(A,B,r)$ be a random bipartite graph with sides $A,B$ where each pair in $A\times B$ forms an edge independently with probability $r$.

We define the \emph{planted cut distribution} as follows, with all three choices made independently:
\begin{enumerate}
\item[(1)] Sample a random edge set $R\sim G(A,B,1/2)$.
\item[(2)] Sample a random edge set $P\sim G(A,B,p)$.
\item[(3)] Sample a uniformly random subset $S\subseteq[n]$.
\end{enumerate}

Write $P=P^{\mathrm{cross}}\sqcup P^{\mathrm{in}}$, where $P^{\mathrm{cross}}$ consists of the edges of $P$ crossing the cut formed by $S$, and $P^{\mathrm{in}}$ consists of the rest of the edges in $P$, those with both endpoints on the same side of the cut. Define $\hat G,\hat H$ on $[n]$ by
\[
    E(\hat G)=(R\setminus P)\cup P^{\mathrm{cross}},
    \qquad
    E(\hat H)=(R\setminus P)\cup P^{\mathrm{in}}.
\]

\begin{figure}[H]
\centering
\begingroup

\definecolor{plantedblue}{HTML}{0072B2}
\definecolor{plantedorange}{HTML}{C06013}
\definecolor{sharedgray}{HTML}{A0A5AA}

\def\plantedcutpath{%
  (-.52,1.45)
  .. controls (.35,1.66) and (3.10,1.65) .. (3.65,1.38)
  .. controls (4.12,1.13) and (3.96,.44) .. (3.28,.38)
  .. controls (2.75,.35) and (2.62,.36) .. (2.35,.08)
  .. controls (1.86,-.49) and (.35,-.65) .. (-.30,-.47)
  .. controls (-.47,-.43) and (-.55,-.35) .. (-.62,-.27)
  .. controls (-.97,.07) and (-.99,1.15) .. cycle}

\resizebox{\linewidth}{!}{%
\begin{tikzpicture}[
  x=1cm,
  y=.85cm,
  font=\small,
  vertex/.style={
    circle,
    draw=black,
    line width=.55pt,
    fill=white,
    minimum size=4.5pt,
    inner sep=0pt
  },
  input edge/.style={
    draw=black!62,
    line width=.9pt
  },
  shared/.style={
    draw=sharedgray,
    line width=1pt
  },
  planted cross/.style={
    draw=plantedblue,
    line width=1.35pt
  },
  planted in/.style={
    draw=plantedorange,
    line width=1.35pt,
    dash pattern=on 3.2pt off 2pt
  },
  cut/.style={
    draw=black!65,
    line width=.65pt,
    dash pattern=on 4pt off 2.8pt
  },
  flow/.style={
    draw=black!65,
    line width=.7pt,
    -{Latex[length=2.2mm,width=1.6mm]}
  },
  every node/.style={inner sep=2pt}
]

\foreach \panel/\dx in {R/0,P/5.2,S/10.4} {
  \begin{scope}[xshift=\dx cm]
    \node at (0,1.90) {$A$};
    \node at (3.2,1.90) {$B$};
    \foreach \i/\y in {1/1,2/0,3/-1,4/-2} {
      \coordinate (\panel-a\i) at (0,\y);
      \coordinate (\panel-b\i) at (3.2,\y);
    }
  \end{scope}
}

\node[font=\normalsize] at (1.6,2.65) {$R$};
\node[font=\normalsize] at (6.8,2.65) {$P$};
\node[font=\normalsize] at (12.0,2.65) {$S$};

\foreach \a/\b in {1/2,1/4,2/1,2/3,2/4,3/2,3/3,4/4}
  \draw[input edge] (R-a\a) -- (R-b\b);

\foreach \a/\b in {1/4,2/1,3/1,3/3,4/2}
  \draw[input edge] (P-a\a) -- (P-b\b);

\begin{scope}[xshift=10.4cm]
  \fill[black!3] \plantedcutpath;
  \draw[cut] \plantedcutpath;
  \node at (.75,1.15) {$S$};
\end{scope}

\draw[flow] (6.8,-2.45) -- (6.8,-3.75);

\draw[shared] (-.65,-5.85) -- (-.15,-5.85);
\node[anchor=west] at (0,-5.85) {$R\setminus P$};

\draw[planted cross] (-.65,-6.85) -- (-.15,-6.85);
\node[anchor=west] at (0,-6.85) {$P^{\mathrm{cross}}$};

\draw[planted in] (-.65,-7.85) -- (-.15,-7.85);
\node[anchor=west] at (0,-7.85) {$P^{\mathrm{in}}$};

\node[font=\normalsize] at (3.8,-4.20) {$\widehat G$};
\node[font=\normalsize] at (9.8,-4.20) {$\widehat H$};

\foreach \panel/\dx in {G/2.2,H/8.2} {
  \begin{scope}[shift={(\dx,-6.6)}]
    \fill[black!3] \plantedcutpath;
    \draw[cut] \plantedcutpath;
    \node at (.75,1.15) {$S$};
    \node at (0,1.90) {$A$};
    \node at (3.2,1.90) {$B$};

    \foreach \i/\y in {1/1,2/0,3/-1,4/-2} {
      \coordinate (\panel-a\i) at (0,\y);
      \coordinate (\panel-b\i) at (3.2,\y);
    }
  \end{scope}

  \foreach \a/\b in {1/2,2/3,2/4,3/2,4/4}
    \draw[shared] (\panel-a\a) -- (\panel-b\b);
}

\foreach \a/\b in {1/4,3/1}
  \draw[planted cross] (G-a\a) -- (G-b\b);

\foreach \a/\b in {2/1,3/3,4/2}
  \draw[planted in] (H-a\a) -- (H-b\b);

\foreach \panel in {R,P,S,G,H} {
  \foreach \i in {1,2,3,4} {
    \node[vertex] at (\panel-a\i) {};
    \node[vertex] at (\panel-b\i) {};
    \node[anchor=east,xshift=-5pt]
      at (\panel-a\i) {$a_{\i}$};
    \node[anchor=west,xshift=5pt]
      at (\panel-b\i) {$b_{\i}$};
  }
}

\end{tikzpicture}%
}

\endgroup
\caption{Top: the three inputs, $R,P,S$ from which $\hat G$ and $\hat H$ are constructed. Bottom: the resulting graphs $\hat G$ and $\hat H$. On $R \setminus P$, $\hat G$ and $\hat H$ agree with $R$. The graph $\hat G$ receives the edges of $P$ crossing the cut, while $\hat H$ receives the edges of $P$ which are internal. If $P^{\mathrm{cross}}\neq \emptyset$, then $\hat G$ $S$-dominates $\hat H$.}
\label{fig:planted-cut}
\end{figure}
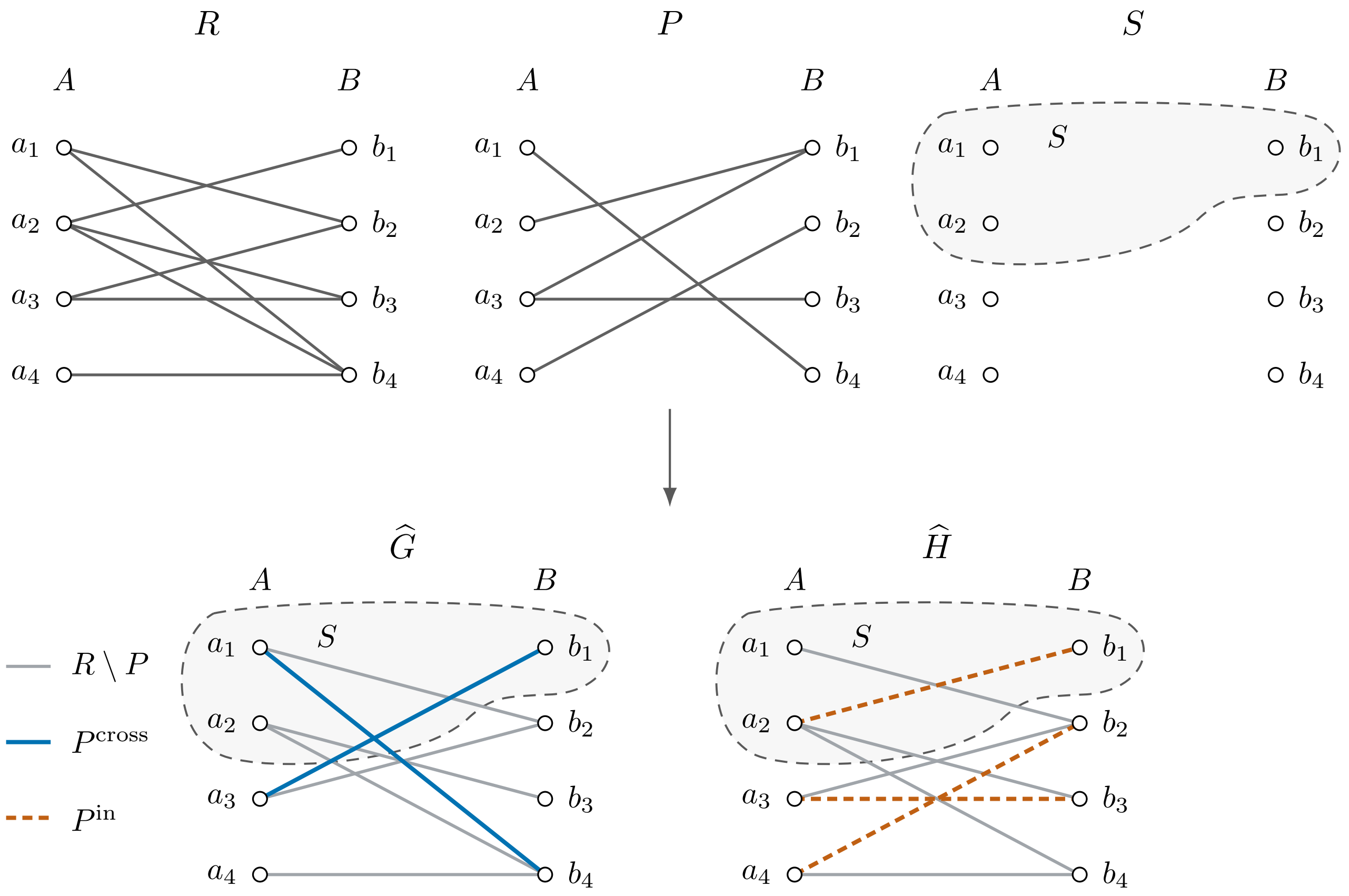

Hence the set $R$ stands for the shared random structure between $\hat G$ and $\hat H$, while $P$ consists of \emph{planted edges} on which $\hat G$ and $\hat H$ will disagree \emph{by construction}. Pairs of vertices in $(A\times B)\setminus P$ are either edges or nonedges in both graphs according to $R$, while pairs in $P$ crossing the hidden cut $S$ belong only to $\hat G$, and pairs in $P$ with both endpoints on the same side of the cut belong only to $\hat H$. See \Cref{fig:planted-cut} for an illustration.

Denote the resulting distribution by $\mu_p$ and its coupling matrix by $K_p$. 

\begin{claim}\label{cl:confusing-distribution}
The distribution $\mu_p$ is confusing.
\end{claim}
\begin{proof}
Fix an unequal pair $(G,H)\in\operatorname{supp}(\mu_p)$ and a realization of $R,P,S$ producing it. If $P^{\mathrm{cross}}\ne\varnothing$, then $G$ $S$-dominates $H$. Otherwise, $E(G)\subsetneq E(H)$, so choosing any $uv\in P^{\mathrm{in}}$, we see that $H$ $\{u\}$-dominates $G$. In either case, \Cref{obs:domination-encoding} gives $G\sim_{X_n}H$.
\end{proof}

\begin{claim}\label{cl:sampler-equality}
For $(\hat G,\hat H)\sim\mu_p$, we have $\Pr[\hat G=\hat H]=(1-p)^{n^2/4}$.
\end{claim}
\begin{proof}
The graphs are equal exactly when $P=\varnothing$, which occurs with probability $(1-p)^{n^2/4}$.
\end{proof}

To prove the lower bound, use \Cref{cl:collision-lower-bound}, which gives $\chi(X_n)\ge\frac1{\Pr[\hat G=\hat H]}$, and by the above claim we would like $p$ to be as large as possible for the best lower bound. On the other hand, to apply \Cref{cl:collision-lower-bound}, the coupling matrix $K_p$ must be PSD. This is clear for $p=0$ ($K_0$ is $2^{-m}$ times the identity), but as $p$ gets larger, positive semidefiniteness may fail.
The main remaining step is to prove that $K_p$ is positive semidefinite. We first show that it is symmetric.

\begin{claim}\label{cl:coupling-symmetry}
The matrix $K_p$ is symmetric. Consequently, the two marginals of $\mu_p$ coincide.
\end{claim}
\begin{proof}
Observe that by the law of total probability, it is enough to prove that for every choice of $R,P$ and any pair $(G,H)$,
\[
    \Pr[(\hat G,\hat H)=(G,H)\mid R,P]
    =\Pr[(\hat G,\hat H)=(H,G)\mid R,P].
\]
Indeed, every edge of $P$ joins $A$ to $B$, so replacing $S$ by $S\mathbin{\triangle}A$ exchanges $P^{\mathrm{cross}}$ and $P^{\mathrm{in}}$, and hence exchanges the two output graphs. This transformation is a bijection between the subsets $S$ producing $(G,H)$ and those producing $(H,G)$. Since $S$ is uniform conditional on $R,P$, the two probabilities are equal. Averaging over $R,P$ implies that $K_p$ is symmetric.
\end{proof}

\begin{rem}
The bipartite restriction is essential for symmetry. If instead we sample $R\sim G(n,1/2)$ and $P\sim G(n,p)$, and $n\geq 3$, symmetry would not hold. For example, take $n=3$, $G$ a triangle, and $H$ the empty graph on $3$ vertices. Then $\Pr[(\hat G,\hat H)=(G,H)]=0$, since no cut crosses three edges forming a triangle, whereas $\Pr[(\hat G,\hat H)=(H,G)]=p^3/4$. This happens precisely when all three edges are in $P$, and $S$ is either $\varnothing$ or $[3]$.
\end{rem}

It remains to prove the following theorem to conclude the lower bound.

\begin{theorem}\label{cl:coupling-positivity}
There is an absolute constant $c>0$ such that, for every sufficiently large even $n$, the matrix $K_p$ is positive semidefinite for $p=c/(\sqrt n\log(2n))$.
\end{theorem}
We reduce \Cref{cl:coupling-positivity} to an analytic positivity statement in \Cref{subsec:analytic-reduction}; that statement is proved in \Cref{sec:positivity-proof}.

\begin{proof}[Proof of \Cref{thm:lower-bound} from \Cref{cl:coupling-positivity}]
For sufficiently large even $n$, choose $p$ as in \Cref{cl:coupling-positivity}. By \Cref{cl:confusing-distribution,cl:coupling-symmetry}, the distribution $\mu_p$ is confusing and has equal marginals. \Cref{cl:collision-lower-bound,cl:sampler-equality} therefore give
\[
    \log_2\chi(X_n)
    =\Omega(pn^2)
    =\Omega\!\left(\frac{n^{3/2}}{\log(2n)}\right).
\]
This proves \Cref{thm:unbounded-bound}, with odd $n$ handled by adding an isolated vertex. The desired space bound follows from the reduction in \Cref{cl:bounded-fault-reduction} and \Cref{cl:coloring-equivalence}.
\end{proof}

\subsection{From matrix positivity to an analytic inequality}\label{subsec:analytic-reduction}

Recall that, since we are in a bipartite setup, choosing a uniformly random cut is equivalent to choosing independent uniform sign vectors $\alpha\in\{\pm1\}^{A}$ and $\beta\in\{\pm1\}^{B}$. This is the cut between the vertices assigned $+1$ and those assigned $-1$. Set $\gamma=\alpha\beta^{\mathsf T}$ for the sign matrix of the cut: $\gamma_{ab}=-1$ if $ab$ crosses the cut, and $\gamma_{ab}=1$ if its endpoints lie on the same side.

To study $\mu_p$, we reverse the order in which we reveal the random choices: we first reveal the cut, and then $R$ and $P$. Since these choices are independent, this does not change the distribution. We study the conditional coupling matrix of $(\hat G,\hat H)$ given $\gamma$, and ask when its average over the random cut is positive semidefinite.

In this subsection, we restrict the coupling matrices to rows and columns indexed by graphs with edges in $A\times B$. All omitted rows and columns are zero, so this restriction preserves positive semidefiniteness. Our argument thus lower bounds the chromatic number of the subgraph of $X_n$ induced by these graphs, and hence also $\chi(X_n)$.

We first consider the toy example of a single edge $ab\in A\times B$. With rows indexed by its edge indicator in $\hat G$ and columns by its edge indicator in $\hat H$, in the order $0,1$, the conditional coupling matrices are
\[
    L_+=\begin{pmatrix}(1-p)/2&p\\0&(1-p)/2\end{pmatrix},
    \qquad
    L_-=\begin{pmatrix}(1-p)/2&0\\p&(1-p)/2\end{pmatrix},
\]
according to whether $\gamma_{ab}=1$ or $-1$. Their average is
\[
    \frac{L_++L_-}{2}
    =\frac12\begin{pmatrix}1-p&p\\p&1-p\end{pmatrix},
\]
whose eigenvalues are $1/2$ and $(1-2p)/2$. Thus it is easy to check directly that the single-edge coupling is positive semidefinite exactly when $p\le1/2$. For multiple edges, checking positive semidefiniteness will be harder, so it is instructive to solve this example in a way that extends to the general case.

Suppose we find an invertible matrix $W$ such that both $W^*L_+W$ and $W^*L_-W$ are diagonal, where $*$ denotes conjugate transpose. Their average is then diagonal as well, and $(L_++L_-)/2$ is positive semidefinite if and only if each diagonal entry of this transformed average is nonnegative. 

For $0<p<1/2$, choose $\theta\in(0,\pi/2)$ with $\sin\theta=p/(1-p)$. We take \footnote{Write $L_\pm=S\pm A$ with $S=\frac12\bigl(\begin{smallmatrix}1-p&p\\p&1-p\end{smallmatrix}\bigr)$ and $A=\frac12\bigl(\begin{smallmatrix}0&p\\-p&0\end{smallmatrix}\bigr)$. Since $S\succ0$ and $\mathrm iA$ is Hermitian, they can be diagonalized simultaneously by congruence, as $W$ does. Then $W^*SW$ is real and $W^*AW$ is purely imaginary, so $W^*L_-W=\overline{W^*L_+W}$.}
\[
    W=\begin{pmatrix}
    1&1\\
    -\sin\theta+\mathrm i\cos\theta&-\sin\theta-\mathrm i\cos\theta
    \end{pmatrix}.
\]
By direct multiplication, we get
\begin{equation}\label{eq:local-congruence}
    W^*L_\pm W=\sqrt{1-2p}\,
    \operatorname{diag}(e^{\pm\mathrm i\theta},e^{\mp\mathrm i\theta}).
\end{equation}
The columns of $W$ are linearly independent, so $W$ is invertible. Averaging the two transformed matrices gives
\[
    W^*\frac{L_++L_-}{2}W=\frac{1-2p}{1-p}I_2,
\]
which is positive definite for $0<p<1/2$.

For multiple edges, the entries of $\gamma$ are correlated, so we cannot take a tensor product of the single-edge averages. Conditional on the cut, however, the pair of edge indicators in $\hat G,\hat H$ at each $ab\in A\times B$ depends only on whether $ab$ belongs to $R$ and to $P$. These membership choices are independent across different pairs of nodes, and independent of the choice of the cut. Let $K_{p,\gamma}$ denote the conditional coupling matrix, conditioned on the cut having sign matrix $\gamma=\alpha\beta^{\mathsf T}$. Then
\[
    K_{p,\gamma}=\bigotimes_{a\in A,b\in B}L_{\gamma_{ab}}.
\]
Consequently, $K_p=\mathbb{E}_\gamma K_{p,\gamma}$, where $\gamma$ is induced by independent uniform $\alpha$ and $\beta$. 

Proceeding as in the single-edge example, we apply $W$ to each edge coordinate simultaneously, using the invertible matrix $W^{\otimes m}$. For each pair $ab$, the cut fixes $\gamma_{ab}$ and hence the factor $W^*L_{\gamma_{ab}}W$. By~\Cref{eq:local-congruence}, its first and second diagonal entries are $\sqrt{1-2p}\,e^{\mathrm i\theta\gamma_{ab}}$ and $\sqrt{1-2p}\,e^{-\mathrm i\theta\gamma_{ab}}$, respectively. We use a separate sign $\varepsilon_{ab}$ to denote which entry is selected: $\varepsilon_{ab}=1$ for the first entry and $\varepsilon_{ab}=-1$ for the second. Thus each sign matrix $\varepsilon\in\{\pm1\}^{A\times B}$ indexes one diagonal entry of the tensor product, obtained by multiplying the selected entries.

\begin{claim}\label{cl:fixed-cut-reduction}
Let $0<p<1/2$, and let $\theta$ and $W$ be as above. For every cut sign matrix $\gamma$, the matrix $(W^{\otimes m})^*K_{p,\gamma}W^{\otimes m}$ is diagonal. Under the indexing above, its diagonal entry corresponding to $\varepsilon\in\{\pm1\}^{A\times B}$ is
\[
    (1-2p)^{m/2}\exp\!\left(\mathrm i\theta
    \sum_{a\in A,b\in B}\varepsilon_{ab}\gamma_{ab}\right).
\]
\end{claim}
\begin{proof}
The tensor-product multiplication rule and~\Cref{eq:local-congruence} give
\[
    (W^{\otimes m})^*K_{p,\gamma}W^{\otimes m}
    =\bigotimes_{a\in A,b\in B}(W^*L_{\gamma_{ab}}W).
\]
Each factor is diagonal, so their tensor product is diagonal. By~\Cref{eq:local-congruence}, the entry selected by $\varepsilon_{ab}$ in the factor corresponding to $ab$ is $\sqrt{1-2p}\,e^{\mathrm i\theta\varepsilon_{ab}\gamma_{ab}}$. Multiplying over the $m$ pairs gives the factor $(1-2p)^{m/2}$, while the exponents add to give the stated formula.
\end{proof}

Averaging over cuts gives the following criterion for positive semidefiniteness.

\begin{corollary}\label{cl:analytic-reduction}
Let $0<p<1/2$ and $\sin\theta=p/(1-p)$, where $\theta\in(0,\pi/2)$. If, for every $\varepsilon\in\{\pm1\}^{A\times B}$,
\[
    \mathbb{E}_{\alpha,\beta}\exp\!\left(\mathrm i\theta
    \sum_{a\in A,b\in B}\varepsilon_{ab}\alpha_a\beta_b\right)\ge0,
\]
where $\alpha\in\{\pm1\}^{A}$ and $\beta\in\{\pm1\}^{B}$ are independent uniform sign vectors, then $K_p\succeq0$.
\end{corollary}

\begin{proof}
Set $U=W^{\otimes m}$. By \Cref{cl:fixed-cut-reduction}, the matrix $U^*K_pU=\mathbb{E}_\gamma[U^*K_{p,\gamma}U]$ is diagonal, and its diagonal entry indexed by $\varepsilon$ is exactly the expectation above multiplied by $(1-2p)^{m/2}$. The expectation is real: replacing $\alpha$ by $-\alpha$ changes the sign of the exponent and hence conjugates the integrand. Since $(\alpha,\beta)$ and $(-\alpha,\beta)$ have the same distribution, the expectation equals its complex conjugate. Thus the hypothesis makes every diagonal entry nonnegative, and invertibility of $U$ gives $K_p\succeq0$.
\end{proof}

To prove the positivity of $K_p$, we prove that the diagonal entries of $U^*K_p U$ are positive. In fact, we prove a stronger analytic statement that applies to general real matrices. When applied with $T=\theta \varepsilon^{\mathsf T}$, the result implies the positivity of $K_p$.
\begin{theorem}\label{prop:cosine-positivity}
There is an absolute constant $c_0>0$ such that, if every row and column of $T\in\mathbb R^{B\times A}$ has Euclidean norm at most $c_0/\log(2n)$, then
\[
    \mathbb{E}_{\alpha,\beta}\exp\!\left(\mathrm i\beta^{\mathsf T}T\alpha\right)>0,
\]
where $\alpha\in\{\pm1\}^{A}$ and $\beta\in\{\pm1\}^{B}$ are independent uniform sign vectors.
\end{theorem}

We prove \Cref{prop:cosine-positivity} in \Cref{sec:positivity-proof}.

\begin{proof}[Proof of \Cref{cl:coupling-positivity} from \Cref{prop:cosine-positivity}]
Choose $p=c/(\sqrt n\log(2n))$ for a sufficiently small absolute constant $c>0$, and set $\theta=\arcsin(p/(1-p))$. For every $\varepsilon\in\{\pm1\}^{A\times B}$, set $T=\theta\varepsilon^{\mathsf T}$. Each row and column of $T$ has norm $\theta\sqrt{n/2}\le4p\sqrt{n/2}$, which is at most $c_0/\log(2n)$ for sufficiently small $c$. Since $\beta^{\mathsf T}T\alpha=\theta\sum_{a\in A,b\in B}\varepsilon_{ab}\alpha_a\beta_b$, \Cref{prop:cosine-positivity} verifies the hypothesis of \Cref{cl:analytic-reduction}, giving $K_p\succeq0$.
\end{proof}

\subsection{Proof of positivity}
\label{sec:positivity-proof}
We now prove \Cref{prop:cosine-positivity}. For brevity, we suppress the sets $A,B$ from the notation and prove the following theorem:
\begin{theorem}\label{thm:bi-positivity}
Let $T\in\mathbb R^{n\times n}$, $n\ge1$, and
$\tau=1/(12\log(2n))$, with natural logarithms. If every row
and column of $T$ has Euclidean norm at most $\tau$, then
\[
    Z(T):=\mathbb{E}_{\alpha,\beta}
    e^{\mathrm i\beta^{\mathsf T}T\alpha}>0,
    \qquad
    \alpha,\beta\text{ independent and uniform in }\{\pm1\}^n.
\]
\end{theorem}

Applied with matrix dimension $|A|=|B|$, this theorem gives \Cref{prop:cosine-positivity} with $c_0=1/12$ and concludes the proof of the lower bound on connectivity oracles.

For $x\in\mathbb R^d$, define $K(x)=\prod_{j=1}^d\cos x_j$. We begin with the following observation:

\begin{observation}
\[
    Z(T)=\mathbb{E}_\alpha K(T\alpha).
\]
\end{observation}
\begin{proof}
For every real number $t$ and a uniform sign $\beta_j\in\{\pm1\}$, we have $\mathbb{E}_{\beta_j}e^{\mathrm i\beta_jt}=(e^{\mathrm it}+e^{-\mathrm it})/2=\cos t$. Thus, for $x\in\mathbb R^n$, independence of the coordinates of $\beta$ gives
\[
    \mathbb{E}_\beta e^{\mathrm i\beta^{\mathsf T}x}
    =\prod_{j=1}^n\mathbb{E}_{\beta_j}e^{\mathrm i\beta_jx_j}
    =\prod_{j=1}^n\cos x_j=K(x).
\]
Applying this identity with $x=T\alpha$ and averaging over $\alpha$ gives $Z(T)=\mathbb{E}_\alpha K(T\alpha)$.
\end{proof}

Since $\cos\theta>0$ for $\theta\in(-\pi/2,\pi/2)$, the function $K$ is positive on the box $Q_0=(-\pi/2,\pi/2)^n$. Let $\phi\sim N(0,\tau I_n)$ be independent of $\alpha$, and let $\rho$ be the density of $T\alpha+\phi$. Explicitly, define
\begin{equation}\label{eq:bi-density}
    \rho(x)=(2\pi\tau)^{-n/2}\mathbb{E}_\alpha
    e^{-\|x-T\alpha\|^2/(2\tau)}.
\end{equation}
The following observation shows that this smoothing preserves the sign of $Z(T)$.

\begin{observation}\label{obs:bi-smoothing}
We have
\begin{equation}\label{eq:bi-smoothing}
    \int_{\mathbb R^n}\rho(x)K(x)\,dx=e^{-n\tau/2}Z(T).
\end{equation}
In particular, $Z(T)>0$ if and only if $\int\rho K>0$.
\end{observation}
\begin{proof}
For every real number $t$, the cosine addition formula gives
\[
    \mathbb{E}_{\phi_j}\cos(t+\phi_j)
    =\cos t\,\mathbb{E}_{\phi_j}\cos\phi_j
    -\sin t\,\mathbb{E}_{\phi_j}\sin\phi_j
    =e^{-\tau/2}\cos t,
\]
since $\phi_j\sim N(0,\tau)$ satisfies $\mathbb{E}_{\phi_j}\cos\phi_j=e^{-\tau/2}$ and, by symmetry, $\mathbb{E}_{\phi_j}\sin\phi_j=0$. By independence of the coordinates of $\phi$, it follows that $\mathbb{E}_\phi K(x+\phi)=e^{-n\tau/2}K(x)$ for every $x\in\mathbb R^n$. Averaging with $x=T\alpha$ gives
\[
    \int_{\mathbb R^n}\rho(x)K(x)\,dx
    =\mathbb{E}_{\alpha,\phi}K(T\alpha+\phi)
    =e^{-n\tau/2}\mathbb{E}_\alpha K(T\alpha)
    =e^{-n\tau/2}Z(T).
\]
\end{proof}

In light of the above, to prove that $Z(T)$ is positive, it suffices to prove that
$\int_{\mathbb R^n}\rho(x)K(x)\,dx>0$.
We show that the positive contribution from $Q_0$ dominates the total
absolute contribution from $Q_0^c$. Concretely, we prove that
\[
    \int_{Q_0^c}\rho(x)|K(x)|\,dx
    \le\frac14\int_{Q_0}\rho(x)K(x)\,dx.
\]
Since $\rho K>0$ on $Q_0$, this implies the desired positivity.

\subsubsection{Analytic preliminaries}\label{sec:bi-prelims}

We collect three estimates used later in the proof of \Cref{cl:bi-cube-comparison}.
Throughout, $C\in\mathbb R^{n\times n}$ is symmetric and positive definite with
\[
    c:=\max_jC_{jj}\le\frac1{4\log(2n)},
\]
we write $|x|_C^2=x^{\mathsf T}C^{-1}x$, and $G\sim N(0,C)$. For $k\in\mathbb Z^n$, let $Q_k=\pi k+Q_0$ be the open cube of side length $\pi$ centered at $\pi k$. The cubes are pairwise disjoint and cover $\mathbb R^n$ up to a null set. On $Q_k$, each factor $\cos x_j$ has the constant sign $(-1)^{k_j}$.

\begin{claim}\label{cl:bi-cosine-midpoint}
Let $k\in\mathbb Z^n$, $x\in Q_k$ and $y\in-Q_k$, and set $u=(x+y)/2$ and $v=(x-y)/2$. Then $u\in Q_0$, $v\in Q_k$, and
\[
    \sqrt{|K(x)K(y)|}\le K(u)|K(v)|.
\]
\end{claim}
\begin{proof}
Write $x=\pi k+a$ and $y=-\pi k+b$ with $a,b\in Q_0$. Since $Q_0$ is convex and symmetric, $u=(a+b)/2\in Q_0$ and $v=\pi k+(a-b)/2\in Q_k$. Since $-Q_k=Q_{-k}$, both $\cos x_j$ and $\cos y_j$ have sign $(-1)^{k_j}$, so their product is nonnegative. Using $x=u+v$ and $y=u-v$, and recalling that $\cos(\alpha+ \beta)=\cos(\alpha)\cos(\beta)-\sin(\alpha)\sin(\beta)$, we obtain
\[
    0\le\cos x_j\cos y_j
    =\cos(u_j+v_j)\cos(u_j-v_j)
    =\cos^2u_j\cos^2v_j-\sin^2u_j\sin^2v_j
    \le\cos^2u_j\cos^2v_j.
\]
Taking square roots, using $\cos u_j>0$, and multiplying over $j$ proves the claim.
\end{proof}

The next estimate is a standard consequence of \v{S}id\'ak's inequality, specialized to our setting.

\begin{claim}\label{cl:bi-box}
Let $B=\prod_j(-\delta_j,\delta_j)$ with $0\le\delta_j\le\pi/2$ for every $j$. Then
\[
    \Pr(G\in B)\ge\frac12\prod_j\sin\delta_j.
\]
\end{claim}
\begin{proof}
Recall that \v{S}id\'ak's inequality~\cite{sidak1967} states that for a centered Gaussian vector, the events that its coordinates lie in symmetric intervals are positively correlated: for all $\delta_1,\dots,\delta_n\ge0$,
\[
    \Pr(G\in B)=\Pr\bigl(|G_j|<\delta_j\text{ for all }j\bigr)\ge\prod_j\Pr(|G_j|<\delta_j).
\]
We bound each factor. The function $t\mapsto\Pr(|G_j|<t)$ vanishes at $0$, is increasing, and is concave on $[0,\infty)$, since its derivative is twice the density of $G_j$ at $t$. Hence, using the Gaussian tail bound with $C_{jj}\le c$ and $\min\{\delta_j,1\}\ge\sin\delta_j$, we obtain:
\[
    \Pr(|G_j|<\delta_j)
    \ge\min\{\delta_j,1\}\Pr(|G_j|<1)
    \ge\sin\delta_j\bigl(1-2e^{-1/(2c)}\bigr).
\]
Multiplying over $j$ and applying Bernoulli's inequality gives
\[
    \Pr(G\in B)
    \ge\Bigl(\prod_j\sin\delta_j\Bigr)\bigl(1-2ne^{-1/(2c)}\bigr)
    \ge\frac12\prod_j\sin\delta_j,
\]
since $c\le1/(4\log(2n))$ implies $2ne^{-1/(2c)}\le1/(2n)\le1/2$.
\end{proof}

\begin{claim}\label{cl:bi-gaussian-box}
For every $k\in\mathbb Z^n$ and $v\in Q_k$,
\begin{equation}\label{eq:bi-gaussian-box}
    e^{-|v|_C^2/2}|K(v)|\le2\Pr(G\in Q_k).
\end{equation}
\end{claim}
\begin{proof}
We use the following well-known consequence of the Cameron--Martin formula (see, e.g., \cite[Theorem~3.1]{li-shao2001}): for every symmetric Borel set $B$ and every $v\in\mathbb R^n$,
\[
    \Pr(G\in v+B)\ge e^{-|v|_C^2/2}\Pr(G\in B).
\]
Let $\delta_j=\pi/2-|v_j-\pi k_j|\in(0,\pi/2]$ be the distance from $v$ to the nearest $j$-faces of $Q_k$, and set $B=\prod_j(-\delta_j,\delta_j)$. Then $v+B\subseteq Q_k$ and $|\cos v_j|=\sin\delta_j$. The shift inequality and \Cref{cl:bi-box} give
\[
    \Pr(G\in Q_k)
    \ge\Pr(G\in v+B)
    \ge e^{-|v|_C^2/2}\Pr(G\in B)
    \ge\tfrac12e^{-|v|_C^2/2}|K(v)|.
\]
\end{proof}

\subsubsection{A criterion for domination by \texorpdfstring{$Q_0$}{Q0}}

We now give a criterion on a function $f$ under which the positive contribution $\int_{Q_0}fK$ dominates the total absolute contribution $\int_{Q_0^c}f|K|$. In the next subsubsection, we verify it for the smoothed density $\rho$.

Throughout, $C$, $c$, $G$ and the cubes $Q_k$ are as in \Cref{sec:bi-prelims}, and $f:\mathbb R^n\to(0,\infty)$ is even, measurable and integrable. We say that $f$ satisfies the midpoint condition (M) if
\begin{equation}\label{eq:bi-M}\tag{M}
    f(x+y)f(x-y)\le2f(x)^2e^{-|y|_C^2}\qquad(x,y\in\mathbb R^n).
\end{equation}
Since $|x+y|_C^2+|x-y|_C^2=2|x|_C^2+2|y|_C^2$, the density of $N(0,C)$ satisfies \eqref{eq:bi-M} with equality and without the factor $2$. More generally, without the factor $2$, \eqref{eq:bi-M} says exactly that $f(x)e^{|x|_C^2/2}$ is midpoint log-concave, that is, $f$ is at least as log-concave as $N(0,C)$. The factor $2$ is the slack we need, since $\rho$ is a mixture of Gaussians and need not be log-concave.

Set $h=f|K|$ and $I_k=\int_{Q_k}h$ for $k\in\mathbb Z^n$. Note that $I_0=\int_{Q_0}fK>0$. The key step is that, relative to $I_0$, the mass of $h$ in each cube is dominated up to a constant factor by the Gaussian mass of that cube.

\begin{claim}\label{cl:bi-cube-comparison}
If $f$ satisfies \eqref{eq:bi-M}, then for every $k\in\mathbb Z^n$,
\begin{equation}\label{eq:bi-cube-comparison}
    \frac{I_k}{I_0}\le2\sqrt2\,\Pr(G\in Q_k).
\end{equation}
\end{claim}
\begin{proof}
We compare $h$ on $Q_k$ and $-Q_k$ with $h$ on $Q_0$, using the midpoint form of the Pr\'ekopa--Leindler inequality~\cite{gardner2002}. This inequality states that if measurable and integrable functions $F_0,F_1,F_2:\mathbb R^n\to[0,\infty)$ satisfy
\begin{equation}\label{eq:bi-pl-condition}
    F_0\!\left(\frac{x+y}2\right)\ge\sqrt{F_1(x)F_2(y)}
    \qquad(x,y\in\mathbb R^n),
\end{equation}
then
\[\int F_0\ge\sqrt{\int F_1\int F_2}.\]

\noindent Fix $k\in\mathbb Z^n$, and let
\[
    F_0=2\sqrt2\,\Pr(G\in Q_k)\,h\mathbf1_{Q_0},
    \qquad F_1=h\mathbf1_{Q_k},
    \qquad F_2=h\mathbf1_{-Q_k}.
\]
Since $h$ is even, $\int F_1=\int F_2=I_k$, while $\int F_0=2\sqrt2\,\Pr(G\in Q_k)I_0$. Hence the claim follows once we verify \eqref{eq:bi-pl-condition}.

The right-hand side of \eqref{eq:bi-pl-condition} vanishes unless $x\in Q_k$ and $y\in-Q_k$, so fix such $x,y$, and set $u=(x+y)/2$ and $v=(x-y)/2$. By \Cref{cl:bi-cosine-midpoint}, $u\in Q_0$, $v\in Q_k$, and $\sqrt{|K(x)K(y)|}\le K(u)|K(v)|$. Applying \eqref{eq:bi-M} at the point $u$ with shift $v$ gives $\sqrt{f(x)f(y)}\le\sqrt2\,f(u)e^{-|v|_C^2/2}$. Multiplying these two bounds, using $K(u)>0$, and applying \Cref{cl:bi-gaussian-box} at $v$ gives
\[
    \sqrt{h(x)h(y)}
    \le\sqrt2\,h(u)\,e^{-|v|_C^2/2}|K(v)|
    \le2\sqrt2\,\Pr(G\in Q_k)\,h(u)
    =F_0(u),
\]
which is \eqref{eq:bi-pl-condition}.
\end{proof}

Summing over the cubes gives us the criterion.
\begin{lemma}\label{lem:bi-midpoint}
If $f$ satisfies \eqref{eq:bi-M}, then
\[
    \int_{Q_0^c}f|K|\le\frac14\int_{Q_0}fK.
\]
\end{lemma}
\begin{proof}
The cubes partition $\mathbb R^n$ up to a null set, so $\int_{Q_0^c}f|K|=\sum_{k\ne0}I_k$. Summing \Cref{cl:bi-cube-comparison} over $k\ne0$, and then applying a union bound and the Gaussian tail bound, gives
\[
    \sum_{k\ne0}I_k
    \le2\sqrt2\,\Pr(G\notin Q_0)\,I_0
    \le2\sqrt2\sum_{j=1}^n\Pr(|G_j|\ge\pi/2)\,I_0
    \le4\sqrt2\,ne^{-\pi^2/(8c)}I_0
    \le\frac14I_0,
\]
since $c\le1/(4\log(2n))$ gives $e^{-\pi^2/(8c)}\le(2n)^{-\pi^2/2}\le1/(16\sqrt2\,n)$.
\end{proof}

\begin{corollary}\label{cor:bi-positive-integral}
If $f$ satisfies \eqref{eq:bi-M}, then
\[
    \int fK\ge\frac34\int_{Q_0}fK>0,
    \qquad
    \int f|K|\le2\int fK.
\]
\end{corollary}
\begin{proof}
By \Cref{lem:bi-midpoint}, $\sum_{k\ne0}I_k\le I_0/4$. Therefore
\[
    \int fK\ge I_0-\sum_{k\ne0}I_k\ge\frac34I_0>0,
    \qquad
    \int f|K|\le\frac54I_0\le\frac53\int fK\le2\int fK.
\]
\end{proof}
\subsubsection{Applying the criterion to the smoothed density}

It remains to show that $\rho$ satisfies \eqref{eq:bi-M} with
\[
    C=C_T:=\tau I_n+\frac2\tau\,TT^{\mathsf T}.
\]
This does not follow from general principles: $\rho$ is a mixture of $2^n$ Gaussians centered at the points $T\alpha$, and such a mixture need not be log-concave. Instead, we use the fact that the centers $T\alpha$ are signed sums of the columns of $T$.

For $D\subseteq[n]$, let $T_D$ be the $n\times|D|$ submatrix of $T$ formed by the columns indexed by $D$, and let
\[
    F_D(y)=\mathbb{E}_{s\in\{\pm1\}^{D}}\,e^{-\|y-T_Ds\|^2/\tau}
    \qquad(y\in\mathbb R^n).
\]
Up to normalization, $F_D$ is a mixture of identical Gaussian bumps centered at the $2^{|D|}$ signed sums $T_Ds$ of the columns in $D$. For example, $F_\varnothing(y)=e^{-\|y\|^2/\tau}$ is a single bump at the origin.

\paragraph{Decoupling the two factors.}
The product $\rho(x+y)\rho(x-y)$, by the definition of $\rho$ in \Cref{eq:bi-density}, involves two independent sign vectors $\alpha,\alpha'$, one for each factor. On the coordinates where $\alpha$ and $\alpha'$ agree, the two copies move together, and these coordinates only interact with $x$. On the coordinates where they disagree, the copies move in opposite directions, and these coordinates only interact with $y$. The next claim makes this precise.

\begin{claim}\label{cl:bi-two-copies}
For every $x,y\in\mathbb R^n$,
\begin{equation}\label{eq:bi-two-copies}
    (2\pi\tau)^n\rho(x+y)\rho(x-y)=\mathbb E_D\bigl[F_{D^c}(x)\,F_D(y)\bigr],
\end{equation}
where $D$ is a uniformly random subset of $[n]$.
\end{claim}
\begin{proof}
Let $\alpha,\alpha'\in\{\pm1\}^n$ be independent and uniform. By \eqref{eq:bi-density},
\[
    (2\pi\tau)^n\rho(x+y)\rho(x-y)
    =\mathbb E_{\alpha,\alpha'}\exp\Bigl(-\frac{\|x+y-T\alpha\|^2+\|x-y-T\alpha'\|^2}{2\tau}\Bigr).
\]
Let $D=\{j:\alpha_j\ne\alpha'_j\}$, $a=(\alpha+\alpha')/2$ and $b=(\alpha-\alpha')/2$. Then $D$ is a uniformly random subset of $[n]$. Conditional on $D$, the vectors $a$ and $b$ are independent uniform sign vectors on $D^c$ and $D$, respectively, and zero elsewhere. Since $\alpha=a+b$ and $\alpha'=a-b$, the parallelogram identity $\|p+q\|^2+\|p-q\|^2=2\|p\|^2+2\|q\|^2$ with $p=x-Ta$ and $q=y-Tb$ turns the exponent into $-\|x-Ta\|^2/\tau-\|y-Tb\|^2/\tau$. Conditional on $D$, the two resulting factors are independent, with expectations $F_{D^c}(x)$ and $F_D(y)$.
\end{proof}

Applying \eqref{eq:bi-two-copies} once at $y$ and once at $y=0$ expresses both sides of the desired inequality $\rho(x+y)\rho(x-y)\le2\rho(x)^2e^{-|y|_{C_T}^2}$ as averages over $D$:
\[
    (2\pi\tau)^n\rho(x+y)\rho(x-y)=\mathbb E_D\bigl[F_{D^c}(x)F_D(y)\bigr],
    \qquad
    (2\pi\tau)^n\rho(x)^2=\mathbb E_D\bigl[F_{D^c}(x)F_D(0)\bigr].
\]
Since $F_{D^c}(x)\ge0$, it suffices to compare the two averages term by term, that is, to show that for every $D\subseteq[n]$ and $y\in\mathbb R^n$,
\[
    F_D(y)\le2e^{-|y|_{C_T}^2}F_D(0).
\]
We prove this next.

\begin{lemma}\label{lem:bi-subset-comparison}
For every $D\subseteq[n]$ and $y\in\mathbb R^n$,
\begin{equation}\label{eq:bi-subset-comparison}
    F_D(y)\le2e^{-|y|_{C_T}^2}F_D(0).
\end{equation}
\end{lemma}
\begin{proof}
\emph{Completing the square.}
Fix $D\subseteq[n]$ and $y\in\mathbb R^n$. Extend each $s\in\{\pm1\}^D$ by zero to a vector in $\mathbb R^n$, so that $T_Ds=Ts$ and $\|s\|^2=|D|$. We first extract the factor $e^{-|y|_{C_T}^2}$ from $F_D(y)$ by completing the square in $s$. Since $\|s\|^2=|D|$, we may write
\[
    F_D(y)
    =e^{\tau|D|/2}\,
    \mathbb E_{s\in\{\pm1\}^D}
    \exp\!\left(
        -\frac{\|y-Ts\|^2}{\tau}
        -\frac{\tau\|s\|^2}{2}
    \right).
\]
The factor $e^{\tau|D|/2}$ is independent of $y$ and therefore cancels in the ratio $F_D(y)/F_D(0)$.

We now complete the square in the exponent. Set $\Sigma=\tau I_n+(2/\tau)T^{\mathsf T}T$. Expanding gives
\[
    \frac{\|y-Ts\|^2}{\tau}+\frac{\tau\|s\|^2}{2}
    =\frac{\|y\|^2}{\tau}
      +\frac12s^{\mathsf T}\Sigma s-\frac2\tau s^{\mathsf T}T^{\mathsf T}y.
\]
Since $\Sigma\succeq\tau I_n\succ0$, we may define $w=(2/\tau)\Sigma^{-1}T^{\mathsf T}y$. Then $(2/\tau)s^{\mathsf T}T^{\mathsf T}y=s^{\mathsf T}\Sigma w$, so completing the square yields
\[
    \frac{\|y-Ts\|^2}{\tau}+\frac{\tau\|s\|^2}{2}
    =\frac{\|y\|^2}{\tau}-\frac12w^{\mathsf T}\Sigma w
      +\frac12(s-w)^{\mathsf T}\Sigma(s-w).
\]
To identify the term independent of $s$, note that $\frac12w^{\mathsf T}\Sigma w=\frac2{\tau^2}y^{\mathsf T}T\Sigma^{-1}T^{\mathsf T}y$. The Woodbury identity gives
\[
    C_T^{-1}=\frac1\tau I_n-\frac2{\tau^2}T\Sigma^{-1}T^{\mathsf T},
\]
so $\|y\|^2/\tau-\frac12w^{\mathsf T}\Sigma w=y^{\mathsf T}C_T^{-1}y=|y|_{C_T}^2$. Thus
\begin{equation}\label{eq:bi-complete-square}
    \frac{\|y-Ts\|^2}{\tau}+\frac{\tau|D|}{2}
    =|y|_{C_T}^2+\frac12(s-w)^{\mathsf T}\Sigma(s-w).
\end{equation}

For $z\in\mathbb R^n$, define
\[
    g_D(z):=\mathbb E_{s\in\{\pm1\}^D}
    e^{-\frac12(s-z)^{\mathsf T}\Sigma(s-z)}.
\]
Negating and exponentiating \eqref{eq:bi-complete-square}, and then averaging over $s$, gives
\[
    F_D(y)=e^{\tau|D|/2}e^{-|y|_{C_T}^2}g_D(w).
\]
Since $w=0$ when $y=0$, we also have $F_D(0)=e^{\tau|D|/2}g_D(0)$. The common factor $e^{\tau|D|/2}$ cancels in the ratio $F_D(y)/F_D(0)$. As $w$ need not be supported on $D$, it suffices to prove that $g_D(z)\le2g_D(0)$ for every $z\in\mathbb R^n$.

\emph{Bounding the Gaussian mixture.}
If $D=\varnothing$, then $g_D(z)=e^{-z^{\mathsf T}\Sigma z/2}\le1=g_D(0)$, so assume $D\ne\varnothing$. Let $G'\sim N(0,\Sigma)$ and write $G'_D=(G'_j)_{j\in D}$. We use the Gaussian characteristic function to express the shift $z$ as a factor of modulus $1$. For every $u\in\mathbb R^n$, we have $\mathbb E_{G'}e^{\mathrm i\langle u,G'\rangle}=e^{-u^{\mathsf T}\Sigma u/2}$. Applying this identity with $u=z-s$ and averaging over $s$ gives
\[
    g_D(z)
    =\mathbb E_{G'}\!\left[e^{\mathrm i\langle z,G'\rangle}\mathbb E_s e^{-\mathrm i\langle s,G'\rangle}\right]
    =\mathbb E_{G'}\!\left[e^{\mathrm i\langle z,G'\rangle}K(G'_D)\right],
\]
where the second equality uses the independence of the signs $s_j$ and the identity $\mathbb E_{s_j}e^{-\mathrm is_jt}=\cos t$. In particular, $g_D(0)=\mathbb E K(G'_D)$. Since $g_D(z)>0$ and the phase has modulus $1$, taking absolute values gives $g_D(z)\le\mathbb E|K(G'_D)|$.

To compare this last expectation with $g_D(0)$, we apply \Cref{cor:bi-positive-integral} to the Gaussian density of $G'_D$. Its covariance $\Sigma_{DD}:=(\Sigma_{ij})_{i,j\in D}$ is a principal submatrix of $\Sigma$, and hence is positive definite. Its density is even, positive and integrable. As noted after \eqref{eq:bi-M}, it satisfies the midpoint condition with $C=\Sigma_{DD}$, even without the factor $2$. By the column assumption, for every $j\in D$,
\[
    \Sigma_{jj}
    =\tau+\frac2\tau\sum_iT_{ij}^2
    \le3\tau
    =\frac1{4\log(2n)}
    \le\frac1{4\log(2|D|)}.
\]
Thus \Cref{cor:bi-positive-integral} applies in dimension $|D|$ and gives
\[
    g_D(z)
    \le\mathbb E|K(G'_D)|
    \le2\,\mathbb E K(G'_D)
    =2g_D(0).
\]
The factor $2$ enters only when we replace the signed cosine product by its absolute value. Substituting this bound into the expression for $F_D(y)$ proves \eqref{eq:bi-subset-comparison}.
\end{proof}

\begin{proof}[Proof of \Cref{thm:bi-positivity}]
For every $x,y\in\mathbb R^n$, we have
\[
    \rho(x+y)\rho(x-y)
    \overset{\text{(1)}}{=}\frac{\mathbb E_D[F_{D^c}(x)F_D(y)]}{(2\pi\tau)^n}
    \overset{\text{(2)}}{\le}\frac{2e^{-|y|_{C_T}^2}\mathbb E_D[F_{D^c}(x)F_D(0)]}{(2\pi\tau)^n}
    \overset{\text{(3)}}{=}2e^{-|y|_{C_T}^2}\rho(x)^2,
\]
where (1) follows from \Cref{cl:bi-two-copies}, (2) follows from \Cref{lem:bi-subset-comparison} since $F_{D^c}(x)\ge0$, and (3) follows from \Cref{cl:bi-two-copies} at $y=0$. Thus $\rho$ satisfies \eqref{eq:bi-M} with $C=C_T$.

The density $\rho$ is positive and integrable, and it is even by \eqref{eq:bi-density} and the symmetry of $\alpha$. Moreover, $C_T\succeq\tau I_n\succ0$, and the row assumption gives, for every $i\in[n]$,
\[
    (C_T)_{ii}
    =\tau+\frac2\tau\sum_jT_{ij}^2
    \le3\tau
    =\frac1{4\log(2n)}.
\]
Thus \Cref{cor:bi-positive-integral} applies to $\rho$. Together with \Cref{obs:bi-smoothing}, it yields
\[
    Z(T)
    =e^{n\tau/2}\int\rho K
    \ge\frac34e^{n\tau/2}\int_{Q_0}\rho K
    >0,
\]
since $\rho K>0$ on $Q_0$.
\end{proof}

\subsection{Remarks}\label{subsec:remarks}

The argument in \Cref{subsec:coupling-bound} can also be viewed as a certificate for the Lov\'asz theta function. For a finite graph $Y$ with at least one vertex, one may define the theta function using the following semidefinite formulation~\cite[Theorem~4]{Lovasz1979}:
\[
    \vartheta(Y)=\max\left\{\mathbf1^{\mathsf T}M\mathbf1:
    M\succeq0,\ \operatorname{tr} M=1,\
    M_{uv}=0\text{ for }\{u,v\}\in E(Y)\right\}.
\]
Here $M$ is a real symmetric matrix indexed by $V(Y)$, and $\mathbf1$ is the all-ones vector. The following theorem of Lov\'asz shows that $\vartheta$ can be used to lower bound the chromatic number of a graph.

\begin{theorem}[Lov\'asz~\cite{Lovasz1979}]
For every finite graph $Y$ with at least one vertex,
\[
    \alpha(Y)\le\vartheta(Y)\le\chi(\overline Y),
\]
where $\alpha(Y)$ is the maximum size of an independent set in $Y$, and $\overline Y$ is its complement.
\end{theorem}

Let $K$ be the positive semidefinite coupling matrix of a confusing distribution $\mu$ on pairs $(\hat G,\hat H)$. Since $\mu$ is confusing, $K_{G,H}=0$ whenever $G,H$ are distinct and nonadjacent in $X_n$. Also, $\mathbf1^{\mathsf T}K\mathbf1=1$ and $\operatorname{tr} K=\Pr[\hat G=\hat H]>0$, where positivity follows because $K$ is nonzero and positive semidefinite. Thus $K/\operatorname{tr} K$ is feasible for $\vartheta(\overline{X_n})$, and certifies
\[
    \chi(X_n)\ge\vartheta(\overline{X_n})
    \ge\frac{\mathbf1^{\mathsf T}K\mathbf1}{\operatorname{tr} K}
    =\frac1{\Pr[\hat G=\hat H]}.
\]
This recovers the lower bound in \Cref{cl:collision-lower-bound}. We chose the probabilistic exposition to make the argument more self-contained and, in our view, more intuitive.

Since $\vartheta(\overline Y)$ also lower bounds the fractional chromatic number of $Y$~\cite[Theorem~10]{Lovasz1979}, the same asymptotic space lower bound holds for the worst-case expected encoding length of Las Vegas data structures, by an argument similar to that in Assadi, Farach-Colton, and Kuszmaul~\cite[Lemma~3.1]{AssadiFarachColtonKuszmaul2023}.

\paragraph{Bounded faulty degree}

We also remark that the above lower bound applies to bounded-degree faults \cite{BodwinHP24}. For a set of edges $F$ with vertices in $[n]$, define its \emph{faulty degree} by
\[
    d(F)=\max_{v\in[n]}|\{e\in F:v\in e\}|.
\]
In the bounded-degree failure setting, the valid queries are sets $F \subseteq E$, with $d(F)\leq d$, where $d$ is some specified parameter. By noting that the $f$ faults in each block of size $\sqrt f$ used in the reduction of Appendix \ref{sec:LB-reduction} have bounded degree at most $\sqrt{f}$, one gets the following theorem: 

\begin{theorem}\label{thm:faulty-degree-lower-bound}
For all sufficiently large $n$ and every integer $1\le d\le n-1$, every deterministic data structure that encodes a simple graph $G$ on $[n]$ and, given any $F\subseteq E(G)$ with $d(F)\le d$, returns the connected-component partition of $G-F$ requires
\[
    \Omega\!\left(\frac{n\sqrt d}{\log(2d)}\right)
\]
bits in the worst case.
\end{theorem}
\noindent The adaptation is straightforward, so its proof is omitted.

\section{Fast queries with $\tilde{O}(\sqrt{f}n)$ space}
\label{sec:fast-query}

We obtain $\widetilde O(f)$ query time using
$\widetilde O(\sqrt f \cdot n)$ bits of space. This improves over the polynomial query time of \cite{KKP206} and improves the space bound from $fn$ to $\sqrt{f}n$ in the classical construction of \cite{PatrascuT07}. Following the decomposition
framework of Khanna, Konrad, and Putterman~\cite{KKP206}, we partition
the edges into high-degree expander pieces and a sparse remainder
$R$. Our improvement over~\cite{KKP206} combines two ingredients:
faster deterministic sparse recovery and a fixed sparse graph
preprocessed for connectivity queries. Specifically, we replace
each expander by a star with an auxiliary center, obtaining a graph
$J$ with $\widetilde O(n\sqrt f)$ edges (that consists of the stars and the subgraph $R \subseteq G$), and preprocess $J$ using
the connectivity oracle of P\u{a}tra\c{s}cu and
Thorup~\cite{PatrascuT07}. Given a failure set $F$, we delete
the failed remainder edges and the star edges at vertices whose
residual degree within their expander is small. Denote this set of deleted edges by $F’\subseteq E(J)$. These $O(f)$ deletions
yield a compact component representation in $J-F'$ in $\widetilde O(f)$ time.
We then recover the surviving edges incident to these low-degree
vertices from their stored sketches and merge the components
joined by the recovered edges. Only $O(f)$ edges are recovered,
and the resulting representation describes the connected
components of $G-F$. Thus each query processes only $O(f)$
changes to the preprocessed graph, taking $\widetilde O(f)$
time in total.

 The preprocessing algorithm
$\Enc$ stores the oracle, and the decoding algorithm
$\operatorname{Dec}$ processes a valid query. 

\begin{theorem}\label{thm:fast-connectivity}
Let $G$ be a simple undirected $n$-vertex graph and
$1\le f\le n^2$. There is an oracle with polynomial deterministic preprocessing
time and $\widetilde O(\sqrt f\,n)$ bits of space such that, for every
$F\subseteq E(G)$ with $|F|\le f$, the decoder computes an implicit
representation of $\operatorname{cc}(G-F)$ in $\widetilde O(f)$ time.
This representation returns the component identifier of any vertex
in $\widetilde O(1)$ time. In particular, a pairwise connectivity
query $(s,t,F)$ is answered in $\widetilde O(f)$ time.
\end{theorem}

Two vertices receive the same identifier exactly when they belong
to the same component. Reporting the entire partition explicitly
takes an additional $\widetilde O(n)$ time. The same sparse
representation also yields a distance oracle.

\begin{theorem}\label{thm:fast-distances}
Let $G$ be a simple undirected $n$-vertex, $m$-edge graph with
positive integer edge lengths bounded by $n^{O(1)}$, and let
$1\le f\le n^2$. There is an oracle with $(m+n)^{1+o(1)}$
preprocessing time and $\sqrt f\,n^{1+o(1)}$ bits of space that
answers every query $(s,t,F)$, where $F\subseteq E(G)$ and
$|F|\le f$, in $f n^{o(1)}$ time with $\operatorname{polylog} n$
multiplicative stretch. The query specifies the length of each
failed edge. Disconnected pairs are reported exactly.
\end{theorem}

\subsection{The ingredients}
\label{subsec:fast-primitives}

We use three ingredients: an expander decomposition, deterministic
sparse recovery, and a component-reporting oracle on a fixed graph.
For a graph $X$, let $d_X(v)=|N_X(v)|$ and
$\operatorname{vol}_X(S)=\sum_{v\in S}d_X(v)$. We call $X$ a
$\phi$-expander if every nonempty proper subset $S\subset V(X)$
satisfies
\[
 |\delta_X(S)|\ge\phi\min\{
   \operatorname{vol}_X(S),
   \operatorname{vol}_X(V(X)\setminus S)\}.
\]

\paragraph{High-degree expander pieces.}
The decomposition supplies edge-disjoint pieces, together with a
short index for locating the piece containing a queried edge.
The remainder $R$ below is an edge set.

\begin{lemma}\label{lem:fast-decomposition}
There is a parameter $\phi=\log^{-O(1)}(2n)$ such that, for every
integer $D\ge1$, a simple $n$-vertex, $m$-edge graph can be
decomposed deterministically in $(m+n)^{1+o(1)}$ time into a
collection $\mathcal X$ of subgraphs and a remainder $R$, with
\[
 E(G)=R\mathbin{\dot\cup}
      \mathop{\dot\bigcup}_{X\in\mathcal X}E(X),
\]
satisfying the following properties:
\begin{enumerate}
 \item Every $X\in\mathcal X$ is a $\phi$-expander with minimum degree at least $D$. 
 \item $|R|=O(nD/\phi)$, and every vertex belongs to
 $O(\log n)$ pieces.
 \item From $O(\log n)$ stored identifiers per vertex, the
 unique piece containing an original edge, or its membership
 in $R$, can be determined in $O(\log n)$ time.
\end{enumerate}
\end{lemma}

\begin{proof}
We use Chuzhoy's deterministic expander
decomposition~\cite{Chuzhoy23}, in the degree-preserving form
stated by Li and Vaughn~\cite[Theorem~13]{LiVaughn26}.
On a graph $J$, it returns a vertex partition with at most
$|E(J)|/2$ crossing edges. For every cluster $U$, the graph
$J\langle U\rangle$, obtained from $J[U]$ by adding loops that
preserve the degrees in $J$, has conductance at least
$\phi=\log^{-9-o(1)}(2n)$. The running time is
$(|E(J)|+|V(J)|)^{1+o(1)}$.

Set $\tau=\lceil D/\phi\rceil$. In a round with residual
graph $R_j$ and $m_j$ edges, stop if $m_j\le2n\tau$.
Otherwise, repeatedly peel vertices of degree less than $\tau$,
removing at most $n\tau$ edges, and decompose the remaining
graph $J$. Retain $J[U]$ for every nonsingleton cluster $U$.
For $v\in U$, both sides of its singleton cut in
$J\langle U\rangle$ have volume at least $\tau$, so
$d_{J[U]}(v)\ge\phi\tau\ge D$.
Removing loops decreases the volumes in the conductance
denominator and preserves the bound $\phi$.

The next residual consists of the peeled edges and the crossing
edges. If $p_j\le n\tau$ edges were peeled, then
\[
 m_{j+1}\le p_j+\frac{m_j-p_j}{2}
 \le\frac{m_j+n\tau}{2}<\frac34m_j.
\]
Thus there are $O(\log n)$ rounds, and the terminal remainder
has $O(nD/\phi)$ edges. A vertex belongs to at most one retained
piece per round. The pieces are edge-disjoint because their
edges are removed from the residual graph, and the geometric
decrease gives the claimed total running time.

For every vertex and round, store its retained-piece identifier,
or a null identifier. The owner of an original edge $\{u,v\}$
is the first round in which its endpoints have the same nonnull
identifier. Until that round the edge remains residual, and in
that round it belongs to the retained induced cluster. If no
such round exists, the edge belongs to $R$.
\end{proof}

The following robustness property explains why a star can
represent the vertices whose residual degree remains large.

\begin{lemma}\label{lem:fast-robustness}
Let $k=\lceil\sqrt f\rceil$, let $0<\phi\le1$, and let $X$
be a simple $n$-vertex $\phi$-expander graph with minimum degree at least $D\ge8k/\phi$.
After at most $f$ edge failures, every two vertices of residual
degree greater than $k$ have distance at most
$L=O(\phi^{-1}\log(2n))$.
\end{lemma}

\begin{proof}
Let $F_X\subseteq E(X)$ be the failures, and set
\[
 B=\{v:d_{F_X}(v)\ge\phi D/4\},\qquad
 U=V(X)\setminus B,\qquad Y=(X-F_X)[U].
\]
Counting failed incidences gives
$|B|\le8f/(\phi D)\le k\le\phi D/8$.
In particular, $U$ is nonempty and $Y$ has minimum degree
at least $D-\phi D/4-|B|\ge D/2$.

We first show that $Y$ has conductance at least $\phi/2$.
Choose the smaller-volume side $S$ of a cut in $Y$, and put
$a=\operatorname{vol}_Y(S)$. Both sides of the corresponding
cut in $X$ have volume at least $a$. At most $f$ crossing
edges fail, and at most $|S||B|$ lead to $B$. Since
$|S|\le2a/D$,
\[
 |\delta_Y(S)|\ge\phi a-f-|S||B|
 \ge\frac{3\phi}{4}a-f.
\]
If $a\ge4f/\phi$, this is at least $\phi a/2$.
Otherwise, $|S|<8f/(\phi D)\le D/8$, and simplicity gives
$2|E_Y(S)|\le |S|^2\le a/4$, where $E_Y(S)$ denotes the
edges with both endpoints in $S$. Hence
$|\delta_Y(S)|\ge3a/4\ge\phi a/2$ in this case as well.

A ball in $Y$ containing at most half its total volume grows
by a factor at least $1+\phi/2$ in one step. Since the total
volume is less than $n^2$, the diameter of $Y$ is
$O(\phi^{-1}\log(2n))$. Every vertex of residual degree
greater than $k$ either lies in $U$ or has a surviving neighbor
in $U$, because $|B|\le k$. Two additional edges therefore
suffice to connect any two such vertices through $Y$.
\end{proof}

\paragraph{Deterministic sparse recovery.}
We use the following interface, based on the sketch
of~\cite{IzumiEWM23}, as a black box. Its proof is given
in Appendix \ref{sec:proof-sketching-oracle}.
For a vertex $v$, write $E_G(v)$ for its incident-edge set.

\begin{lemma}\label{lem:sketching-oracle}\label{lem:fast-sketch}
Let $G=(V,E)$ be a simple $n$-vertex, $m$-edge graph, and let
$1\le k\le n^2$. There is a deterministic preprocessing
algorithm that assigns every vertex $v$ a sketch
$s(v)=\sigma_k(E_G(v))$, where $\sigma_k$ is linear over
$\mathbb F_2$ on edge-incidence vectors, with the following
guarantees.
\begin{enumerate}[label=(\roman*)]
 \item Each sketch uses $O(k\log n)$ bits.
 \item Given a list $F_v\subseteq E_G(v)$ of failed edges
 incident to $v$, if $|N_{G-F_v}(v)|\le k$, the residual
 neighbor set $N_{G-F_v}(v)$ can be reported in
 $(|F_v|+k)(\log n)^{O(1)}$ time.
 \item Under the same sparsity promise, if the residual sketch
 $s(v)+\sigma_k(F_v)=\sigma_k(E_G(v)\setminus F_v)$ has
 already been formed, the decoding time is
 $k(\log n)^{O(1)}$.
 \item All $n$ sketches occupy $O(nk\log n)$ bits and can be
 constructed in $(m+nk)(\log n)^{O(1)}$ total deterministic
 time. In particular, when $m>nk$, construction takes
 $\widetilde O(m)$ time.
\end{enumerate}
\end{lemma}

The decoder has no guarantee when the residual degree exceeds
$k$; in particular, it need not detect a violation of the
promise. We therefore store the degrees and invoke recovery
only after verifying that the residual degree is at most $k$.
The construction is vertexwise: a sketch at $v$ can be formed
in $\widetilde O(d_G(v)+k)$ time. We use this bound when
applying the lemma separately to the expander pieces, with
their inherited vertex labels.

\paragraph{The base connectivity oracle.}
We use \cite{PatrascuT07} through the following interface.
For a fixed graph $J$ with $N$ vertices and $M$ edges, it has
polynomial preprocessing time and $\widetilde O(M+N)$-bit
space. A deletion set $D_J\subseteq E(J)$ is processed in
$\widetilde O(|D_J|+1)$ time, after which the component
identifier of any vertex in $J-D_J$ is available in
$\widetilde O(1)$ time.
To handle additional edge insertions, we query the identifiers
of their endpoints and merge the corresponding components.
Only identifiers touched by inserted edges need to be stored
in this auxiliary computation; see \cite[Section~2.1]{PatrascuT07}.

\subsection{Preprocessing and the query algorithm}
\label{subsec:fast-repairs}

Set $k=\lceil\sqrt f\rceil$ and
$D=\lceil8k/\phi\rceil$, and apply
\Cref{lem:fast-decomposition}. Introduce a distinct auxiliary
vertex $c_X$ for each piece $X\in\mathcal X$, and form
\begin{equation}\label{eq:fast-base}
 J=\left(
 V\cup\{c_X:X\in\mathcal X\},\;
 R\cup\bigl\{\{v,c_X\}:X\in\mathcal X,\ v\in V(X)\bigr\}
 \right).
\end{equation}
Queries concern the original vertices $V$; the centers are
used only to represent connections within the pieces.

\paragraph{Preprocessing.}
The encoding $\Enc(G)$ stores:
\begin{enumerate}
 \item the sparse graph $J$ and its base connectivity oracle;
 \item the per-vertex identifiers from
 \Cref{lem:fast-decomposition}, for locating the owner of
 each queried edge; and
 \item for every incidence $(v,X)$ with $v\in V(X)$, the
 degree $d_X(v)$ and a $k$-sparse-recovery sketch $s_X(v)$
 of the $X$-edges incident to $v$.
\end{enumerate}
After constructing these records, we discard the edge sets
of the expander pieces.

\paragraph{Processing a failure query.}
Given $F\subseteq E(G)$ with $|F|\le f$, do the following.
\begin{enumerate}
 \item Scan $F$ using the owner identifiers. Form the nonempty
 lists $F_X(v)$ of failed $X$-edges incident to $v$, and
 initialize $D_F=F\cap R$ and $I_F=\varnothing$.
 \item For each listed incidence $(v,X)$ with
 $d_X(v)-|F_X(v)|\le k$, add $(v,c_X)$ to $D_F$.
 Recover the surviving $X$-neighbors of $v$ using
 \Cref{lem:sketching-oracle}, and add their incident edges
 to $I_F$.
 \item Process $D_F$ in the base oracle, writing
 $\lambda_0(v)$ for the component identifier of $v$ in
 $J-D_F$. For every $\{u,v\}\in I_F$, union
 $\lambda_0(u)$ and $\lambda_0(v)$ in a union--find structure.
 \item Return the implicit partition whose identifier for
 $v$ is the representative of $\lambda_0(v)$, treating
 untouched identifiers as singleton sets. A pairwise query
 compares the representatives of $s$ and $t$.
\end{enumerate}

\subsection{Correctness and complexity}
\label{subsec:fast-analysis}

Let $J_F=(J-D_F)\cup I_F$ be the conceptual repaired graph;
the decoder stores only its modification lists and the
union--find structure. For every incidence $(v,X)$, write
$r_X(v)=d_X(v)-|F_X(v)|$, taking unlisted failure lists to
be empty. An untouched incidence has residual degree
$d_X(v)\ge D>k$, so the scan finds every incidence requiring
recovery.

\begin{lemma}\label{lem:fast-repairs}
The lists $D_F$ and $I_F$ contain $O(|F|)$ edges and can be
computed in $\widetilde O(|F|+1)$ time. For all $s,t\in V$,
the vertices $s,t$ are connected in $J_F$ if and only if
they are connected in $G-F$.
\end{lemma}

\begin{proof}
Let $b_F$ be the number of incidences $(v,X)$ with
$r_X(v)\le k$. Each such incidence has lost at least $D-k$
edges. Because the pieces are edge-disjoint, each failed
edge is counted at most twice. Therefore
\begin{equation}\label{eq:fast-repair-count}
 b_F\le\frac{2|F|}{D-k},\qquad
 \sum_{(v,X):\,r_X(v)\le k}r_X(v)
 \le b_Fk\le\frac{2|F|k}{D-k}=O(|F|).
\end{equation}
These bounds control the deleted spokes and all recovered
edges, including duplicates. Owner lookup and grouping take
$\widetilde O(|F|+1)$ time. The failed-incidence lists passed
to sparse recovery have total length at most $2|F|$, and
the sum of the recovery capacities is $b_Fk=O(|F|)$.
Thus \Cref{lem:sketching-oracle} gives
$\widetilde O(|F|+1)$ total recovery time. Duplicate recovered
edges may be removed by sorting within the same time bound;
for connectivity, repeated union operations are harmless.

We next compare connectivity. Every surviving edge of $R$
is present in $J_F$. For a surviving edge $(u,v)$ of a
piece $X$, an endpoint of residual degree at most $k$
recovers the edge itself. Otherwise both spokes remain,
giving the path $u,c_X,v$. Hence every connection in $G-F$
is preserved in $J_F$.

Conversely, every original edge in $J_F$ survives in $G-F$.
Every passage through a center $c_X$ joins two vertices of
residual $X$-degree greater than $k$, which are connected
in $X-(F\cap E(X))$ by \Cref{lem:fast-robustness}.
Replacing each such passage by a surviving path proves
the converse.
\end{proof}

\begin{proof}[Proof of \Cref{thm:fast-connectivity}]
There are $O(n\log n)$ vertex--piece incidences and
$O(nD/\phi)=\widetilde O(nk)$ remainder edges. Hence $J$,
its base oracle, and all incidence records occupy
$\widetilde O(nk)=\widetilde O(\sqrt f\,n)$ bits.

The decomposition takes $(m+n)^{1+o(1)}$ time. Sketch
construction takes $\widetilde O(m)$ additional time,
because
\[
 \sum_{X\in\mathcal X}\sum_{v\in V(X)}d_X(v)\le2m,
 \qquad
 \sum_{X\in\mathcal X}k|V(X)|
 \le\frac{2k}{D}\sum_{X\in\mathcal X}|E(X)|=O(m).
\]
The remaining preprocessing, including the base oracle,
is polynomial.

By \Cref{lem:fast-repairs}, a query produces $O(|F|)$
deletions and insertions in $\widetilde O(|F|+1)$ time.
The base oracle processes the deletions within the same
bound. The union--find structure contains $O(|F|)$ identifiers
and performs $O(|F|)$ unions, each merging the components
joined by a recovered edge. Together with untouched identifiers,
its final sets represent $\operatorname{cc}(G-F)$ by
\Cref{lem:fast-repairs}.
Indexing these identifiers in a balanced dictionary and
using union by size gives $\widetilde O(|F|+1)$ total
processing time and $\widetilde O(1)$ time per identifier
query, without changing the stored state during such a query.
\end{proof}

\subsection{Approximate distances}
\label{subsec:fast-distances}

The same modification lists yield a sparse emulator of $G-F$.
Choose a positive integer $L=O(\phi^{-1}\log(2n))$ satisfying
\Cref{lem:fast-robustness}. In an unweighted graph, give every
spoke of $J$ length $L$ and every remainder or recovered edge
length $1$. Then, for original vertices,
\begin{equation}\label{eq:fast-unweighted-emulator}
 d_{G-F}(s,t)\le d_{J_F}(s,t)\le2L\,d_{G-F}(s,t).
\end{equation}
Indeed, a passage through a center costs $2L$ and can be
replaced by a surviving path of length at most $L$.
Conversely, every surviving original edge is represented
by itself or by a two-spoke path of length $2L$.

\paragraph{Weighted graphs.}
For lengths $\ell(e)\in[1,W]$, where $W=n^{O(1)}$, partition
the edges into classes
\[
 E_i=\{e:2^i\le\ell(e)<2^{i+1}\}.
\]
Apply the decomposition and sketch construction to each
nonempty class, using distinct centers and piece identifiers
across classes. A failed edge's supplied length identifies
its class, after which the stored indices identify its owner.
Give each spoke in class $i$ length $L2^{i+1}$, retain the
original lengths of remainder edges, and give recovered
class-$i$ edges length $2^{i+1}$. This rounding allows the
sketch to recover only edge identities, without storing
the lengths of the omitted edges.

The pieces remain edge-disjoint across all classes, so
\Cref{eq:fast-repair-count} still gives $O(|F|)$ changes
and $\widetilde O(|F|+1)$ total recovery time.
The $O(\log n)$ classes add only polylogarithmic factors
to the space bound.
A passage through a class-$i$ center represents a surviving
path of at most $L$ edges, each of length less than $2^{i+1}$.
In the other direction, a surviving class-$i$ edge is
represented with length at most
$2L2^{i+1}\le4L\ell(e)$. Consequently,
\begin{equation}\label{eq:fast-weighted-emulator}
 d_{G-F}(s,t)\le d_{J_F}(s,t)\le4L\,d_{G-F}(s,t).
\end{equation}
All auxiliary edge lengths are positive, integral, and
polynomially bounded.

\begin{proof}[Proof of \Cref{thm:fast-distances}]
Preprocess the weighted graph $J$ with the deterministic
fully dynamic distance oracle of Haeupler, Long, and
Saranurak~\cite[Theorem~12.1 of the full version]{HaeuplerLS24}.
On a graph with $N$ vertices and initial size $M_0$, for
$1/\log^c N\le\varepsilon\le1$, where $c>0$ is a fixed
constant, it has initialization time $O(M_0N^\varepsilon)$,
worst-case update time $O(N^\varepsilon)$, query time
$O(\log\log N/\varepsilon^4)$, and approximation
$2^{\operatorname{poly}(1/\varepsilon)}$.
Here size includes vertex records, so isolated vertices
are accounted for.

We have $N=\widetilde O(n)$ and
$M_0=\widetilde O(\sqrt f\,n)$.
Choose $\varepsilon=(\log\log N)^{-a}$ for a sufficiently
small fixed $a>0$, handling bounded $N$ directly.
Then the approximation is $\operatorname{polylog} n$ and
$N^\varepsilon=n^{o(1)}$.
Given $(s,t,F)$, compute and deduplicate the modification lists,
apply the $O(|F|)$ deletions and insertions to the dynamic oracle,
and query $(s,t)$ in the resulting graph $J_F$.
The total time is $(|F|+1)n^{o(1)}\le f n^{o(1)}$.
By \eqref{eq:fast-weighted-emulator}, the answer has
$\operatorname{polylog} n$ stretch for $G-F$ and reports
disconnection exactly.

The decompositions and sketches take $(m+n)^{1+o(1)}$
total time. Moreover, the number of spokes over all classes
is at most $2m/D$, since the pieces are edge-disjoint and
have minimum degree $D$. The remainder has at most $m$
edges, so $M_0=O(m+n)$ as well. Initialization therefore
takes $(m+n)^{1+o(1)}$ time.

The initialized RAM state occupies at most its
initialization time times the word length, hence
$\sqrt f\,n^{1+o(1)}$ bits. We record and undo all memory changes made during each query,
restoring the initialized state when the query finishes.
This preserves the query-time bound and uses only
$(|F|+1)n^{o(1)}$ additional bits, within the claimed space bound.
\end{proof}

The fast query bound comes from processing only the changes
to a preprocessed sparse graph: the expander stars represent
all vertices of large residual degree, while sparse recovery
supplies the few edges needed at the remaining incidences.

\bibliographystyle{alpha}
\bibliography{ref}

\appendix

\section{Missing proofs}\label{sec:missing}

\begin{proof}[Proof of \Cref{thm:fromconnectedtocomponents}]
First assume that $G$ is connected. Store a spanning tree $T$ in
addition to the Boolean encoding. Given $F$ with $|F|\le f$, query the
Boolean oracle on every subset of $F$. Let $\mathcal C_F$ be the family
of inclusion-minimal sets $C\subseteq F$ for which $G-C$ is
disconnected. For each $C\in\mathcal C_F$, the graph $G-C$ has exactly
two components: restoring any one edge of $C$ must reconnect the graph.
Moreover, $C$ consists precisely of all edges between these two
components.

We can recover the two sides of each such cut from $T$ and $C$.
Fix a root $r$ of $T$, and let $P_T(r,v)$ denote the edge set of the
unique $r$--$v$ path in $T$. Assign each vertex $v$ the bit
\[
 b_C(v)=|P_T(r,v)\cap C|\bmod 2.
\]
Every edge of $C$ on the path switches sides of the cut, while every
other edge of the path lies within one side, so $b_C(v)=0$ exactly when
$v$ lies in the same component of $G-C$ as $r$.

The decoder groups vertices according to their signatures
$(b_C(v))_{C\in\mathcal C_F}$. We claim that these groups are precisely
the components of $G-F$. Vertices connected in $G-F$ remain connected
in $G-C$ for every $C\subseteq F$, so they have identical signatures.
Conversely, suppose $u,v$ are disconnected in $G-F$, and choose an
inclusion-minimal set $C\subseteq F$ separating them. Restoring any
edge of $C$ reconnects $u$ and $v$, so every edge of $C$ joins their
respective components in $G-C$. Since $G$ is connected, $G-C$ has no
other component. Thus restoring any edge of $C$ reconnects the whole
graph, giving $C\in\mathcal C_F$ and $b_C(u)\ne b_C(v)$. This proves
the claim. All oracle queries are subsets of $F$, so they respect the
failure budget $f$. The additional stored information is the tree $T$,
which uses $O(n\log n)$ bits.

Now let $G$ be arbitrary, with components $V_1,\dots,V_c$. Choose a
representative $r_i\in V_i$ for each $i$, and let $D$ be a tree on
$\{r_1,\dots,r_c\}$. Since the $r_i$ lie in distinct components,
$D\cap E(G)=\emptyset$, and $G^+=G\cup D$ is a connected $n$-vertex
graph. Store the Boolean encoding of $G^+$, a spanning tree $T^+$ of
$G^+$ containing $D$, and the identities of the edges of $D$ within
$T^+$. Each component of $T^+-D$ uses only edges of $G$ and therefore
lies inside a single $V_i$. Since $T^+-D$ has exactly $c$ components,
these are precisely $V_1,\dots,V_c$, so the decoder can recover the
original components.

Given $F\subseteq E(G)$ with $|F|\le f$, we have $F\subseteq E(G^+)$.
Apply the connected-input reduction to $G^+$ with failure set $F$,
which yields the components of $G^+-F$. Output their nonempty
intersections with $V_1,\dots,V_c$. To see that this yields the
components of $G-F$, fix $i$. Edges of $G$ have both endpoints in the
same original component, and every edge of $D$ incident to $V_i$ has
endpoint $r_i$. Hence any path that leaves $V_i$ and later returns
must visit $r_i$ twice.
A simple path between two vertices of $V_i$ therefore never leaves
$V_i$. Consequently, two vertices of $V_i$ are connected in $G^+-F$ if
and only if they are connected in $G-F$. Vertices in distinct $V_i$ are
disconnected in $G-F$, so the output is exactly the set of components
of $G-F$.

All oracle queries are subsets of $F$, so the failure budget $f$ is
preserved. The scheme stores the Boolean encoding of $G^+$, which uses $s(n,f)$
bits, together with $T^+$ and the marking of $D$, which use
$O(n\log n)$ bits.
\end{proof}

\begin{proof}[Proof of \Cref{cl:coloring-equivalence}]
Suppose first that $\operatorname{Enc},\operatorname{Dec}$ form a correct encoding. If adjacent inputs $x,y$ had the same encoding, a common valid query witnessing their adjacency would give $\operatorname{Dec}$ identical inputs but require different answers. Hence $x\mapsto\operatorname{Enc}(x)$ is a proper coloring of $X$.

Conversely, suppose that $X$ admits a proper coloring with at most $M$ colors. Fix an ordering $x_1,\ldots,x_N$ of $\mathcal I$, independent of the encoded input, and let $c:\mathcal I\to[M]$ be the proper coloring for which $(c(x_1),\ldots,c(x_N))$ is lexicographically smallest, and set $\operatorname{Enc}(x)=c(x)$.

Given a color $i$ and a query $Q$, $\operatorname{Dec}$ proceeds as follows:
\begin{enumerate}
\item Enumerate all sequences $(c_1,\ldots,c_N)\in[M]^N$ in lexicographic order and take the first that defines a proper coloring of $X$.
\item Find the first input $y\in\mathcal I$ with $c(y)=i$ and $Q\in\mathcal Q(y)$. If none exists, return an arbitrary fixed answer.
\item Output $a(y,Q)$.
\end{enumerate}
For a valid query on $x$, the input $y$ in step~2 exists, and $a(y,Q)=a(x,Q)$ because $x,y$ have the same color. Thus the encoding is correct. The fixed-length bound follows since $b$ bits provide $2^b$ states.
\end{proof}

\section{Constructing the mask family $\mathcal M_n$}\label{sec:mask}

Without loss of generality, assume that $n$ is even. For odd $n$, add
an isolated vertex and restrict the resulting masks to the original
vertices.
To define the family of masks, fix a decomposition of $K_n$ into $n-1$ perfect matchings, each of size
$n/2$, using the standard round-robin construction:
\[
 E(K_n)=\bigsqcup_{i=1}^{n-1}\Pi_i.
\]
For $T\subseteq[n-1]$, define the mask $M_T$ by
$E(M_T)=\bigcup_{i\in T}\Pi_i$, and set $\mathcal M_n=\{M_T:T\subseteq[n-1]\}$.

We now prove \Cref{lem:mask-polynomial}.

\begin{proof}[Proof of (1)]
The matchings are nonempty and edge-disjoint, so distinct subsets $T$
define distinct masks. The characteristic vector of $T$ identifies
$M_T$ and allows its reconstruction in polynomial time.

\textit{Proof of (2) and (3).}
We use Zouzias's deterministic matrix-balancing algorithm~\cite[Theorem~4
and Algorithm~1]{Zouzias2012}.\footnote{In the notation of
\cite[Theorem~4]{Zouzias2012}, use $f_i(\pm1)=\pm A_i$, $t=n-1$,
$\gamma=\rho^2=1$, and
$\varepsilon=\tfrac12\sqrt{\log n/(n-1)}$. Uniform signs give
$\mathbb E f_i=0$ and $\|\mathbb E f_i^2\|_{\mathrm{op}}\le1$.} In the form needed here, its input is
$n-1$ symmetric rational $n\times n$ matrices $A_1,\ldots,A_{n-1}$ with
$\|A_i\|_{\mathrm{op}}\le1$. It finds signs
$\varepsilon_1,\ldots,\varepsilon_{n-1}\in\{-1,1\}$ in polynomial time
such that
\[
 \left\|\sum_{i=1}^{n-1}\varepsilon_i A_i\right\|_{\mathrm{op}}
 =O\bigl(\sqrt{n\log n}\bigr).
\]

Given $G$, let $A_i$ be the adjacency matrix of the graph with edge
set $E(G)\cap\Pi_i$, for each $i\in[n-1]$. Run the algorithm on these
matrices and let $\varepsilon_i$ be the returned signs. Set
$T=\{i:\varepsilon_i=-1\}$. If $|E(G\cap M_T)|>|E(G)|/2$, replace
$T$ by $[n-1]\setminus T$ and negate every $\varepsilon_i$.
Return the mask $M=M_T$.

We verify that $M$ is $O(n^{3/2}\sqrt{\log n})$-good for $G$. Each $A_i$ is the adjacency
matrix of a matching, so it is symmetric and $A_i^2$ is diagonal with
entries in $\{0,1\}$. Thus $\|A_i\|_{\mathrm{op}}\le1$, and the
algorithm applies. Put $X=\sum_{i=1}^{n-1}\varepsilon_i A_i$.
Negating all signs preserves $\|X\|_{\mathrm{op}}$, and
$X=A_G-2A_{G\cap M}$. Therefore, for every $S\subseteq[n]$,
\[
 \bigl||\delta_G(S)|-2|\delta_{G\cap M}(S)|\bigr|
 =\left|\mathbf1_S^{\mathsf T}X\mathbf1_{[n]\setminus S}\right|
 \le\sqrt{|S|(n-|S|)}\,\|X\|_{\mathrm{op}}
 \le\frac n2\|X\|_{\mathrm{op}}=O(n^{3/2} \sqrt{\log n}),
\]
where we used $\sqrt{|S|(n-|S|)}\le n/2$. The halving condition holds by construction, so $M$ is
$O(n^{3/2}\sqrt{\log n})$-good for $G$. All steps take polynomial time since Zouzias's algorithm takes polynomial time.
\end{proof}

\section{From unbounded to bounded failures}\label{sec:LB-reduction}

We assume that $f>100$, since the case of $f\le100$, or any other fixed bound on $f$, follows from the elementary $\Omega(n)$ lower bound. Our approach is based on the equivalence between confusion graph colorings, and encodings supporting valid queries, described in \Cref{sec:prelim}. We show that an unbounded-fault lower bound of $\widetilde\Omega(n^{1+\alpha})$ bits implies a lower bound of $\widetilde\Omega(nf^{\alpha/2})$ bits for at most $f$ faults. 

Recall that the \emph{OR product} of graphs $T_1,T_2$, denoted $T_1\vee T_2$, has vertex set $V(T_1)\times V(T_2)$, with $(u_1,u_2)$ adjacent to $(v_1,v_2)$ if $u_1v_1\in E(T_1)$ or $u_2v_2\in E(T_2)$. We write $T^{\vee k}$ for the OR product of $k$ copies of $T$; two $k$-tuples are adjacent precisely when their entries are adjacent in at least one coordinate.

\begin{claim}\label{cl:bounded-fault-reduction}
Fix $0<\alpha\le1$. If $\log_2\chi(X_n)=\widetilde\Omega(n^{1+\alpha})$, then $\log_2\chi(X_{n,f})=\widetilde\Omega(nf^{\alpha/2})$.
\end{claim}
\begin{proof}
We may assume that $f$ is a perfect square: otherwise replace $f$ by the largest square below it. Likewise, assume that $k=n/\sqrt f$ is an integer; otherwise replace $n$ by the largest multiple of $\sqrt f$ that is smaller than or equal to $n$. These roundings affect the desired lower bound by at most a constant factor.

We claim that $X_{\sqrt f}^{\vee k}$ is isomorphic to an induced subgraph of $X_{n,f}$. Partition $[n]$ into $k$ fixed blocks of $\sqrt f$ vertices and map each tuple $(G_1,\ldots,G_k)$ to the disjoint union of its graphs on these blocks.

To verify the claim, consider two tuples $(G_1,\ldots,G_k)$ and $(H_1,\ldots,H_k)$. If they are adjacent in the OR product, then $G_i,H_i$ admit a common valid distinguishing query $F$ for some $i$. Since $|F|\le\binom{\sqrt f}{2}\le f$, the same query distinguishes the two disjoint unions. Conversely, suppose a common valid query $F$ distinguishes the disjoint unions. As every connected component lies within a single block, their component partitions must differ on some block $i$. The restriction of $F$ to that block is then a common valid query distinguishing $G_i,H_i$, so the tuples are adjacent in the OR product.

It is well known that for some absolute constant $C$, we have $\chi(T^{\vee k})\ge(\chi(T)/(C\log(2|V(T)|)))^k$ (see properties~(i),~(ii) in~\cite[Section~2]{AlonEtAl2008}). Since $|V(X_{\sqrt f})|=2^{\binom{\sqrt f}{2}}$, we get:
\begin{align*}
    \log_2\chi(X_{n,f})
    &\ge \log_2\chi(X_{\sqrt f}^{\vee k})\\
    &\ge k\bigl(\log_2\chi(X_{\sqrt f})-O(\log f)\bigr)\\
    &=\widetilde\Omega\!\left(\frac n{\sqrt f}\ \cdot (\sqrt{f})^{(1+\alpha)}\right)
    =\widetilde\Omega(nf^{\alpha/2}).
\end{align*}
\end{proof}

\section{Deterministic Sparse-Recovery Sketches}
\label{sec:deterministic-sparse-recovery}
\label{sec:proof-sketching-oracle}

The proof uses two deterministic algebraic algorithms as black boxes.

\begin{lemma}[Algebraic black boxes]
\label{lem:algebraic-black-boxes}
The following deterministic algorithms are available.
\begin{enumerate}[label=(\roman*)]
    \item \emph{Fast recurrence recovery.}
    Let $a_1,\ldots,a_{2K}$ be elements of a field $\mathbb K$.  If the
    shortest recurrence of this sequence has order $t\leq K$,\footnote{For
    the finite sequence $a_1,\ldots,a_{2K}$, a recurrence of order $d$ is a
    monic polynomial $C(z)=z^d+c_{d-1}z^{d-1}+\cdots+c_0$ satisfying
    $a_{j+d}+c_{d-1}a_{j+d-1}+\cdots+c_0a_j=0$ for
    $j=1,\ldots,2K-d$.  A shortest recurrence has minimum order.} then the
    half-gcd implementation of Berlekamp--Massey returns $t$ and the
    recurrence coefficients in $\widetilde O(K)$ operations over $\mathbb K$;
    see Massey~\cite{Massey69} and von zur Gathen and
    Gerhard~\cite[Section~5.9 and Chapter~11]{GathenGerhard13}.

    \item \emph{Root finding over a small-characteristic extension.}
    A degree-$d$ polynomial $P\in\mathbb F_{p^r}[z]$ \emph{splits} over
    $\mathbb F_{p^r}$ if $P(z)=c\prod_{i=1}^d(z-a_i)$ for
    $c\in\mathbb F_{p^r}^*$ and $a_1,\ldots,a_d\in\mathbb F_{p^r}$; it is
    \emph{squarefree} if the $a_i$'s are distinct.  For fixed $p$, Petit's
    deterministic
    successive-resultants algorithm\footnote{Petit conditions determinism on
    the constituent resultant and small-degree root-finding routines.  We use
    deterministic resultant computation.  For our $p=2$ instantiation, each
    nonlinear small-root problem is a squarefree quadratic $z^2+az+c$ with
    $a\ne0$.  Substituting $z=ay$ reduces it to $y^2+y=c/a^2$.  Over
    $\mathbb F_{2^r}$, the map $y\mapsto y^2+y$ is $\mathbb F_2$-linear
    with kernel $\{0,1\}$.  A binary linear solver is precomputed in
    $O(r^3)$ operations; each subsequent solve costs $O(r^2)$ and yields
    at most two candidates.  The recovery tree for $d$ distinct roots has
    at most $d-1$ quadratic branching calls; linear polynomials are solved
    directly.  Thus this deterministic implementation preserves the stated
    $\widetilde O(dr^2+r^3)$ bound~\cite[Sections~3.2 and~3.5]{Petit14}.} lists all
    roots of a squarefree polynomial that splits over $\mathbb F_{p^r}$ in
    $\widetilde O(dr^2+r^3)$ operations over $\mathbb F_p$; see
    Petit~\cite[Section~4.3]{Petit14}.
\end{enumerate}
\end{lemma}

\begin{proof}[Proof of Lemma~\ref{lem:sketching-oracle}]
We first define the linear sketch and prove its injectivity on sparse sets.
We then give the decoder and, separately, show how to form the sketches
within the claimed preprocessing and query times.

\paragraph{Edge labels.}
Number the vertices by $[n]$ and set
\[
    b=\left\lceil\log_2(n^2+1)\right\rceil,
    \qquad
    \mathbb F=\mathbb F_{2^b}.
\]
Represent $\mathbb F$ in a polynomial basis, which can be constructed
deterministically in $(\log n)^{O(1)}$ time using Shoup's algorithm \cite{Shoup90}.  For an
edge $e=\{x,y\}$, define
\begin{equation}
    \operatorname{idx}(e)
      = (\min\{x,y\}-1)n+\max\{x,y\}\in[n^2],
    \label{eq:canonical-edge-coordinate}
\end{equation}
and let $\alpha_e\in\mathbb F^*$ be the field element whose $b$-bit
polynomial-basis representation is the binary expansion of $\operatorname{idx}(e)$.  The inequality
$2^b>n^2$ ensures that these labels are distinct and nonzero.  Moreover,
if $i=\operatorname{idx}(e)$, then the endpoints $x=\min e$ and $y=\max e$
are recovered by
\[
    x=\left\lfloor\frac{i-1}{n}\right\rfloor+1,
    \qquad
    y=((i-1)\bmod n)+1.
\]
Thus a label can be decoded into its edge in $O(1)$ word operations.  Since
$b=O(\log n)$, a field element occupies $O(1)$ machine words, and each
field operation takes $(\log n)^{O(1)}$ word operations.

\paragraph{The sketch and its injectivity.}
For $A\subseteq E$ and $j\geq1$, define
\begin{equation}
    S_j(A)=\sum_{e\in A}\alpha_e^j,
    \qquad
    \sigma_k(A)=\bigl(S_1(A),\ldots,S_{2k}(A)\bigr).
    \label{eq:power-sum-sketch}
\end{equation}
The use of positive powers and nonzero labels ensures injectivity on sparse
sets.  Indeed, let $C\subseteq E$ be nonempty and let
$r=|C|\leq2k$.  Its first $r$ power sums satisfy
\[
    \bigl(S_1(C),\ldots,S_r(C)\bigr)^{\mathsf T}
      =M_C\mathbf 1,
    \qquad
    M_C=(\alpha_e^j)_{j\in[r],e\in C}.
\]
Writing
\[
    M_C=V_C\operatorname{diag}(\alpha_e:e\in C),
\]
where $V_C$ is the Vandermonde matrix with exponents $0,\ldots,r-1$, shows
that $M_C$ is nonsingular: the labels are distinct and nonzero.  Hence
$M_C\mathbf 1\neq0$, so no nonempty set of at most $2k$ edges has an
all-zero sketch.

Because $\mathbb F$ has characteristic two,
\[
    \sigma_k(A)+\sigma_k(B)=\sigma_k(A\mathbin{\triangle}B).
\]
It follows that $\sigma_k(A)=\sigma_k(B)$ and
$|A\mathbin{\triangle}B|\leq2k$ imply $A=B$.  In particular,
$\sigma_k$ is injective on all edge sets of size at most $k$.

For each vertex $v$, store
\[
    s(v)=\sigma_k(E_G(v)),
\]
where $E_G(v)$ is the set of edges incident to $v$.  Viewing $\mathbb F$ as
a $b$-dimensional vector space over $\mathbb F_2$ makes $s(v)$ an
$\mathbb F_2$-linear sketch of the incidence vector of $E_G(v)$.  Given a
set $F_v\subseteq E_G(v)$ of failed incident edges, let
$R=E_G(v)\setminus F_v$.  Characteristic-two cancellation gives
\begin{equation}
    \sigma_k(R)=s(v)+\sigma_k(F_v).
    \label{eq:residual-power-sum-sketch}
\end{equation}

\paragraph{Decoding.}
Assume the promise $t:=|R|\leq k$.  If the residual sketch is zero, the
injectivity proved above implies $R=\varnothing$.  Suppose therefore that
$t\geq1$, and define the locator polynomial
\begin{equation}
    P_R(z)=\prod_{e\in R}(z-\alpha_e)
      =z^t+c_{t-1}z^{t-1}+\cdots+c_0.
    \label{eq:residual-locator-polynomial}
\end{equation}
For every $1\leq j\leq2k-t$, evaluating $P_R$ at its roots yields
\begin{equation}
    S_{j+t}(R)+c_{t-1}S_{j+t-1}(R)+\cdots+c_0S_j(R)
      =\sum_{e\in R}\alpha_e^jP_R(\alpha_e)=0.
    \label{eq:residual-power-sum-recurrence}
\end{equation}
Thus $P_R$ is a recurrence for the $2k$ stored power sums.  We next show
that it is their unique shortest recurrence.

Suppose that a polynomial $Q$ of degree $d<t$ were a recurrence.  Since
$d<t\leq k$, we have $2k-d\geq t$, so its recurrence equations are available
for $j=1,\ldots,t$.  Substituting the power sums gives
\[
    \sum_{e\in R}\alpha_e^jQ(\alpha_e)=0,
    \qquad j=1,\ldots,t.
\]
The coefficient matrix $(\alpha_e^j)_{j\in[t],e\in R}$ is the nonsingular
scaled Vandermonde matrix above.  Hence $Q(\alpha_e)=0$ for every $e\in R$,
contradicting $\deg Q<t$.  Likewise, the difference of two monic degree-$t$
recurrences has degree less than $t$ and satisfies the same $t$ equations,
because $2k-t\geq t$.  The two recurrences must therefore coincide.

By Lemma~\ref{lem:algebraic-black-boxes}(i), Berlekamp--Massey recovers
$P_R$ from the residual sketch using $\widetilde O(k)$ field operations.
The polynomial $P_R$ splits over $\mathbb F$, and it is squarefree because
the labels $\alpha_e$ are distinct.  Lemma~\ref{lem:algebraic-black-boxes}(ii),
with $p=2$, $r=b=O(\log n)$, and $d=t\leq k$, lists its roots in
$k(\log n)^{O(1)}$ deterministic time.  These roots are exactly the labels
of the residual edges.  Inverting~\eqref{eq:canonical-edge-coordinate} and
taking the endpoint opposite $v$ reports $N_{G- F_v}(v)$.  Thus, if
$\deg_{G- F_v}(v)\leq k$, the decoder reports all surviving neighbors
of $v$; no guarantee is required otherwise.

\paragraph{Forming the sketches.}
It remains to justify the claimed time bounds for constructing $s(v)$ during
preprocessing and for computing $\sigma_k(F_v)$ at query time.  Let
$\mathsf M(d)$ denote the number of field operations needed to multiply two
degree-$d$ polynomials.  Fast polynomial multiplication gives
$\mathsf M(d)=\widetilde O(d)$~\cite[Ch.~8]{GathenGerhard13}.  For any edge
set $A$, construct
\[
    Q_A(z)=\prod_{e\in A}(1-\alpha_e z)
\]
with a balanced product tree.  At each level, the degrees of the factors sum
to $|A|$, so the level costs $\widetilde O(|A|)$ field operations.  Summing
over the $O(\log |A|)$ levels gives $\widetilde O(|A|)$ field operations in
total.  In characteristic two, the logarithmic derivative satisfies
\[
    \frac{Q_A'(z)}{Q_A(z)}
      =\sum_{e\in A}\frac{\alpha_e}{1-\alpha_e z}
      =\sum_{j\geq0}S_{j+1}(A)z^j.
\]
To compute this series modulo $z^{2k}$, note that $Q_A(0)=1$ and hence
$Q_A$ is invertible as a formal power series.  Starting with $R_1=1$, the
Newton update
\[
    R_{2d}=R_d(2-Q_AR_d)\bmod z^{2d},
    \qquad R_d\equiv Q_A^{-1}\pmod{z^d},
\]
doubles the precision; indeed,
$1-Q_AR_{2d}\equiv(1-Q_AR_d)^2\pmod{z^{2d}}$, so the update remains valid
in characteristic two.  Each update costs $O(\mathsf M(d))$ field
operations, and summing over the geometrically increasing precisions gives
$O(\mathsf M(k))=\widetilde O(k)$ field operations.  Multiplying the
resulting inverse by $Q_A'$ modulo $z^{2k}$ has the same cost.  The first
$2k$ coefficients of the product are $S_1(A),\ldots,S_{2k}(A)$, so
$\sigma_k(A)$ can be formed in
\[
    (|A|+k)(\log n)^{O(1)}
\]
word operations.

Applying this procedure with $A=F_v$ and then using
\eqref{eq:residual-power-sum-sketch} proves the
$(|F_v|+k)(\log n)^{O(1)}$ query bound in part~(ii).  If the residual sketch
has already been formed, only the $k(\log n)^{O(1)}$ decoding time remains,
which proves part~(iii).  For preprocessing, applying the procedure once to
each incident-edge set gives total time
\[
    \sum_{v\in V}\bigl(|E_G(v)|+k\bigr)(\log n)^{O(1)}
      =(2m+nk)(\log n)^{O(1)}
      =(m+nk)(\log n)^{O(1)}.
\]
Finally, each sketch contains $2k$ elements of $\mathbb F_{2^b}$, each represented by $b=O(\log n)$ bits, and therefore uses $O(k\log n)$ bits.  All $n$ sketches use
$O(nk\log n)$ bits.  This proves parts~(i) and~(iv) and completes the proof.
\end{proof}

\end{document}